\documentclass[aps,prb,amsmath,amssymb,floatfix,
superscriptaddress,nofootinbib,twocolumn]{revtex4-2}

\usepackage{graphicx}
\usepackage{bm}
\usepackage{hyperref}
\usepackage{braket}
\usepackage{color} 
\usepackage[ampersand]{easylist}
\usepackage{xcolor}
\usepackage{outlines}

\newcommand{\cO}{\mathcal{O}}

\newcommand{\cH}{\mathcal{H}}
\newcommand{\Tr}{\mathrm{Tr}}

\newcommand{\oL}{\Lambda}
\newcommand{\oT}{\Omega}

\newcommand{\bg}{\bar{g}}
\newcommand{\bh}{\bar{h}}
\newcommand{\bk}{\bar{k}}

\newcommand{\bs}{\bar{s}}
\newcommand{\bq}{\bar{q}}

\newcommand{\fp}{\mathfrak{p}}

\begin{document}

\title{Symmetry-twisted multi-entropies: Order parameters for 2D SPT Phases}

\author{Ramanjit Sohal}
\affiliation{Pritzker School of Molecular Engineering, University of Chicago, Chicago, IL 60637, USA}
\author{Michael Levin}
\affiliation{Leinweber Institute for Theoretical Physics, University of Chicago, Chicago, IL 60637, USA}
\author{Ruben Verresen}
\affiliation{Pritzker School of Molecular Engineering, University of Chicago, Chicago, IL 60637, USA}

\date{\today}

\begin{abstract}
Although symmetry-protected topological phases (SPTs) can be distinguished by their entanglement properties, it has been unclear how to extract this information directly from 
expectation values beyond the 1D case. 
Here, we close this gap and propose a pair of nonlocal order parameters that can detect and distinguish all bosonic SPTs in 2D protected by internal discrete, Abelian unitary symmetries. The desired topological invariants are extracted by these quantities by effectively simulating the SPT path integral on topologically non-trivial spacetime manifolds. 
Our order parameters are defined in terms of expectation values of partial symmetry and permutation operations acting on fixed numbers of replicas of the system in finite spatial regions. These expectation values correspond to symmetry-twisted versions of 
multipartite entanglement quantities known as \emph{multi-entropies}. 
We show explicitly that our two order parameters detect symmetry-protected four-party and six-party entanglement, respectively, and we constrain possible ``spurious" contributions. We analytically test our proposal in fixed-point lattice models. 
Our results suggest multipartite entanglement to be a defining feature of SPTs; indeed, we expect our methods to generalize to fermionic and higher-dimensional systems.
\end{abstract}

\maketitle

\section{Introduction}

The study of many-body entanglement is now central to our understanding of quantum phases of matter \cite{wen2007,zeng2019,sachdev2023}. 
Indeed, quantitative probes of entanglement often serve as \emph{nonlocal} order parameters for phases lying beyond the Landau paradigm with the topological entanglement entropy serving as the quintessential example \cite{kitaev2006,levin2006}. 
Surprisingly, symmetry protected topological phases (SPTs) \cite{gu2009,pollmann2010,chen2011,schuch2011,pollmann2012,chen2013,senthil2015} have proven resistant to characterizations of this nature, in spatial dimensions greater than one.
These are short-range correlated phases of matter which, nevertheless, cannot be adiabatically connected to a completely trivial state, provided the protecting symmetry is not broken. Although the boundary of an SPT is ``anomalous", supporting long-range entangled order like gapless excitations, its bulk is completely 
invisible to any local correlators. 
This presents a puzzle: the bulk must  ``know" it is in an SPT phase and it must also know the \emph{topological invariants} that define it, but it is not clear how to extract this information.

At the same time, the fact that SPTs can only be connected to trivial states via symmetry breaking evolution implies they are characterized by symmetry-protected (short-range) entanglement. 
The question to be answered is then two-fold: can one construct bulk order parameters, which must be \emph{nonlocal} by the above, that (i) probe the topological invariants of an SPT and (ii) quantify this entanglement?
In 1D the answer is known to be yes: 
string order parameters and other nonlocal order parameters \cite{perezgarcia2008,haegeman2012,SOP_1D,shiozaki2017,inamura2020} can be constructed, which detect SPT order and probe the resulting symmetry-protected degeneracies in the entanglement spectrum \cite{pollmann2010}.
In 2D, this paradigm of nonlocal order parameters has been extended to only certain classes of SPTs, namely,
crystalline SPTs, i.e., those protected by spatial symmetries \cite{shiozaki2017a,zhang2023,manjunath2024,kobayashi2025,kobayashi2025a,calvera2025,manjunath2026} and SPTs protected by antiunitary symmetries \cite{shapourian2017,shiozaki2018}.

\begin{figure}
  \centering
    \includegraphics[width=0.9\linewidth]{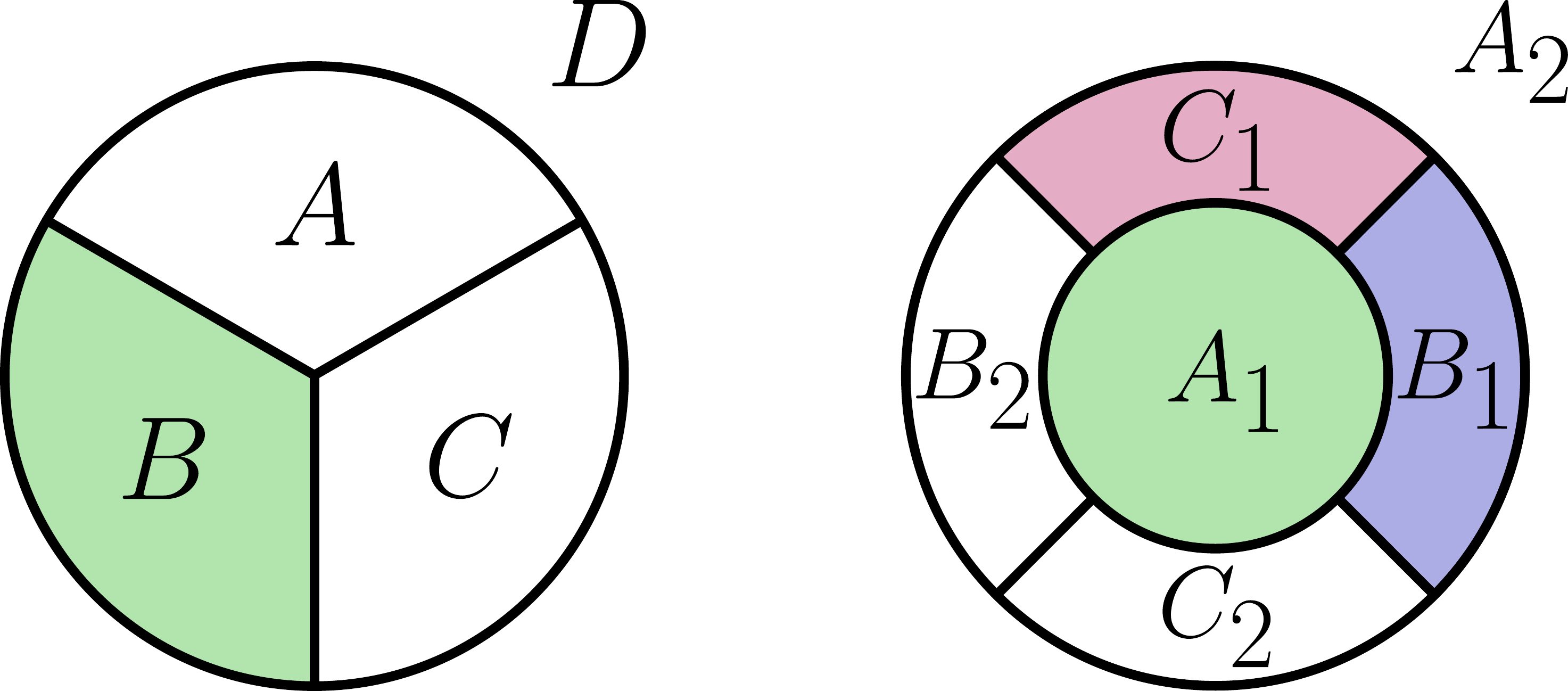}
    \caption{Multipartitions and regions (indicated by color) where partial symmetry operations are applied for the (Left) Type-I/II and (Right) Type-III order parameters $\oL_g^r$ and $\oT_{g,h,k}$, respectively. }\label{fig:partitions}
\end{figure}

Surprisingly, the seemingly more basic situation of 2D SPTs protected by \emph{internal} symmetries remains poorly understood.
Although a complete, finite set of topological invariants characterizing SPTs protected by an arbitrary, on-site symmetry is not known, such a finite set does exist for discrete, Abelian symmetry groups \cite{zaletel2014,wang2015}; as such, one might hope that a complete set of order parameters for Abelian SPTs could be constructed. Indeed, prior works have proposed schemes to extract these invariants using explicit flux insertions \cite{zaletel2014}, membrane order parameters \cite{perezgarcia2008}, and combinations of partial on-site and crystalline symmetries \cite{turzillo2025}, among others \cite{matsuura2016,marvian2017,cian2021,dehghani2021,zhang-entangler2023,liu2025,xu2025,sala2025}. 
However, these proposals suffer from a range of issues; in particular, most of these probes do not exhaustively identify the SPT class, and the ones that do require modifying the Hamiltonian and are thus not purely wavefunction-based.

In the present work, we resolve these issues by introducing a pair of order parameters capable of distinguishing \emph{all} 2D bosonic SPTs protected by internal Abelian unitary symmetries, taking as input the wavefunction on a disk, as shown in Fig.~\ref{fig:partitions}, and the protecting symmetries. 
The key physical insight is that the topological invariants characterizing 2D Abelian bosonic SPTs correspond to the braiding statistics of symmetry defects. Heuristically, our order parameters are constructed to effectively simulate these braidings. More precisely, our order parameters simulate the SPT partition function on nontrivial spacetime manifolds, allowing us to extract the topological response governing the braiding statistics without actually explicitly inserting a flux.
Our construction is thus partly in the spirit of Refs.~\onlinecite{SOP_1D,shiozaki2017,inamura2020,shiozaki2017a,zhang2023,manjunath2024,kobayashi2025,kobayashi2025a,calvera2025,shapourian2017,shiozaki2018,cian2021,dehghani2021}, which used expectation values of nonlocal operators
to simulate SPT partition functions.  

The essential technical advance that allows us to go beyond these prior works and probe 2D SPTs protected by \emph{internal} symmetries is to make use of a \emph{replica permutation} symmetry. 
Analogous to how R\'enyi entanglement entropies are computed by taking multiple copies, or \emph{replicas}, of a wave function \cite{holzhey1994,calabrese2009,hastings2010,islam2015}, our order parameters are expressed as expectation values of replica permutation and partial symmetry operations in a replicated wavefunction. 
This may be viewed as a natural generalization of Refs.~\onlinecite{haegeman2012,SOP_1D,shiozaki2017}, which applied this perspective to 1D SPTs.
Interestingly, our order parameters correspond to \emph{symmetry-twisted} versions \cite{goldstein2018,azsez2020} of recently introduced multipartite entanglement probes known as \emph{multi-entropies} \cite{gadde2022,penington2023,gadde2026}.
Indeed, we will show how our order parameters allow us to establish that 2D bosonic Abelian SPTs are characterized by specific forms of \emph{four}-party and \emph{six}-party entanglement, in a sense we will make quantitatively precise.

Our work thus adds to a growing body of literature demonstrating the utility of multipartite entanglement probes in characterizing many-body states, particularly 2D topologically ordered phases of matter 
\cite{siva2022,liu2022,kim2022,kim2022a,fan2022,fan2022a,fan2023,sohal2023,liu2024,liu2025,sheffer2025,sheffer2026,gass2026,delzotto2026,sheffer2026a}. In particular, recent work has shown that universal data such as the chiral central charge \cite{kim2022,fan2022,fan2023,sheffer2026,gass2026}, electric Hall conductance \cite{fan2022a}, and anyon braiding statistics \cite{sheffer2025,delzotto2026} is hidden in the multipartite correlations of a wavefunction in a topological phase. It is through the consideration of symmetry-twisted multi-entropy probes that we are able to extend this story to 2D SPTs protected by internal symmetries; as we explain, we expect our framework to extend to fermionic and higher-dimensional SPTs. 
Our work thus suggests that symmetry-protected multipartite entanglement is a defining feature of SPTs protected by internal symmetries. More broadly, our work may be viewed as the initiation of the study of symmetry-twisted multi-entropies more generally, which we anticipate will serve as useful probes of generic many-body systems, beyond just gapped phases of matter.

In the balance of this paper, we first motivate the structure of symmetry-twisted multi-entropies and introduce our order parameters in Section~\ref{sec:order-parameters}. In Section~\ref{sec:multipartite}, we quantitatively demonstrate the sense in which they probe symmetry protected multipartite entanglement. We then substantiate our claim that these order parameters simulate the SPT partition function and extract the corresponding topological invariants using topological quantum field theory (TQFT) methods in Section~\ref{sec:tqft}. In Section~\ref{sec:lattice}, we analytically verify our claims in a set of lattice models realizing all SPTs of interest. We conclude with a discussion of our results in Section~\ref{sec:discussion}. We include several Appendices containing technical details.

\section{Symmetry-Twisted Multi-entropies as SPT Order Parameters}\label{sec:order-parameters}

\subsection{SPTs and Symmetry-Twisted Multi-Entropies}

Our aim is to construct nonlocal order parameters for SPTs realized in two-dimensional bosonic systems with a discrete, Abelian onsite symmetry $G$; we may always write such a symmetry group as a product of cyclic factors, $G = \prod_{i=1}^D \mathbb{Z}_{N_i}$, where $D$ and $N_i$ are integers.
It is well-established that the classification of such SPTs is provided by the \emph{group cohomology} of the symmetry group \cite{chen2011}; explicitly, distinct SPT phases are labeled by inequivalent group cohomology classes $[\omega] \in \cH^3[G,U(1)]$ (see Appendix~\ref{app:group-cohomology} for a review).
In the present case in which $G$ is Abelian, the group cohomology can be explicitly computed as,
\begin{align}
	\cH^3[G,U(1)] = \prod_i \mathbb{Z}_{N_i} \times \prod_{i<j} \mathbb{Z}_{N_{ij}} \times \prod_{i<j<k} \mathbb{Z}_{N_{ijk}}, \label{eq:kunneth}
\end{align}
where $N_{ij} = \mathrm{gcd}(N_i,N_j)$ and $N_{ijk} = \mathrm{gcd}(N_i,N_j,N_k)$. The three factors in this expression correspond to the classification of three distinct classes of SPTs called Type-I, II, and III SPTs, respectively \cite{propitius1997,wangsantos2015}. In particular, note that a Type-I SPT is protected by a $\mathbb{Z}_N$ symmetry, a Type-II SPT by a $\mathbb{Z}_N \times \mathbb{Z}_M$ symmetry, and similarly a Type-III SPT requires a $\mathbb{Z}_N \times \mathbb{Z}_M \times \mathbb{Z}_L$ symmetry.

Despite being defined in rather abstract terms, these SPT classes are physically distinguished by their \emph{topological} responses. In particular, symmetry fluxes in these SPTs acquire non-trivial braiding statistics \cite{levin2012}. 
Specifically, a $\mathbb{Z}_N$ flux in a Type-I $\mathbb{Z}_N$ SPT acquires a non-trivial spin, while $\mathbb{Z}_M$ and $\mathbb{Z}_N$ fluxes acquire Abelian mutual braiding statistics in a Type-II $\mathbb{Z}_M \times \mathbb{Z}_N$ SPT \cite{cheng2014,wen2014}; in this sense, symmetry fluxes in Type-I/II SPTs are often described as trapping fractional symmetry charge. Finally, in a Type-III SPT characterized by a non-trivial element of $\mathbb{Z}_{N_{ijk}}$, a $\mathbb{Z}_{N_i}$ flux traps a \emph{zero-mode} which transforms projectively under $\mathbb{Z}_{N_j}\times \mathbb{Z}_{N_k}$ (likewise for permutations of $i,j,k$) and hence acquires non-Abelian braiding statistics \cite{chen2014,wangsantos2015}.

The braiding statistics of symmetry fluxes serve as topological invariants characterizing the three types of Abelian SPTs \cite{wang2015}. 
Indeed, 
the measurement of these braiding phases
provides a physical protocol for detecting and characterizing these SPTs \cite{zaletel2014}. However, it must be emphasized that the flux insertion itself requires modifying the \emph{Hamiltonian}.
We expect that information about the SPT phase should be encoded in the wavefunction itself; our goal is to extract this data with expectation values of (nonlocal) operators.
To that end, we introduce two such order parameters below which we will argue accomplish precisely this, requiring information about the wavefunction only on a finite spatial patch larger than the correlation length. As we later argue, these order parameters may be understood as simulating the above flux response without an explicit flux insertion. Importantly, our order parameters are independent of the topology of the space on which the wavefunction is defined.

Before introducing these order parameters explicitly, let us begin by explaining the general structure behind their construction and the connection to existing probes of multipartite entanglement. 
Indeed, our order parameters will correspond to specific examples of what we call \emph{symmetry-twisted multi-entropies}. 
These are constructed by partitioning space into multiple subregions, taking multiple replicas (i.e. copies) of the state in question, and computing the expectation value of \emph{replica permutation} and \emph{partial symmetry} operators applied to the different subregions.
Explicitly, consider a state $\ket{\psi}$, invariant under the symmetry group $G$ with unitary representation $U_g$ for $g\in G$. Suppose we partition space into $K$ regions, $A_i$, with $i=1, \dots , K$ (with two such examples given in Fig.~\ref{fig:partitions}) and take $R$ replicas of the wavefunction. We define a symmetry-twisted multi-entropy to be an expectation value of the form,
\begin{align}
    \cO(\psi) = \bra{\psi}^{\otimes R} \bigotimes_{i=1}^K \pi_{A_i}^{i} \bigotimes_{a=1}^R \bigotimes_{i=1}^K U_{g_{i,a}}^{A_i} \ket{\psi}^{\otimes R}. \label{eq:stme}
\end{align}
Here, $i$ and $a$ index the different spatial subregions and replicas, respectively. 
A partial symmetry operator $U_g^A$ is the truncation of the global symmetry operator $U_g$, with $g\in G$, to subregion $A$. We restrict ourselves to onsite symmetries, i.e. we can write $U_g = \prod_{x} U_{g,x}$ with $x$ denoting a lattice site, such that there always exists a canonical choice of such a truncation to $A$ as $U_g^A = \prod_{x\in A} U_{g,x}$. The operators $\pi^{i}$ are replica permutation operators, acting on the replica index of the replicated wavefunction with elements of the permutation symmetry group $S_{R}$; similarly, $\pi_A^i$ is a permutation operator restricted to region $A$. 

Note that the R\'enyi entanglement entropy is a special instance of Eq.~\eqref{eq:stme}, with no partial symmetries and a single cyclic permutation acting on one subregion. 
For this reason, quantities of the form of Eq.~\eqref{eq:stme} without partial symmetries are referred to as \emph{multi-entropies}, serving as multipartite generalizations of the R\'enyi entropies.\footnote{Note that expectation values of general permutation operators in replicated wavefunctions are also called multi-invariants \cite{gadde2026}, with the multi-entropy referring to a particular choice of permutations \cite{penington2023,gadde2022}; we will follow the convention in the condensed matter literature \cite{sheffer2025,sheffer2026} of referring to all such quantities as multi-entropies.}
The study of multi-entropies (for a particular set of permutations) was initiated in Refs.~\onlinecite{penington2023,gadde2022} for the purpose of studying multipartite entanglement in holography and later generalized to arbitrary permutation operators \cite{gadde2026}. These quantities have found utility in probing universal features of chiral and topologically ordered phases to which probes of bipartite entanglement, like the entanglement entropy, are insensitive \cite{sheffer2025,sheffer2026,gass2026}. Much as how one may study symmetry-resolved versions of the entanglement entropy \cite{goldstein2018,azsez2020}, the symmetry-twisted multi-entropies we have introduced in Eq.~\eqref{eq:stme} yield a further generalization of the R\'enyi entropy, which we expect to provide symmetry-resolved probes of multipartite entanglement. 
Indeed, we will show how two specific choices of such quantities are sensitive to 2D SPT order. 

\subsection{The Order Parameters}

\begin{figure}
  \centering
    \includegraphics[width=\linewidth]{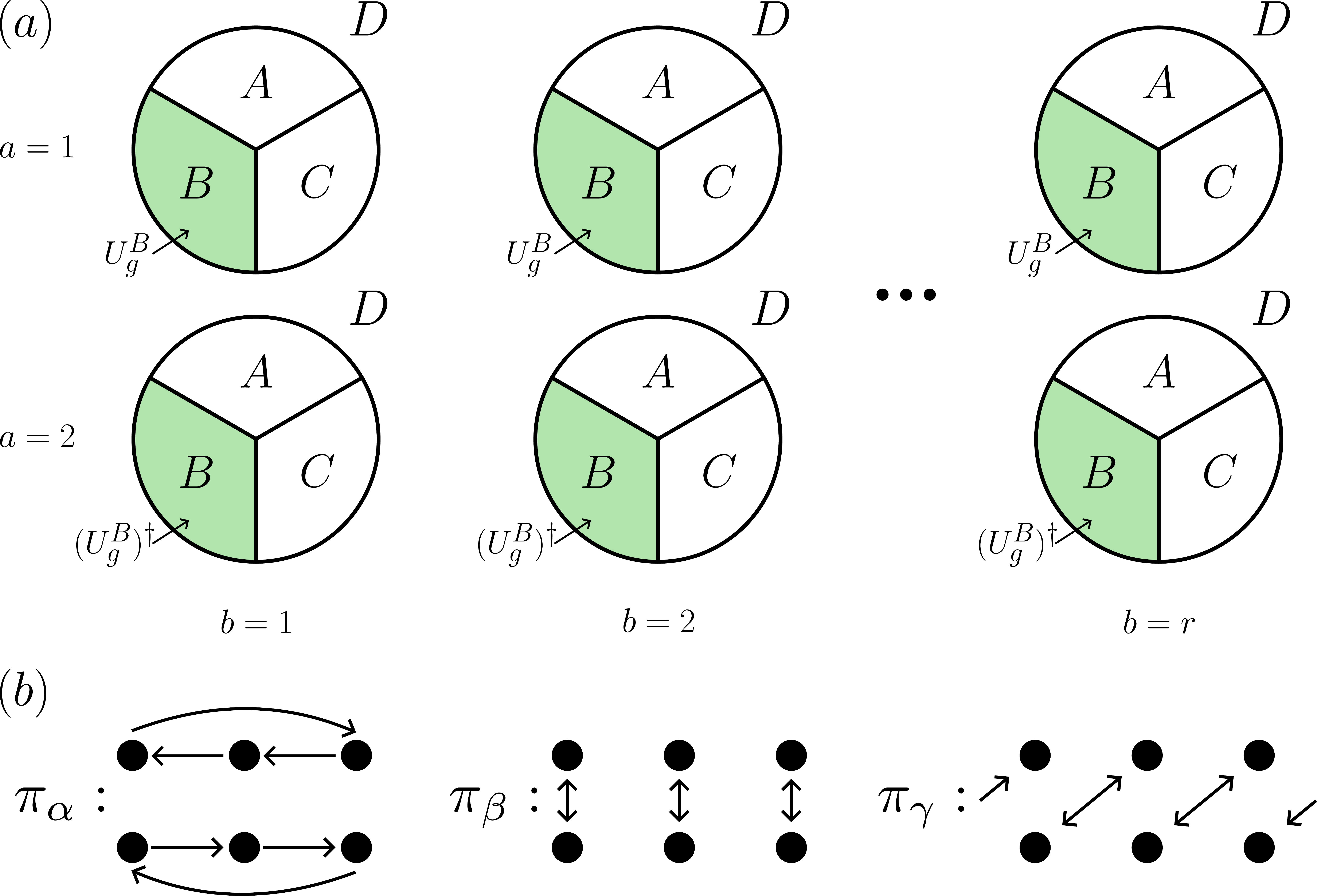}
    \caption{(a) The $2r$ replicas, labeled by $(a,b)$ and the partial symmetry actions in the definition of $\oL_g^r$. 
    (b) A graphical representation of the permutations in Eq.~\eqref{eq:type-I-permutations} in the case with $r=3$. 
    }\label{fig:type-I}
\end{figure}

\begin{figure}
  \centering
    \includegraphics[width=0.95\linewidth]{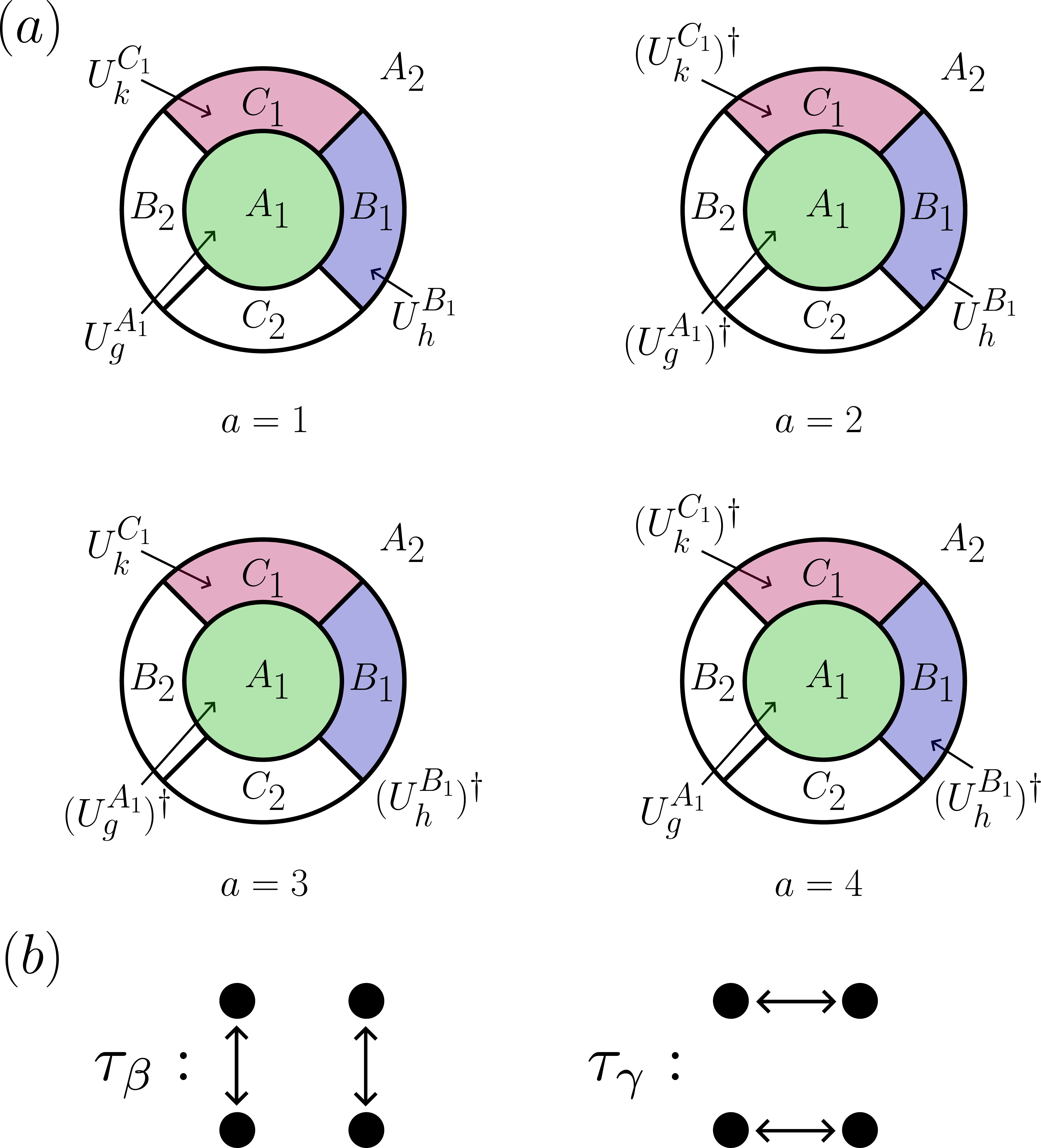}
    \caption{(a) The four replicas, labeled by $a=1,\dots, 4$ 
    and the partial symmetry operations in the definition of $\oT_{g,h,k}$. 
    (b) The action of the permutations in Eq.~\eqref{eq:type-III-permutations}. 
    }\label{fig:type-III}
\end{figure}

Having provided the general context,
we now introduce two specific symmetry-twisted multi-entropies, which we claim serve as nonlocal order parameters for 2D Abelian SPTs.
The first order parameter is computed by first partitioning space into four regions, as in Fig.~\ref{fig:partitions}, such that $ABC$ forms a disk with radius larger than the correlation length. We then consider $2r$ replicas of the wavefunction, with $r\in\mathbb{Z}^+$, and define the order parameter as the expectation value of the following combination of replica permutation operators and partial symmetries acting on the various subregions in this replicated space: 
\begin{align}
    \oL^r_g(\psi) \equiv \bra{\psi}^{\otimes 2r} \pi_\alpha^A \pi_\beta^B \pi_\gamma^C \bigotimes_{b=1}^r (U_g^B)_{1,b} (U_g^B)^\dagger_{2,b} \ket{\psi}^{\otimes 2r} . \label{eq:type-I}
\end{align}
Here, we index the $2r$ replicas as $(a,b)$ with $a=1,2$ and $b=1, \dots , r$. The partial symmetry operator $(U_g^B)_{a,b}$ acts on region $B$ of replica $(a,b)$. 
These replica permutation operators are explicitly given by,
\begin{align}
    \begin{split}
    \pi_\alpha(1,b) &= (1,b-1), \quad \pi_\alpha(2,b) = (2,b+1) \\
    \pi_\beta(a,b) &= (a+1,b) \\
    \pi_\gamma(1,b) &= (2,b-1), \quad \pi_\gamma(2,b) = (1,b+1) \, \,
    \end{split} \label{eq:type-I-permutations}
\end{align}
where the indices $a$ and $b$ are interpreted cyclically modulo two and $r$, respectively.
As for the onsite symmetries, the superscripts $A,B,C$ on the permutation operators indicate truncations of said operators to the corresponding subregion.
The replica structure and action of the partial symmetries are shown in Fig.~\ref{fig:type-I}. 
Connecting to prior work on multipartite entanglement probes, let us note that $\oL_e^r$  is simply the ``lens space multi-entropy"  introduced and studied in Refs.~\onlinecite{sheffer2025,sheffer2026}. Hence, $\oL_g^r$ may be viewed as the symmetry-twisted version of this quantity.

In constructing the second order parameter, we instead consider an annular tripartition into regions $A$, $B$, and $C$ where we further bipartition each region, e.g. $A$, into $A_1$ and $A_2$ as shown in Fig.~\ref{fig:partitions}. We then consider \emph{four} replicas of this system and compute the following expectation value of permutations and partial symmetries in this replicated space:
\begin{widetext}
\begin{align}
	\oT_{g,h,k}(\psi) \equiv \braket{\psi |^{\otimes 4} \tau_\beta^B\tau^C_\gamma [ U_g^{A_1} U_h^{B_1} U_k^{C_1}] \otimes [ (U_g^{A_1})^\dagger U_h^{B_1} (U_k^{C_1})^\dagger]
    \otimes [ (U_g^{A_1})^\dagger (U_h^{B_1})^\dagger U_k^{C_1}] 
    \otimes [ U_g^{A_1} (U_h^{B_1})^\dagger (U_k^{C_1})^\dagger]  | \psi}^{\otimes 4} \, . \label{eq:type-III}
\end{align}
\end{widetext}
Here, each set of partial symmetries enclosed in a pair of square brackets acts on one of the four replicas and $g,h,k \in G$.
The permutations are given by, 
\begin{align}
	\tau_\beta = (13)(24) , \qquad
	\tau_\gamma = (12)(34) \, . \label{eq:type-III-permutations}
\end{align}
Fig.~\ref{fig:type-III} provides a pictorial representation of the replica structure, the actions of the partial symmetries, and the permutations. 
Remarkably, $\oT_{e,e,e}$ 
is precisely the original tripartite (R\'enyi-$2$) multi-entropy introduced in Refs.~\onlinecite{gadde2022,penington2023} as a probe of tripartite entanglement. Once again, we may interpret $\oT_{g,h,k}$ as a symmetry-twisted version of this multipartite entanglement probe.

The central claim of the present work is that $\oL_g^r$  and $\oT_{g,h,k}$ 
serve as order parameters for Type-I/II and Type-III SPTs, respectively. 
Indeed, let $\ket{\psi}$ be a short-range entangled state, symmetric under $G = \prod_i \mathbb{Z}_{N_i}$.
Let $g_i$ be the generator of the symmetry group $\mathbb{\mathbb{Z}}_{N_i}$ and $e$ the identity element. We further set $N^{ij} = \mathrm{lcm}(N_i,N_j)$ and, for symmetry of presentation, $N^i = N_i$. Then we claim that the \emph{phases} of these order parameters are quantized:
\begin{align}
    \begin{split}
    \mathrm{arg}\left[\oL^{N^i}_{g_i}(\psi)\right] &=\vartheta_{i} , \\ 
    \mathrm{arg}\left[\oL^{N^{ij}}_{g_i g_j}(\psi)\right] &= \vartheta_{ij} , \\ 
    \mathrm{arg}\left[\oT_{g_i, g_j, g_k} (\psi)\right] &= \vartheta_{ijk} \, . 
    \end{split}\label{eq:order-param-claims}
\end{align}
Here, $\vartheta_i$, $\vartheta_{ij}$, $\vartheta_{ijk}$ are topological invariants which specify the Type-I, II, and III SPT classes in Eq.~\eqref{eq:kunneth}. Specifically, $\vartheta_i$ and $\vartheta_{ijk}$ determine the Type-I and Type-III classes, respectively, while the Type-II class is given by $\vartheta_{ij} -N^{ij}(\vartheta_i / N_i + \vartheta_j/N_j)$. Concretely, in order to extract all topological invariants of an SPT protected by a $\mathbb{Z}_{N_i} \times \mathbb{Z}_{N_j} \times{Z}_{N_k}$ symmetry using the symmetry-twisted multi-entropies, we need $\mathrm{max}(2N^{ij},2N^{jk},2N^{ki},4)$ replicas of the wavefunction; for instance, all invariants of a $\mathbb{Z}_2^3$ SPT can be determined using four replicas.
Importantly, the magnitudes of the order parameters are real, non-negative, and decay exponentially with the lengths of the entanglement cuts and \emph{do not} depend on the $g_i$. 

Physically, these invariants capture the quantized topological responses discussed above. Indeed, we will argue in Section~\ref{sec:tqft} that these order parameters effectively simulate the corresponding continuum response theory on nontrivial manifolds in a way that extracts the coefficient of the topological term. As shown in Ref.~\cite{tantivasadakarn2017}, these invariants are in exact correspondence with the classification of Abelian SPTs proposed in Ref.~\cite{wang2015} (see also Ref.~\cite{zaletel2014}) in terms of the braiding statistics of vortices in the topologically ordered state obtained by gauging the protecting symmetry. 
Indeed, $\vartheta_i$  is $N^i$ times the spin of a $\mathbb{Z}_{N_i}$ vortex. Similarly, $\vartheta_{ij}$ is $N^{ij}$ times the spin of a composite $\mathbb{Z}_{N_i}$ and $\mathbb{Z}_{N_j}$ vortex; note that $\vartheta_{ij} -N^{ij}(\vartheta_i / N_i + \vartheta_j/N_j)$ is then the phase obtained by braiding a $g_i$ flux around a $g_j$ flux $N^{ij}$ times. Finally, $\vartheta_{ijk}$ is the phase accrued by braiding a $\mathbb{Z}_{N_i}$ vortex clockwise around a $\mathbb{Z}_{N_j}$ vortex and then a $\mathbb{Z}_{N_k}$ vortex followed by braiding the $\mathbb{Z}_{N_i}$ vortex \emph{counter}-clockwise around the $\mathbb{Z}_{N_j}$ vortex and then the $\mathbb{Z}_{N_k}$ vortex. In this sense, our order parameters may be viewed as simulating the flux response of an SPT without an explicit flux insertion, or as simulating the braiding statistics of symmetry defects. These invariants comprise a \emph{complete} set of (Abelian) SPT data, and thus our order parameters can be used to uniquely fix the SPT phase of any given wave function.

It is illuminating to contextualize our order parameters relative to the 1D case. Indeed, we note that
Ref.~\onlinecite{haegeman2012} introduced a two-replica order parameter for 1D SPTs which may be viewed as a symmetry twisted version of the R\'enyi-$2$ entanglement entropy which, in turn, is simply the R\'enyi-2 \emph{bipartite} multi-entropy \cite{gadde2022}. 
It thus seems reasonable to claim that $\oT_{g,h,k}$, 
as a symmetry-twisted R\'enyi-2 tripartite multi-entropy, is the natural generalization of the construction in Ref.~\onlinecite{haegeman2012} to 2D. 
The authors of Ref.~\onlinecite{SOP_1D} extended the construction of Ref.~\onlinecite{haegeman2012} to a larger set of similar R\'enyi type quantities, capable of probing \emph{all} bosonic 1D SPTs, and noted they could be interpreted as simulating TQFT partition functions; this TQFT interpretation was formalized in Ref.~\onlinecite{shiozaki2017}. Our work may thus also be seen as a natural, systematic generalization of these prior studies to the 2D setting.
In fact, while our focus is on 2D SPTs, the spirit of our approach is readily extended to higher dimensions and, as we will discuss in Section~\ref{sec:discussion}, we expect that higher-party symmetry-twisted multi-entropies will serve as order parameters for higher-dimensional SPTs.

\section{Symmetry Protected Multipartite Entanglement}\label{sec:multipartite}

While the status of our proposed order parameters as symmetry-twisted multi-entropies already suggests they should probe symmetry-protected multipartite entanglement, we wish to establish this more quantitatively.
In fact, in order for our order parameters to depend on the topological indices in the manner proposed in Eq.~\eqref{eq:order-param-claims}, they must satisfy certain fundamental consistency checks. 
We now turn to verifying these properties, in the course of which we will find that $\oL_g^r$ and $\oT_{g,h,k}$ probe, respectively, symmetry-protected \emph{four}-party and \emph{six}-party entanglement.

(i) First, the topological invariants of SPTs are additive under stacking; that is, the invariants of the state $\ket{\psi}\otimes\ket{\phi}$ are the sums of the corresponding invariants for $\ket{\psi}$ and $\ket{\phi}$. In order to be consistent with Eq.~\eqref{eq:order-param-claims}, this implies that our order parameters must satisfy $\cO(\ket{\psi}\otimes\ket{\phi}) = \cO(\ket{\psi})\cO(\ket{\phi})$. It is readily verified from the explicit expressions of $\oL_g^r$ and $\oT_{g,h,k}$ in Eqs.~\eqref{eq:type-I} and \eqref{eq:type-III} that this is indeed the case. In fact, \emph{all} symmetry-twisted multi-entropies [Eq.~\eqref{eq:stme}] satisfy this property.

(ii) Second, as a probe of symmetry-protected entanglement, the order parameters should be invariant under symmetric local unitaries, as such operations cannot change correlations between subregions nor do they change the global symmetry properties of the state; here we define a symmetric local unitary to be a unitary operator with support strictly within a given subregion involved in the multipartition and which commutes with the global symmetries.
One readily confirms from Eqs.~\eqref{eq:type-I} and \eqref{eq:type-III} that our order parameters satisfy $\cO(V^\alpha \ket{\psi}) = \cO(\ket{\psi})$ for any unitary $V^\alpha$ with support entirely in subregion $\alpha$ (with, e.g. $\alpha = A,B,C,D$ in Eq.~\eqref{eq:type-I}) and satisfying $[V^\alpha,U_g] = 0$ for any $g \in G$. 
Again, this is also a property satisfied by all symmetry-twisted multi-entropies.
This also motivates the expectation that our order parameters should decay with a perimeter law in a gapped state, as all degrees of freedom entangling the different subregions should be ``localized" to the entanglement cuts in such a state.

(iii) Finally, our proposed order parameters should only be sensitive to multipartite entanglement. More precisely, our order parameters are evaluated on $K$-partite wavefunctions ($K=4$ and $K=6$ for $\oL_g^r$ and $\oT_{g,h,k}$,
respectively, taking into account the bipartitions of, e.g., $A$ into $A_1$ and $A_2$ in the latter). 
We demand that their phases vanish on any 
state which does not have support on all $K$ subregions. It is in this sense that our order parameters probe $K$-partite entanglement.
For instance, we claim that $\oL_g^r(\psi)$ must be real and non-negative when evaluated on a state, $\ket{\psi} = \ket{\psi_{ABC}}$, supported only on $ABC$. In particular, this means specifically that $\oL_g^r$ and $\oT_{g,h,k}$ probe symmetry-protected four-party and six-party entanglement, respectively. Combined with the claims of Eq.~\eqref{eq:order-param-claims}, this in turn implies that Type-I/II and Type-III SPTs are characterized by such tetrapartite and hexapartite entanglement. Note that it is \emph{not} a given that a generic symmetry-twisted multi-entropy of the form Eq.~\eqref{eq:stme} will satisfy an analogous condition.

Before establishing property (iii), let us briefly comment on its significance. Concretely, property (iii) means that tensoring in a $G$-invariant ``0D" entangled state such as a Bell pair entangling two regions or a tripartite entangled state of qudits entangling three regions meeting at a trijunction cannot affect the phases of the order parameters. This is certainly a \emph{necessary} condition for any valid SPT order parameter, since tensoring with such 0D states should not change the nature of the SPT phase. Therefore the fact that our order parameters satisfy this property can be interpreted as non-trivial evidence of their validity. 

At the same time, it is important to keep in mind that our order parameters likely do not work in complete generality: we expect that they can take on incorrect ``spurious'' values for certain finely tuned states. Indeed, it is known that other entanglement-based order parameters for topological phases, like the topological entanglement entropy, can receive spurious contributions from short-ranged 1D entangled states lying along portions of the entanglement cut
\cite{zou2016,williamson2019,kato2020,gass2024}. 
Nevertheless, such examples are finely tuned and it is expected that such entanglement-based quantities should be faithful order parameters for ``generic" states. While we expect $\oL_g^r$
will suffer from similar spurious 1D contributions, our arguments will in fact allow us to partially constrain the possible spurious 1D contributions to $\oT_{g,h,k}$. 

\subsection{Establishing Non-Negativity}

We now describe our approach to establishing property (iii). 
Central to our argument is the fact that we can express each order parameter as a fully contracted tensor network with $2R$ tensors, corresponding to the kets and bras of the $R$ replicas of the wavefunction, which we represent with black and white circles, respectively. Each tensor has $K$ differently colored legs, each corresponding to one of the subregions in the multipartition. Explicitly, for a tripartite state $\ket{\psi}$ supported on regions $A$, $B$, and $C$, we have,
\begin{align}
\begin{gathered}
    \includegraphics[height=4.5em]{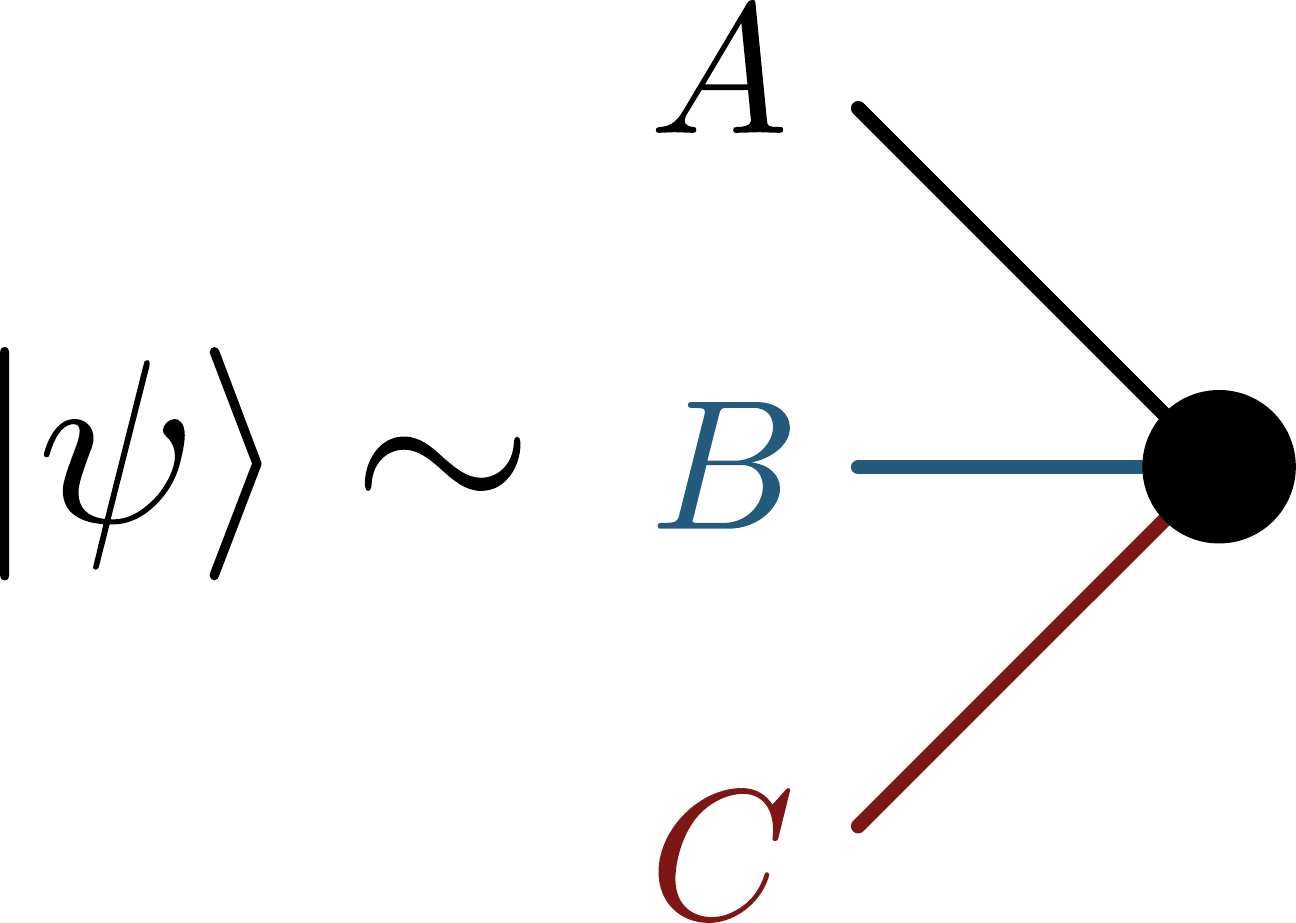}
\end{gathered}\,\,; \qquad 
\begin{gathered}
    \includegraphics[height=4.5em]{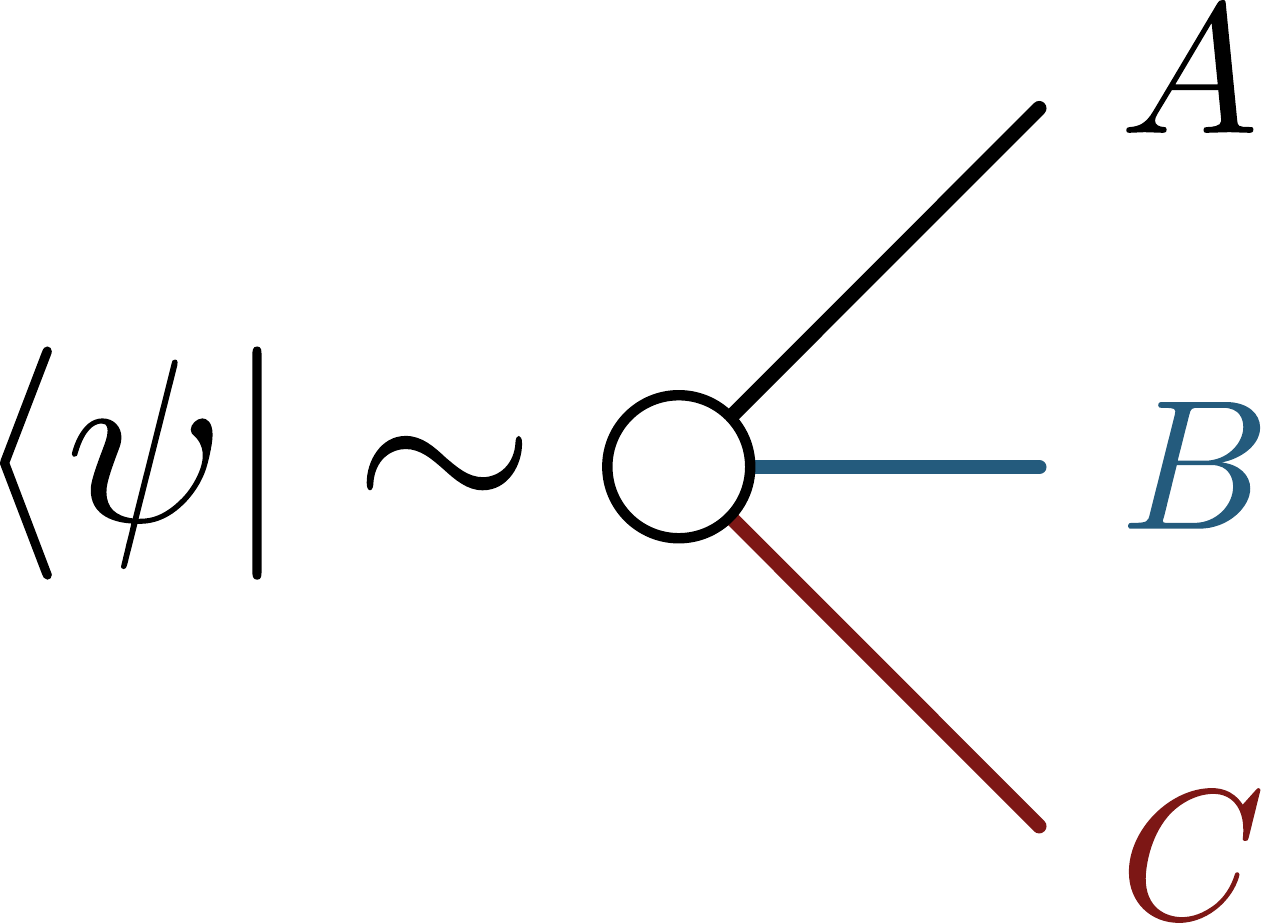}
\end{gathered}
\end{align}
The permutations dictate how the legs of the tensors are contracted, while the partial symmetries decorate the legs as operator insertions represented by colored circles. 
For instance, we depict the partial symmetries $U_g^A$ and $(U_g^A)^\dagger$ acting on the above state with filled-in and empty circles, respectively,
\begin{align}
\begin{gathered}
    \includegraphics[height=4.5em]{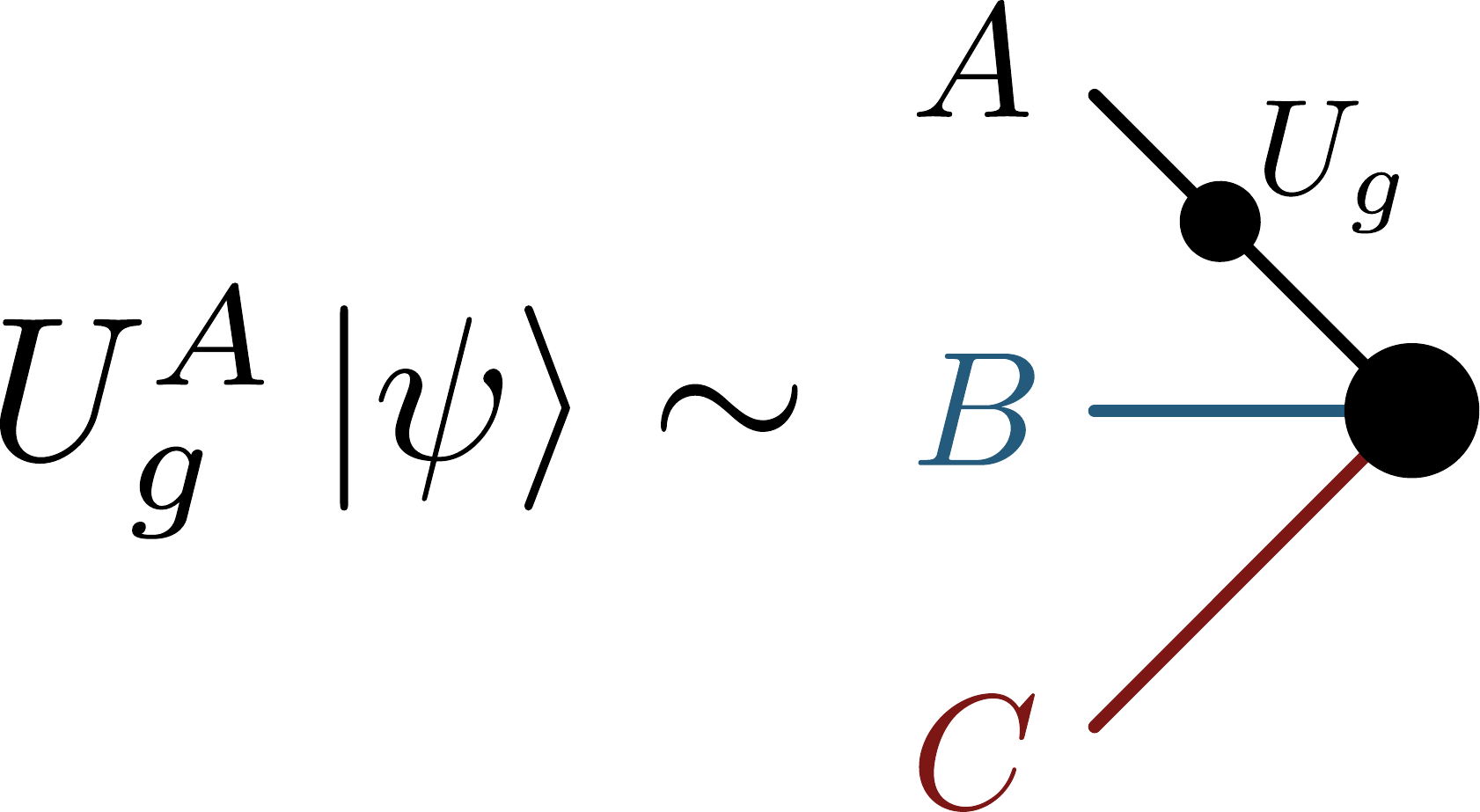}
\end{gathered}\,\,; \qquad 
\begin{gathered}
    \includegraphics[height=4.5em]{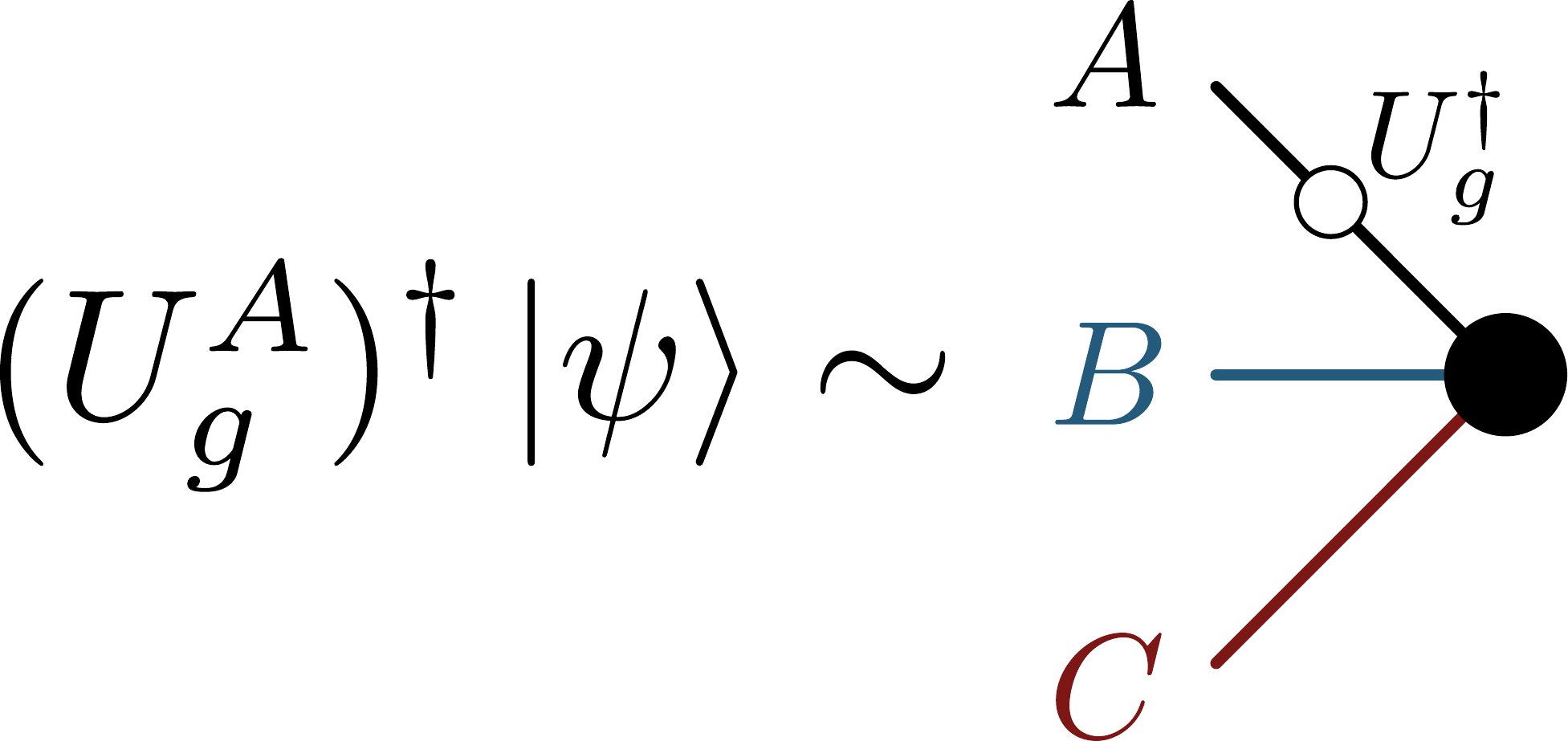}
\end{gathered}
\end{align}
Finally, we will always restrict ourselves to globally symmetric states, which implies the diagrammatic relation,
\begin{align}
\begin{gathered}
    \includegraphics[height=4.5em]{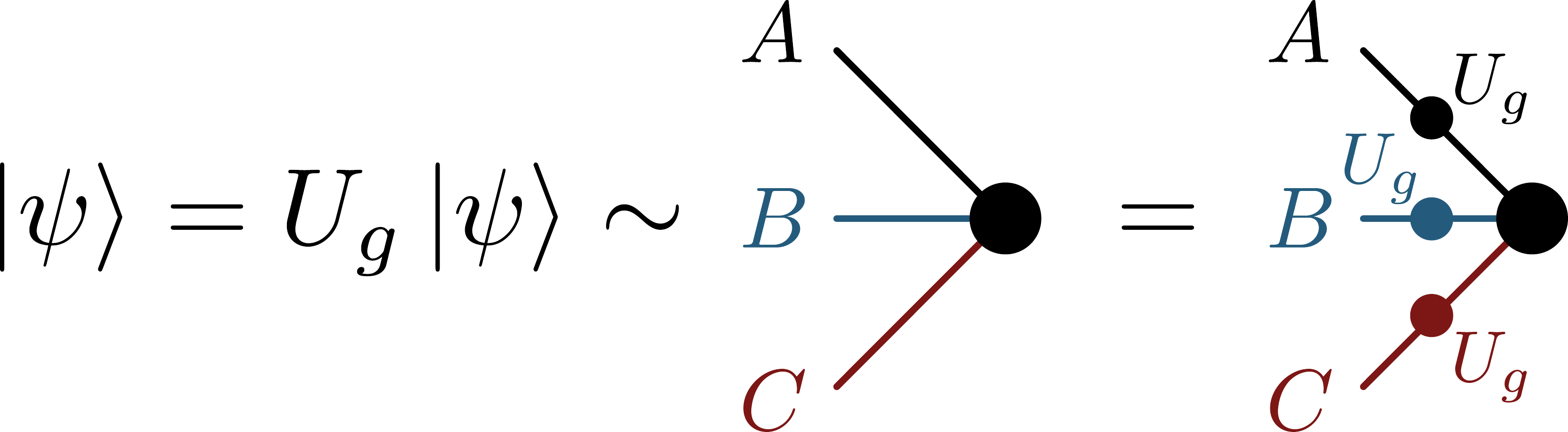} \label{eq:tensor-gauge}
\end{gathered}
\end{align}
In other words, the global symmetry of the state $\ket{\psi}$ implies the tensor network is invariant under the action of symmetry operations on all legs surrounding any given tensor (irrespective of whether it is a ket or a bra); we may view this as a gauge transformation of the tensor network. In particular, we may use this gauge freedom to move and cancel out partial symmetry operators in the tensor network.

As we will show explicitly, when evaluated on a state not supported on all $K$ subregions, the tensor network representations of our order parameters will be such that we can eliminate all partial symmetry operator insertions, using the gauge freedom of Eq.~\eqref{eq:tensor-gauge}.
The crux of our argument is that the resulting tensor network will possess a \emph{reflection symmetry} with respect to some reflection plane.\footnote{See also Ref.~\cite{gadde2026} for a related discussion, not including partial symmetries.} 
Concretely, we say that a tensor network, constructed using the rules described above and after all partial symmetries have been eliminated, is symmetric under reflection with respect to a (hyper-)plane cutting through the legs of the tensor network if:
\begin{enumerate}
	\item Under reflection, each ket is mapped to a bra and vice versa.
	\item A leg crossing the reflection plane is mapped to itself under reflection.
	\item A leg not crossing the reflection plane is mapped to another leg of the same type (i.e. color).
\end{enumerate}
A tensor network satisfying the above can be interpreted as the norm squared of some vector $\ket{T}$, $\braket{T|T} \geq 0$, which is a manifestly real and non-negative quantity.\footnote{The norm may of course vanish, but this would require $\ket{T} = 0$;
we expect that generic perturbations to such a state would result in the norm being non-vanishing.} The vector $\ket{T}$ and its dual $\bra{T}$ correspond to the two tensors obtained by cutting the tensor network along the reflection plane. We emphasize that these criteria provide \emph{sufficient} but not \emph{necessary} criteria for the order parameters to be non-negative.\footnote{In particular, one may define a modified notion of reflection symmetry, which still implies positivity, applicable to tensor networks for which not all partial symmetry insertions have been eliminated. Namely, one only requires that legs crossing the reflection plane have no partial symmetries while legs dressed by a symmetry, e.g. $U_g$, are mapped to legs with the Hermitian conjugate, $U_g^\dagger$. Since we find we can always gauge away the partial symmetries in the cases of interest, for uniformity of treatment, we use the more restrictive notion of reflection symmetry in the main text. \label{fnote:reflection}}

Let us first apply this argument in a minimal example. 
Consider the bipartite state $\ket{\psi_{AB}}$, symmetric under $U_g$ for $g \in G$.
Writing $\ket{\psi_{AB}} = \sum_{im} c_{im} \ket{im}$, where $\ket{i}$ and $\ket{m}$ denote complete bases of states for $A$ and $B$, respectively, we may construct the following two-replica quantity:
\begin{align}
\tilde{\cO}[\ket{\psi_{AB}}] \equiv \begin{gathered}
	\includegraphics[height=5em]{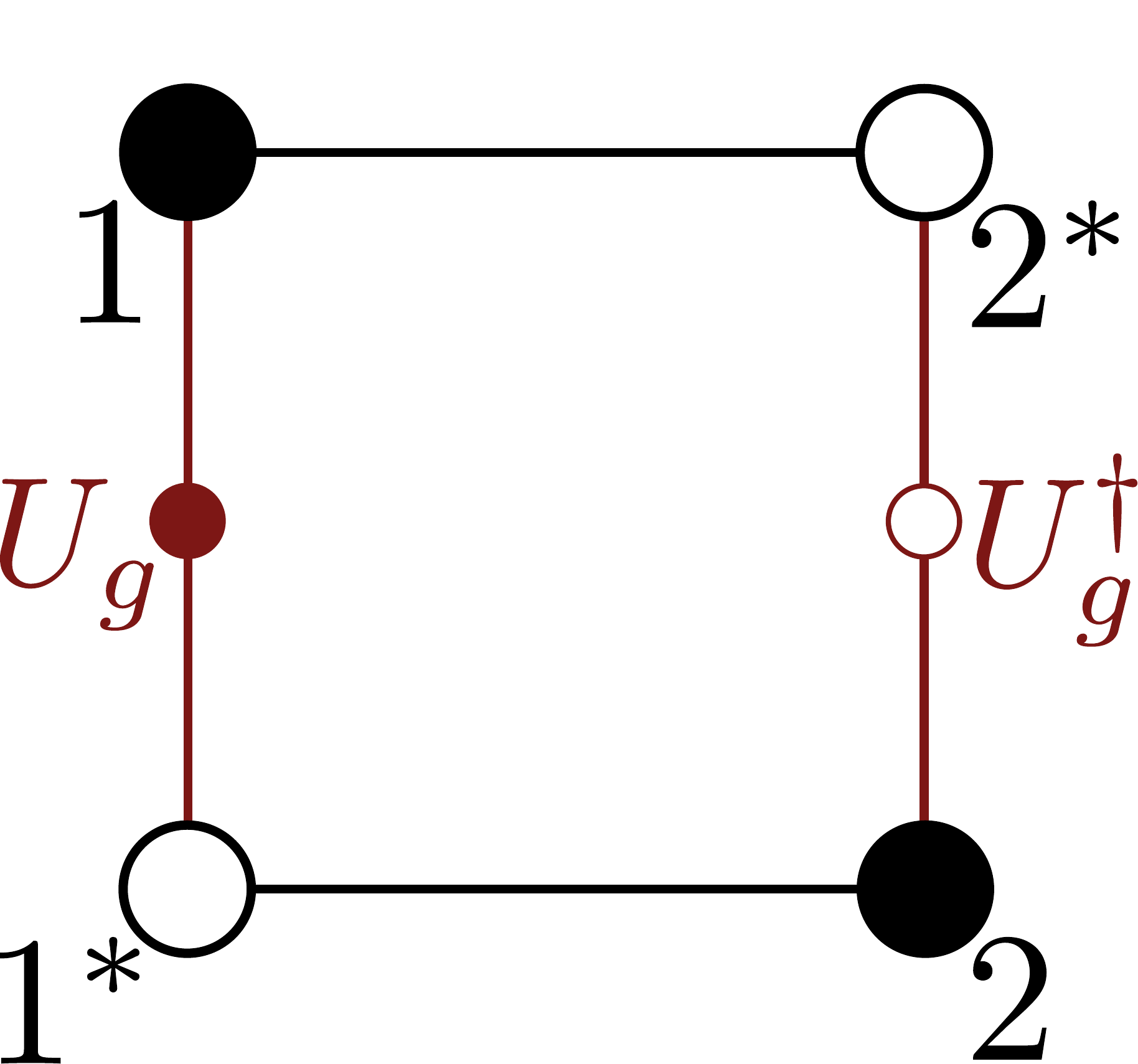}
    \end{gathered} =
    \begin{gathered}
    \includegraphics[height=5em]{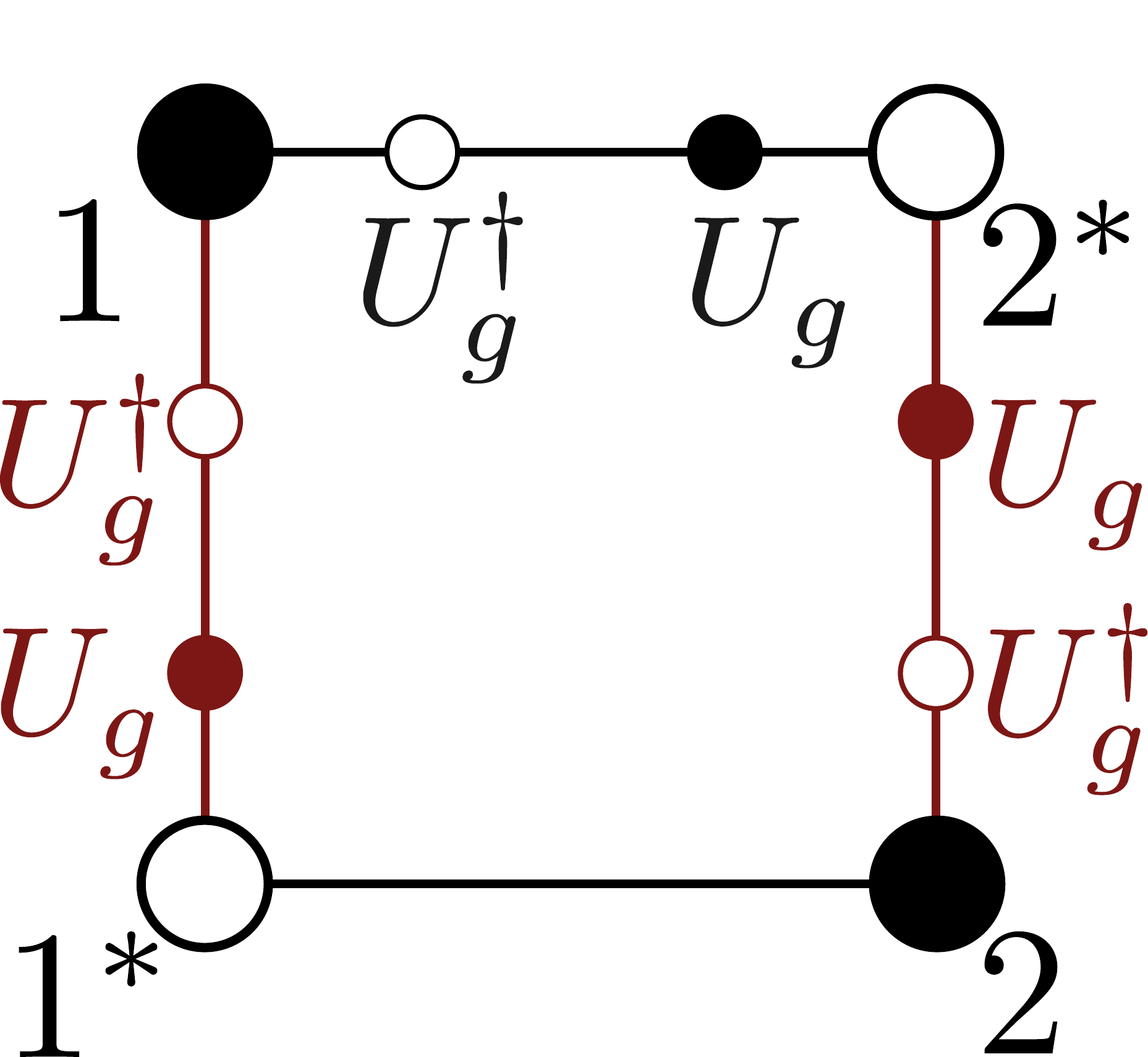} 
    \end{gathered} =
    \begin{gathered}
    \includegraphics[height=5em]{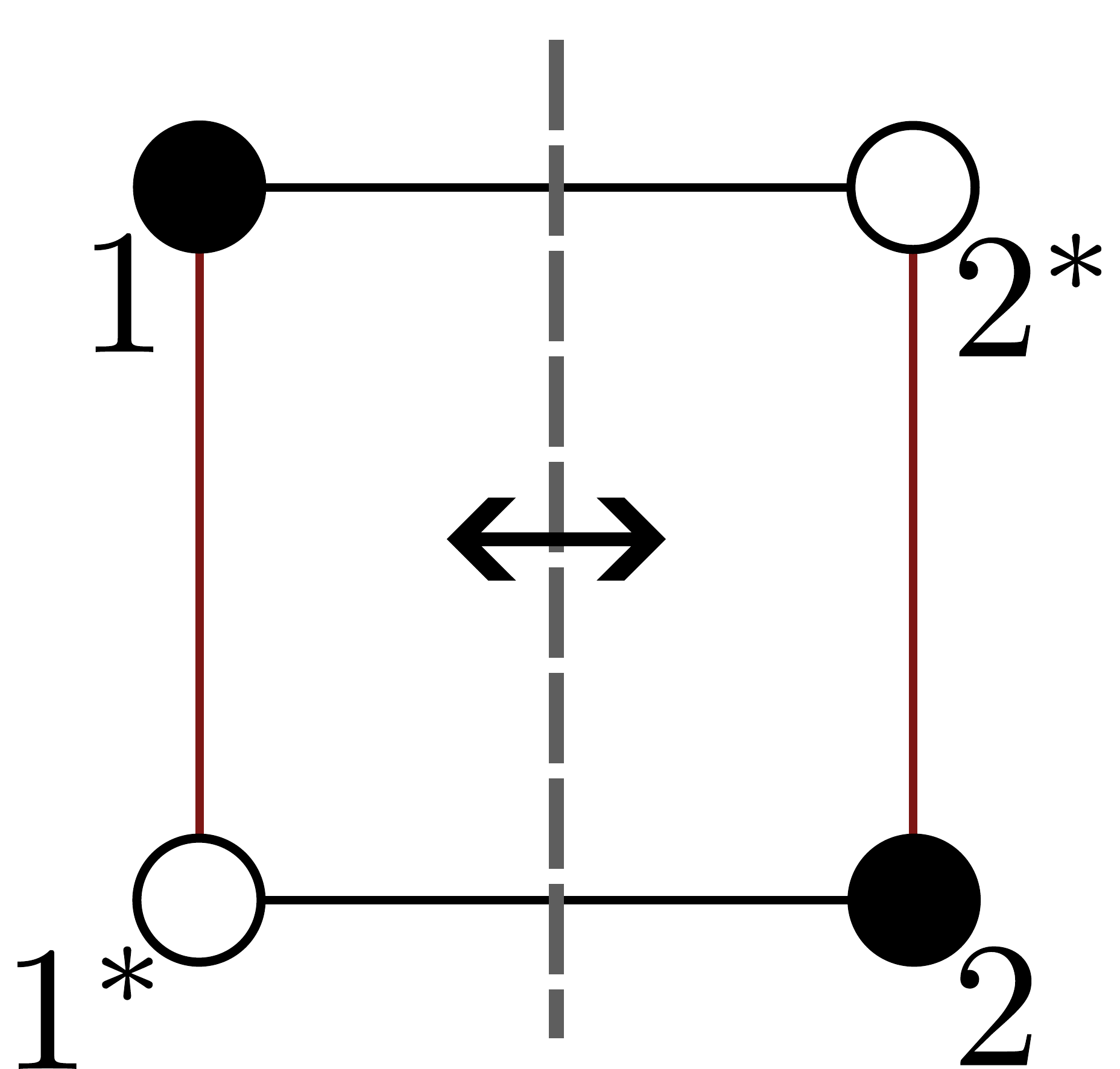}
\end{gathered}
\end{align}
The tensors are labeled by their replica number, with an asterisk denoting a bra. The red and black legs denote subregion $A$ and $B$ indices, respectively, and the $U_g$ ($U_g^\dagger$) insertions on each leg are partial symmetries acting on the subregions associated to the leg. In the second equality, we used the gauge freedom of the tensor network to introduce $U_g$ ($U_g^\dagger$) insertions on the legs surrounding the $1$ ($2^*$) tensor. In the third equality, the partial symmetries cancel out, yielding a reflection symmetry plane satisfying the above three criteria, indicated by the dashed line. 
Writing out the last expression explicitly, we find,
\begin{align}
    \notag \tilde{\cO}[\ket{\psi_{AB}}] &= \sum_{m, n} \left(\sum_{i} c^*_{i n} c_{i m} \right) \left(\sum_{j} c^*_{j m}  c_{j n} \right) \\
	&\equiv \sum_{m,n} T_{m,n} T_{m,n}^* \geq 0 \, ,
\end{align}
which is manifestly non-negative. Here, the vector $T$ and its dual are those obtained by 
blocking the indices $m$ and $n$ together. From this minimal example, one readily verifies that the above conditions for reflection symmetry can be used to deduce positivity of a general tensor network constructed using the rules laid out above.

\subsection{Non-Negativity of the Order Parameters}

\begin{figure}
  \centering    \includegraphics[width=\linewidth]{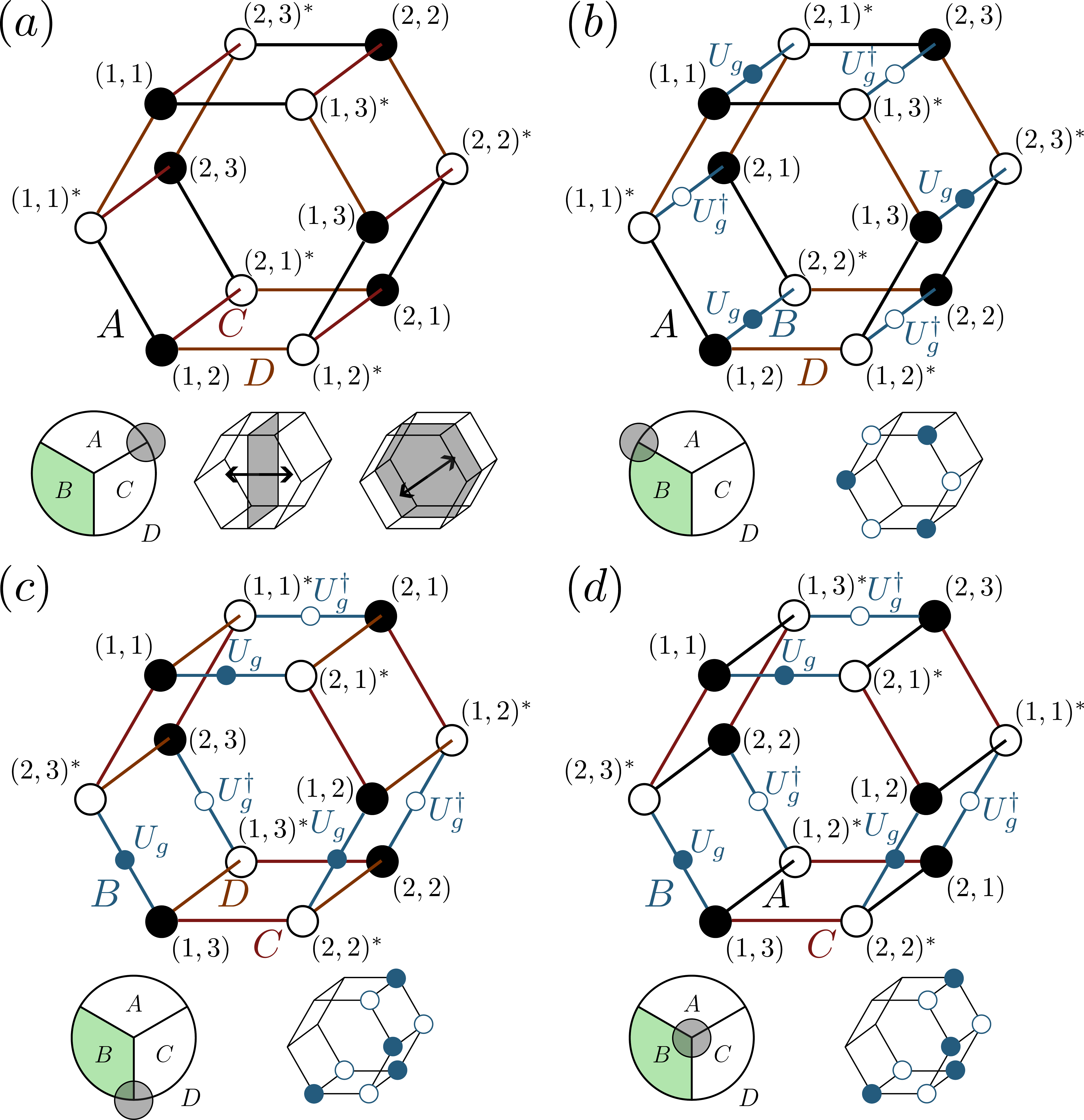}
    \caption{Tensor network expressions for $\oL_g^r$ 
    evaluated on states of the form (a) $\ket{\psi_{ACD}}$, (b) $\ket{\psi_{ABD}}$, (c) $\ket{\psi_{BDC}}$, and (d) $\ket{\psi_{ABC}}$.
    We additionally label each tensor with its replica index $(a,b)$, with an asterisk denoting a bra. 
    Below each tensor network we highlight the trijunction of the subregions which the state entangles. In (a) we also depict the reflection symmetries of the network and in (b)-(d) the tensors to which we apply global symmetries to eliminate all partial symmetries. For a solid (empty) dot, we apply $U_g$ ($U_g^\dagger$) on all surrounding legs. }\label{fig:type-I-positivity}
\end{figure}

We now apply the above argument to verify property (iii) for our order parameters. Let us begin with the Type-I order parameter, $\oL_g^r$, which we must show is real and non-negative on a state which entangles at most three of the subregions. The key observation is that, for such a state, we can arrange the corresponding tensor network representation into a $2r$-gonal prism; the tensors lie on the vertices of the prism, while the legs form the edges, some of which are dressed by partial symmetries. 
We find that we can always gauge away these partial symmetries, such that the resulting tensor network has reflection symmetries, from which the result follows.
We explicitly illustrate this construction in Fig.~\ref{fig:type-I-positivity} for the four possible tripartite entangled states in the case $r=3$, for which the tensor network forms a hexagonal prism. 
The tensor network for $\ket{\psi_{ACD}}$ manifestly possesses reflection planes, whereas for the other states, we must gauge away the partial symmetries, as indicated in the insets of each subfigure (but see also footnote~\ref{fnote:reflection}). Specifically, by acting with $U_g$ ($U_g^\dagger$) on all legs surrounding a solid (empty) circle, one can eliminate all partial symmetries such that the resulting tensor networks are manifestly reflection symmetric.
The construction for general $r$ follows a similar pattern.

\begin{figure}
  \centering
    \includegraphics[width=0.9\linewidth]{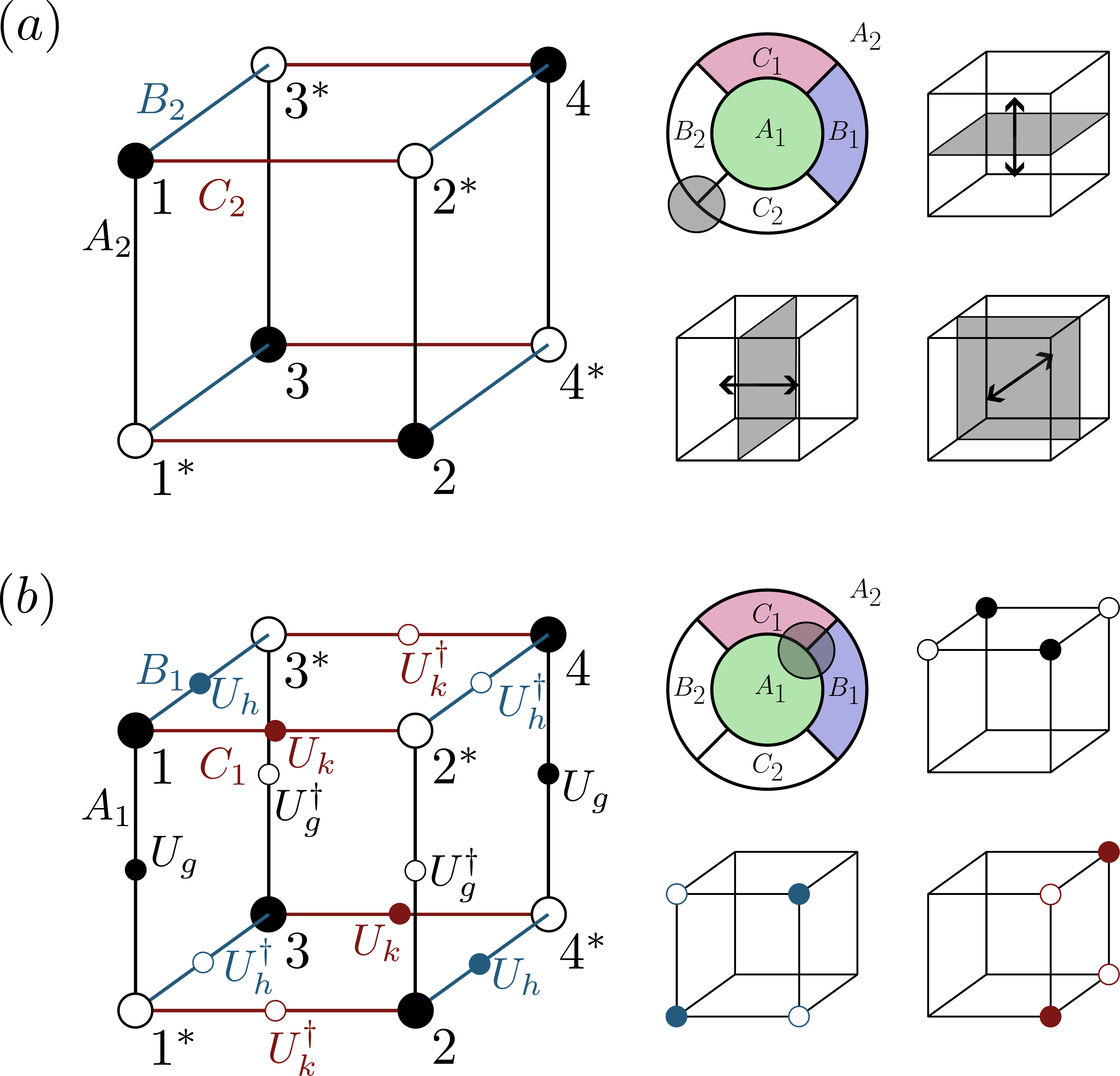}
    \caption{Tensor network representations of $\oT_{g,h,k}$     evaluated on the tripartite states (a) $\ket{\psi_{A_2 B_2 C_2}}$ and (b) $\ket{\psi_{A_1 B_1 C_1}}$. The tensors are labeled by the replica index $a=1, \dots , 4$, with an asterisk denoting a bra. Next to each network, we indicate the trijunction between the three subregions which this state entangles, and the pattern of global symmetries that can be applied to cancel out all partial symmetries. For a solid (empty) black, blue, or red dot, we apply $U_{g,h,k}$ ($U_{g,h,k}^\dagger$) on all surrounding legs, respectively. 
    }\label{fig:type-III-positivity}
\end{figure}

We next turn to the Type-III order parameter, $\oT_{g,h,k}$. 
As it involves a hexapartition, we wish to show that this order parameter has a trivial phase for any state entangling at most five of the subregions. For illustrative purposes, it will be convenient to first show that the phase vanishes on states entangling at most three adjacent subregions, i.e.,  
states with support on $A_i B_j C_k$, with $i,j,k \in \{ 1, 2\}$, corresponding to the eight trijunctions. Once again, the tensor network for $\oT_{g,h,k}$ 
evaluated on states of this form naturally describes a polyhedron, in this case a cube, as depicted in Fig.~\ref{fig:type-III-positivity}. For the $A_2B_2C_2$ trijunction, there are no partial symmetries acting on any of the subregions, and thus the tensor network has all the reflection symmetries of the cube [see Fig.~\ref{fig:type-III-positivity}(a)]. 
In contrast, for the $A_1 B_1 C_1$ trijunction, as depicted in Fig.~\ref{fig:type-III-positivity}(b), we must use the gauge freedom of the tensor network to cancel out all the partial symmetries to make manifest the same reflection symmetry planes. The tensor networks for $\Omega_{g,h,k}$ evaluated on states on all remaining trijunctions have the same structure as Fig.~\ref{fig:type-III-positivity}(b), but with only a subset of the partial symmetries applied; hence, a subset of the same gauge transformations can be used to eliminate the partial symmetries. This establishes the presence of reflection symmetries of the tensor networks and hence non-negativity of the order parameter when evaluated on these states.

While this allows us to exclude spurious 0D entanglement contributions to $\Omega_{g,h,k}$, we can go a step further and show that this order parameter evaluated on any state entangling at most five subregions also has a trivial phase. The argument is similar and we leave the details to Appendix~\ref{app:spurious}. This most general result allows us to exclude certain forms of spurious 1D contributions \cite{zou2016,williamson2019,kato2020,gass2024}, namely, 1D states encircling any one of the six subregions must yield a trivial phase, as these entangle at most five subregions. 
Note that this does not rule out all spurious 1D contributions; for instance, a state encircling $A_1$ and $B_1$ can entangle all six subregions and hence may yield a nontrivial phase for $\Omega_{g,h,k}$.
Nevertheless, this gives us additional reason to believe the phase of $\oT_{g,h,k}$ should be stable to generic perturbations.

\section{Topological Quantum Field Theory Analysis \label{sec:tqft}}

We now provide a topological quantum field theory (TQFT) argument to justify the claims of Eq.~\eqref{eq:order-param-claims} and explain the sense in which the order parameters effectively simulate the braiding of symmetry defects. 
Assuming a valid continuum limit, the low energy physics of an SPT phase coupled to background gauge fields for the protecting symmetry is governed by a TQFT for the response fields. 
This TQFT governs the flux responses reviewed above. 
Moreover, as shown in Ref.~\cite{tantivasadakarn2017} (and reviewed in Appendix~\ref{app:group-cohomology}), the three topological invariants of Eq.~\eqref{eq:order-param-claims} can be extracted by computing the TQFT partition function on certain non-trivial spacetime manifolds with appropriately chosen holonomies for the gauge fields. Our claim is that 
$\oL_g^r$ and $\oT_{g,h,k}$ 
simulate these partition functions. 

More precisely, our order parameters (and symmetry-twisted multi-entropies more generally) are expectation values of partial symmetry and permutation operators in a replicated space. In the continuum limit, a state is described by an imaginary time path integral on a spacetime whose boundary is the spatial surface on which the state is defined; e.g. the partition function on the three-ball prepares a state on the two-sphere. The permutation operators indicate how to ``glue" these different manifolds together, yielding a closed manifold with potentially non-zero genus. The partial symmetry operators introduce symmetry \emph{defects} in spacetime. Our order parameters are constructed such that the symmetry defects glue together to induce the desired holonomies of the background gauge field. That is to say, we will show that our order parameters satisfy, schematically,
\begin{align}
    \cO(\psi) \propto \mathcal{Z}[M_{\{g_i\}}],
\end{align}
where $\cO$ is one of $\oL_g^r$ and $\oT_{g,h,k}$ while $\mathcal{Z}[M_{\{g_i\}}]$ is the TQFT partition function and is evaluated on the spacetime manifold $M$ with symmetry fluxes $\{g_i\}$ along appropriate cycles. Assuming $\ket{\psi}$ is non-chiral, the proportionality constant is expected to be real and non-negative. In particular, as the partial symmetries generate \emph{topological} defects, 
we additionally expect this constant to be independent of these defect insertions.

\subsection{Type-I/II: Simulating the Lens Space}

\begin{figure}
  \centering    \includegraphics[width=0.9\linewidth]{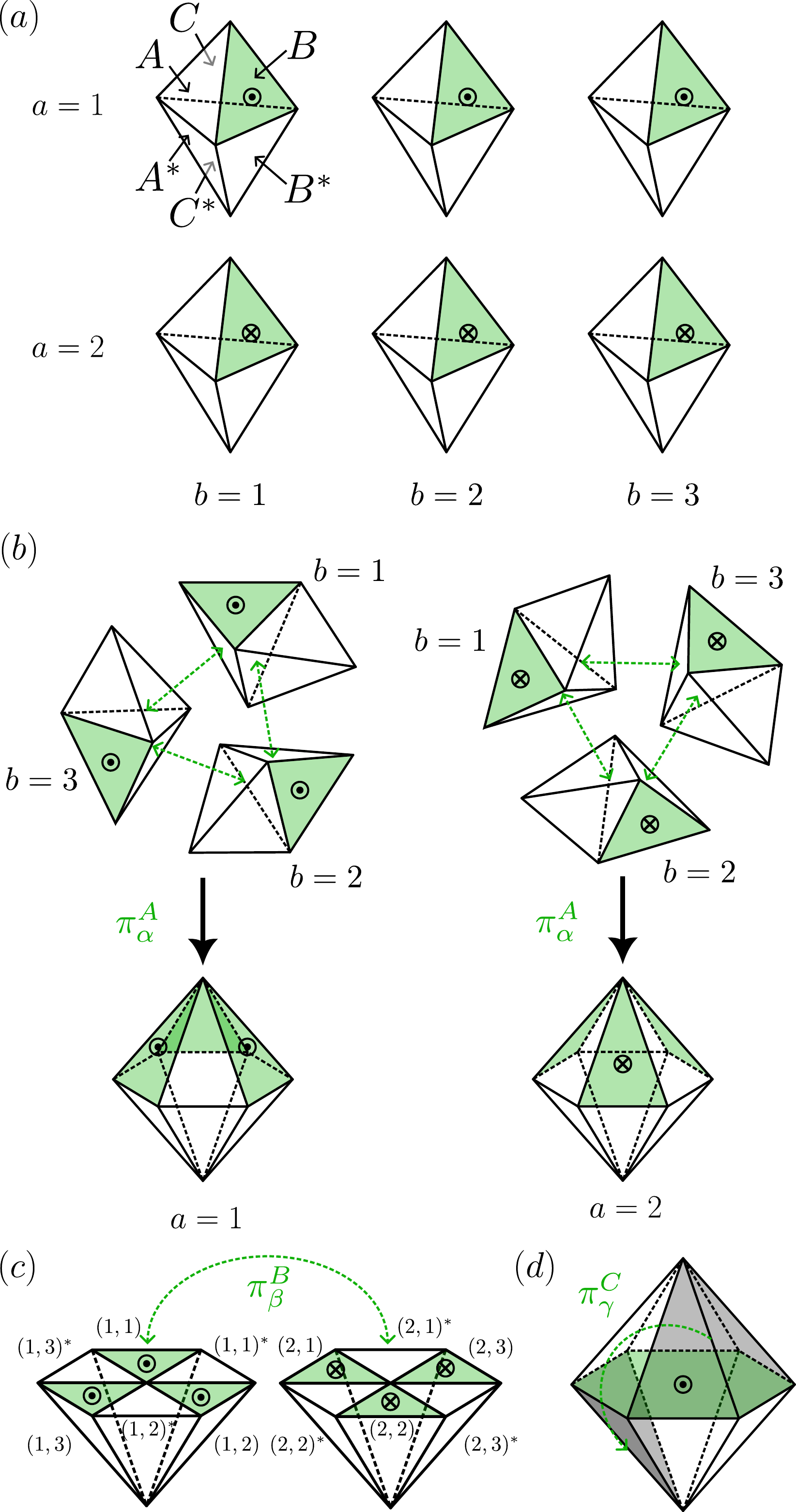}
    \caption{Gluing procedure for $\oL_g^{r=3}$. 
    (a) The reduced density matrices after tracing out $D$. The shaded green faces indicate where partial symmetries are applied with orientation out of/into the surface indicated by a dotted/crossed circle. (b) Gluing the $A$ faces yields two $2r$-gonal bipyramids with symmetry defects on certain faces. (c) Flattening them for clarity, gluing the $B$ faces yields a single bipyramid with a symmetry defect spanning the base of the pyramid. (d) Gluing the $C$ faces yields the lens space $L(3,1)$, with a non-trivial $g$ holonomy along the non-contractible cycle. }\label{fig:lens}
\end{figure}

Let us begin with $\oL_g^r$. Given an SPT protected by a $\mathbb{Z}_{N_i}$ symmetry with generator $g_i$, the invariant $\vartheta_i$ of Eq.~\eqref{eq:order-param-claims}, which fixes the Type-I SPT class, 
is given by the TQFT partition function on the \emph{lens space}, $L(N^i,1)$, with a symmetry flux (i.e. a holonomy) of $g_i$ along the single non-contractible cycle \cite{tantivasadakarn2017}. 
Similarly, for an SPT protected by the symmetry $\mathbb{Z}_{N_i} \times \mathbb{Z}_{N_j}$ with generators $g_i$ and $g_j$, the invariant $\vartheta_{ij}$ 
is given by the TQFT partition function on the lens space $L(N^{ij},1)$ with a $g_i g_j$ flux insertion. Recall that the Type-II class is determined by $\vartheta_{ij} -N^{ij}(\vartheta_i / N_i + \vartheta_j/N_j)$. 
The lens space $L(r,1)$ is obtained by taking the three-ball, rotating the northern hemisphere clockwise by an angle $2\pi/r$ around the central axis, and then gluing the surface of the northern hemisphere to that of the southern hemisphere, as shown in Fig.~\ref{fig:lens}(d) (see Appendix~\ref{app:group-cohomology} for more details). This twist yields a single noncontractible cycle, corresponding to the path connecting the north and south poles, which are identified. Our claim is that Eq.~\eqref{eq:type-I} exactly simulates the TQFT partition function on this manifold.

It has recently been shown in Ref.~\cite{sheffer2025} that $\oL_{g=e}^r$ simulates the lens space partition function. Our new result is that including the partial symmetry operations for general $g$ leads to a $g$-holonomy along the non-contractible cycle of the lens space, as shown in Fig.~\ref{fig:lens} for $r=3$. First, after tracing out $D$, we obtain a triangular bipyramid 
for each replica. The faces of the northern (southern) pyramid correspond to the subregions on the ket (bra). There is a $U_g$ ($U_g^\dagger$) symmetry defect spanning the $B$ ($B^*$) face on the $a=1$ ($a=2$) replicas [Fig.~\ref{fig:lens}(a)]. The $\pi_\alpha^A$ permutation glues the bipyramids together, yielding two $2r$-gonal bipyramids with $U_g$ ($U_g^\dagger$) symmetry defects decorating alternating portions of the northern pyramid for $a=1$ ($a=2$) [Fig.~\ref{fig:lens}(b)]. The $\pi_\beta^B$ permutation then glues these two bipyramids along their northern pyramids, yielding a single $2r$-gonal bipyramid, where the symmetry defects have stitched together into a single membrane with boundary along the equator. Finally, the $\pi_\gamma^C$ permutation glues the northern pyramid to the southern pyramid with a $2\pi / r$ twist, yielding the desired lens space with nontrivial holonomy [Fig.~\ref{fig:lens}(c,d)]. 

\subsection{Type-III: Simulating the Three-Torus}

\begin{figure}
  \centering
    \includegraphics[width=\linewidth]{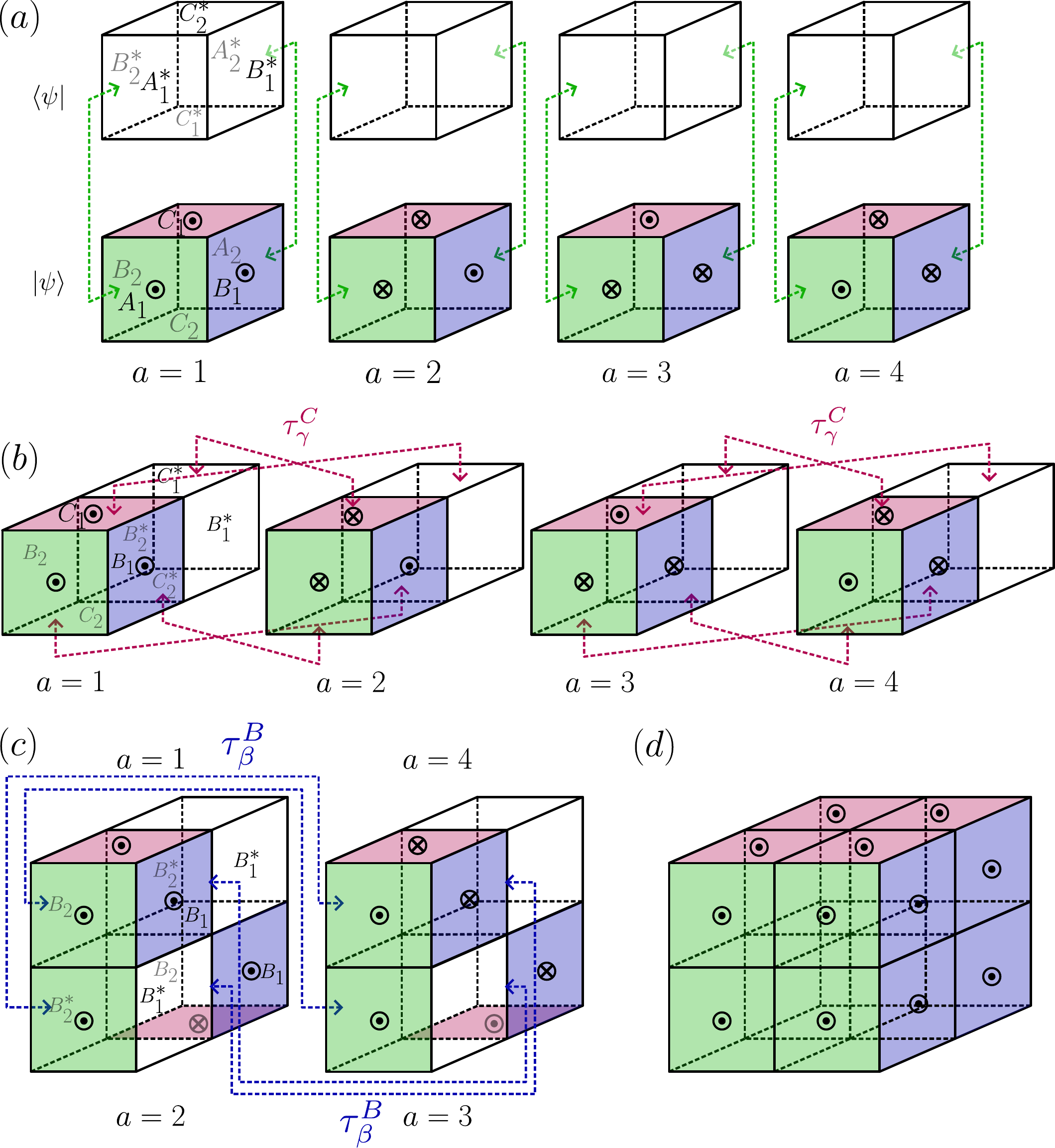}
    \caption{Gluing procedure for $\oT_{g,h,k}$. 
    (a) The bottom (top) row corresponds to the kets (bras). The index $a$ is the replica label. The green, blue, and pink faces correspond to $g$, $h$, and $k$ symmetry defects, with their orientation specified by the filled-in/crossed circles. 
    (b) Gluing together the $A$ subregions yields four filled-in two-tori.
    We drop the labels for the faces that have been glued and identified. 
    (c) Gluing of the $C$ faces.
    (d) Finally, gluing the $B$ faces yields the three-torus, with symmetry defects $g, h,$ and $k$ on the surfaces dual to each of three non-contractible cycles, respectively. 
    }\label{fig:type-III-gluing}
\end{figure}

Next, we turn to $\oT_{g,h,k}$. 
Given an SPT protected by a $\mathbb{Z}_{N_i} \times \mathbb{Z}_{N_j} \times \mathbb{Z}_{N_k}$ symmetry with generators $g_{i,j,k}$, respectively, the invariant $\vartheta_{ijk}$, which fixes the Type-III class, 
is given by the corresponding TQFT partition function on the three torus, $T^3$, with flux insertions of $g_{i,j,k}$ along the three cycles \cite{wang-gu-wen2015,tantivasadakarn2017}. As we show in Fig.~\ref{fig:type-III-gluing}, $\oT_{g,h,k}$ precisely simulates this partition function. 
Let us momentarily consider the case where the spatial manifold is the two-sphere. In the TQFT picture, we then begin with eight filled-in cubes (from four kets and four bras), where each face  corresponds to one of the six subregions and the $A_1,B_1,C_1$ faces on the kets are dressed with $U_g$, $U_h$, and $U_k$ symmetry defects, with orientation dictated by whether we act with, e.g. $U_g$ or $U_g^\dagger$. On performing the $A$ trace, we glue the two cubes in a given replica along their $A$ faces, yielding four filled-in tori. Note that, had we started with a state on an arbitrary spatial manifold, tracing out $A$ would land us at this step. The $\tau_\gamma^C$ permutation glues the $a=1$ ($a=4$) filled-in torus to the $a=2$ ($a=3$) filled-in torus, yielding two manifolds with two non-contractible cycles. Finally, the $\tau_\beta^B$ permutation glues these two manifolds together, yielding the desired three-torus. Keeping track of the positions and orientations of the symmetry defects, we find that the symmetry defects are stitched together to yield three closed, non-contractible membranes, which induce the desired $g$, $h$, and $k$ holonomies along the cycles of the three-torus. 

Let us emphasize that in order for these partition functions to yield the desired SPT invariants, we must assume that the state in question is an SPT, i.e. a short-range entangled phase. Indeed, these partition functions can acquire additional contributions if the system in question is chiral or has intrinsic topological order (i.e. supports anyons). We will comment on these potential contributions in Section~\ref{sec:discussion}. 

Finally, we note that, in the above arguments, we made the (natural) choice that a partial symmetry acting on a state has the effect of inserting, in the TQFT picture, a symmetry defect with normal pointed towards the exterior of the three-ball. Had we instead chosen the normal to point inwards (i.e. flipped the directions of the fluxes), this would have left the lens space partition function unchanged but complex conjugated the three-torus partition function (as can be deduced from Figs.~\ref{fig:lens} and \ref{fig:type-III-gluing}, or from the explicit expressions in Appendix~\ref{app:group-cohomology}). As we will see in the next section, to match with the lattice computations, it seems necessary to use this latter convention. This suggests there is some ambiguity in the choice of the normal direction of the symmetry defects in the TQFT picture, which we do not resolve in this work. Nevertheless, our main claim that $\oL_g^r$ and $\Omega_{g,h,k}$ simulate the appropriate partition functions still stands. In Eq.~\eqref{eq:order-param-claims} and the paragraph following it, we choose the latter convention to be consistent with the lattice computations.

\section{Testing the Proposal in Exactly Solvable Lattice Models}\label{sec:lattice}

\begin{figure}
    \centering
    \includegraphics[width=0.85\linewidth]{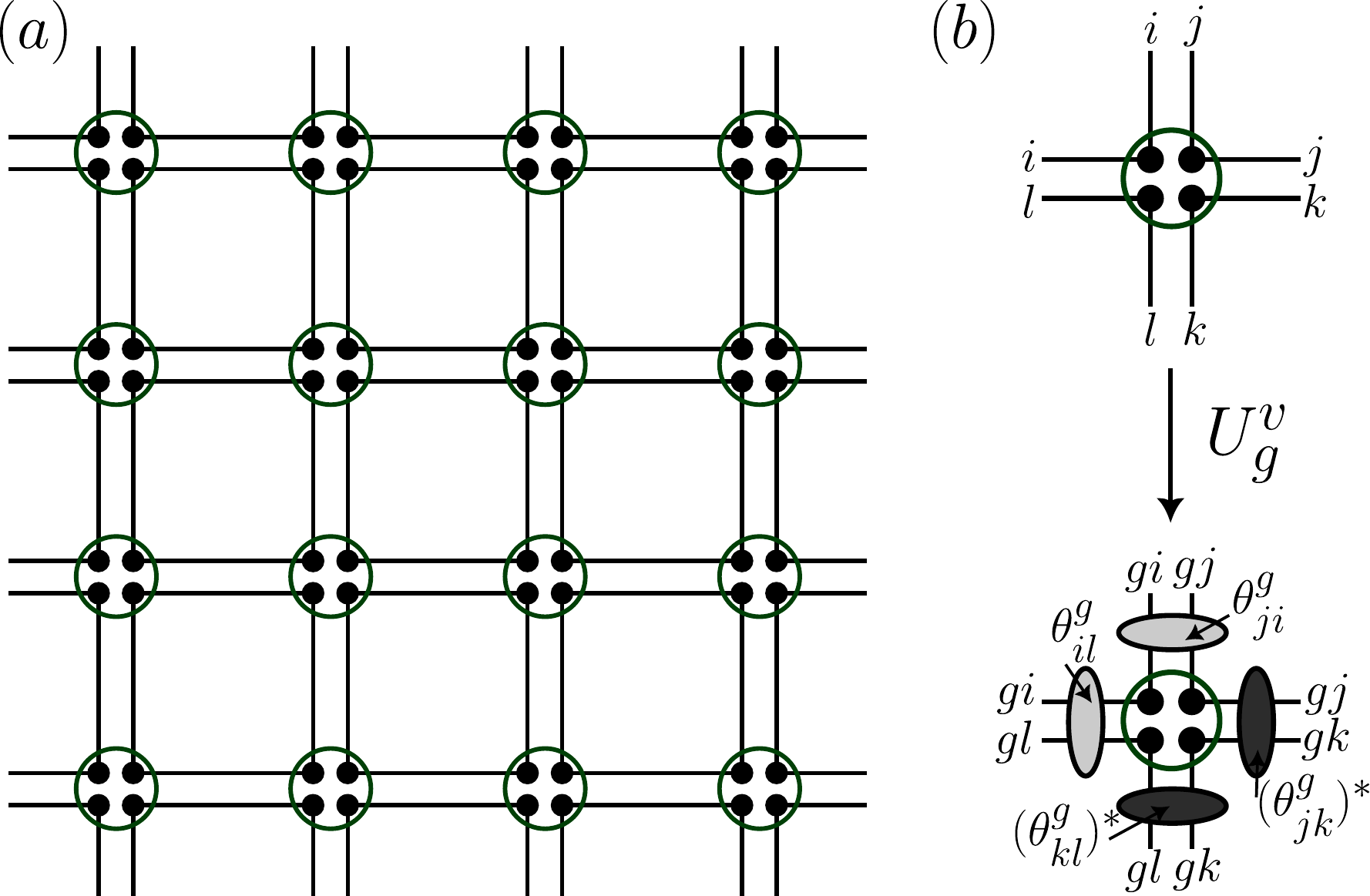}
    \caption{(a) A pictorial representation of the ground state of a lattice group cohomology model for symmetry group $G$. 
    (b) The on-site symmetry action of unitary operator $U_g$ on a single site of the ground state; the ellipses indicate phase factors depending on the states of adjacent sites.}
    \label{fig:cohomology-lattice-model}
\end{figure}

While our TQFT arguments provide a compelling case for the claims of Eq.~\eqref{eq:order-param-claims}, it is important to confirm their validity for explicit SPT wavefunctions. 
To that end,
we now test our proposed order parameters in the (Abelian) lattice group cohomology models introduced in Ref.~\cite{chen2013}. This family of models provides fixed-point lattice realizations of all bosonic SPTs of interest. As a consistency check, we are able to analytically verify that the order parameters indeed probe the desired topological invariants.  
As the computations themselves are quite tedious, we only sketch them out here and relegate the
details to Appendix~\ref{app:lattice-details}. 

Let us first review the construction of the lattice group cohomology fixed-point wavefunctions. We consider a square lattice and place four $G$-valued qudits at each vertex. 
In the following, we will always take $G$ to be an \emph{Abelian} group.
We displace each qudit from its corresponding vertex such that each lies in one of the four adjacent plaquettes. We then place the four qudits from the four different vertices around a given plaquette, $p$, in a GHZ state $\ket{\phi}_p = |G|^{-1/2} \sum_{g \in G} \ket{gggg}$, such that the many-body state is given by,
\begin{align}
	\ket{\phi} = \bigotimes_p \ket{\phi}_p \, . \label{eq:grp-coh-state}
\end{align}
This state is $G$-symmetric, with the representation of $G$ given by,
\begin{align}
	U_g = \bigotimes_v U_g^v
\end{align}
where $g \in G$ and $v$ runs over all vertices. The on-site representation of $U_g$ is, for $v_i$ labelling the states of the four qudits at a vertex,
\begin{align}
	U_g^v\ket{v_1, v_2, v_3, v_4} = \frac{\theta_{v_1,v_2}^g \theta_{v_2,v_3}^g}{\theta_{v_4,v_3}^g \theta_{v_1,v_4}^g}\ket{gv_1, g v_2, g v_3, g v_4} \, . 
\end{align}
Here, the $v_i$ are ordered counter-clockwise around the vertex, with $v_1$ being the northeast-most site, and, 
\begin{align}
	\theta_{ab}^g = \omega(\bar{a}b, \bar{b}\bar{g}, g) \, ,
\end{align}
where an overbar denotes the group inverse and $\omega \in \cH^3[G,U(1)]$ is a three-cocycle which is precisely the cocycle determining the SPT phase in which the state lies. One can further show that this forms a linear representation of the symmetry group $G$. 

\begin{figure}
    \centering
    \includegraphics[width=0.7\linewidth]{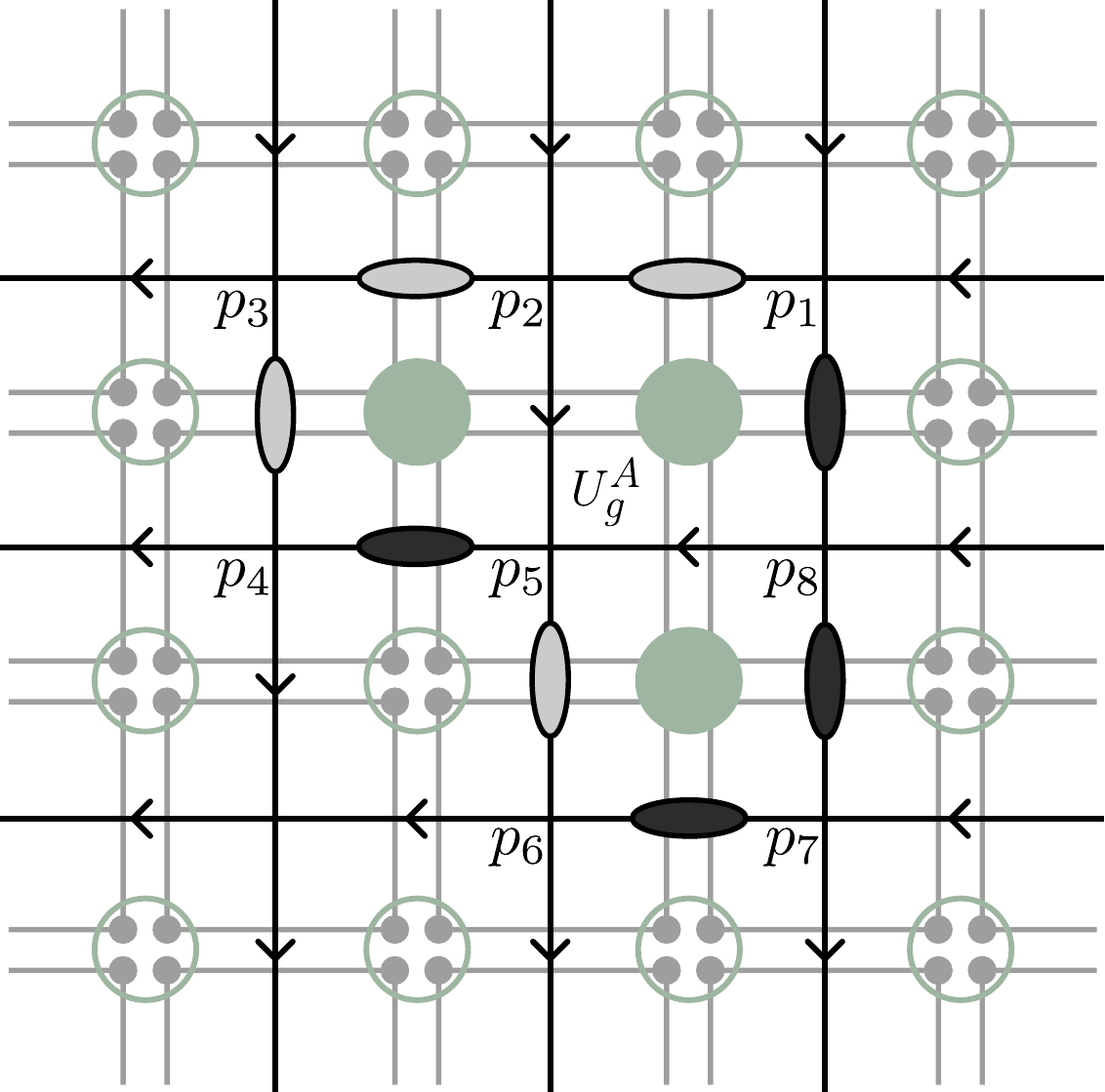}
    \caption{The dual lattice for the lattice model, with oriented bonds connecting plaquettes on the original lattice, and a partial symmetry action acting on the filled in green circles. The corresponding phases indicated by the ellipses, following the convention of Fig.~\ref{fig:cohomology-lattice-model}, are those in Eq.~\eqref{eq:MPU}. 
    }
    \label{fig:lattice-phases}
\end{figure}

It is convenient to represent the ground states of these models diagrammatically as in Fig.~\ref{fig:cohomology-lattice-model}(a). Here, four qudits connected by solid lines are placed in a GHZ state. The on-site symmetry action is represented diagrammatically as in Fig.~\ref{fig:cohomology-lattice-model}(b); a white (black) ellipse on two legs associated to qudits $i$ and $j$ indicates the presence of the phase factor $\theta_{ab}^g$ ($(\theta_{ab}^g)^{-1}$) in the symmetry-transformed state. This diagrammatic representation makes clear that if we consider a partial symmetry operation on some region $A$, all induced phase factors cancel out in the bulk, such that the GHZ states which are entirely contained in $A$ remain invariant.

More explicitly, consider the dual lattice
and assign an orientation to each dual bond, using the convention in Fig.~\ref{fig:lattice-phases}. We may then express a partial symmetry operation acting on region $A$ of a group cohomology state as an operator acting purely on the boundary of $A$ \cite{chen2013}:
\begin{align}
    U_g^A \ket{\phi} = \left(\sum_{\{p_l \}} T_{\{ p_l\}}^g L_{\{ p_l\}}^g \right) \ket{\phi} \, . \label{eq:MPU}
\end{align}
Here, $l$ labels the plaquettes in sequential, counter-clockwise order around the boundary of $A$, while $p_l$ labels the basis of states for plaquette $l$; see Fig.~\ref{fig:lattice-phases}. We have defined,
\begin{align}
    L_{\{ p_l\}}^g = \bigotimes_{l \in \partial A} \bigotimes_{i \in l \cap A} \ket{g p_l}_i\bra{p_l}_i \, ,
\end{align}
as the group action on the qudits lying on the boundary.
The index $i$ labels the vertices belonging to the plaquette $l \in \partial A$ that also lie in $A$. 
The phase factors are contained in,
\begin{align}
    T_{\{ p_l\}}^g = \prod_{l \in \partial A} T_{p_l , p_{l+1}}^g \, .
\end{align}
For two adjacent plaquettes $p_l$ and $p_{l+1}$ we have defined, 
\begin{align}
	T^g_{p_l,p_{l+1}} = \begin{cases}
		\theta^g_{p_l , p_{l+1}} &\text{if the bond points from $l$ to $l+1$}, \\
		(\theta_{p_{l+1} , p_l}^g)^* &\text{if the bond points from $l+1$ to $l$}
	\end{cases} \label{eq:T-definition}
\end{align}
This representation will allow us to reduce the computations of the order parameters to computations on the entanglement cuts. 
In general, a partial symmetry action on an SPT state in region $A$ evaluates to an effective, anomalous symmetry action on the boundary $\partial A$ with anomaly index given by the bulk SPT cocycle \cite{else2014}. The following computations may thus also be seen as a means of extracting gauge invariant data determining the anomaly cocycle of the effective boundary symmetry action, Eq.~\eqref{eq:MPU}.

\subsection{Type-I/II Computation}

\begin{figure}
    \centering
    \includegraphics[width=\linewidth]{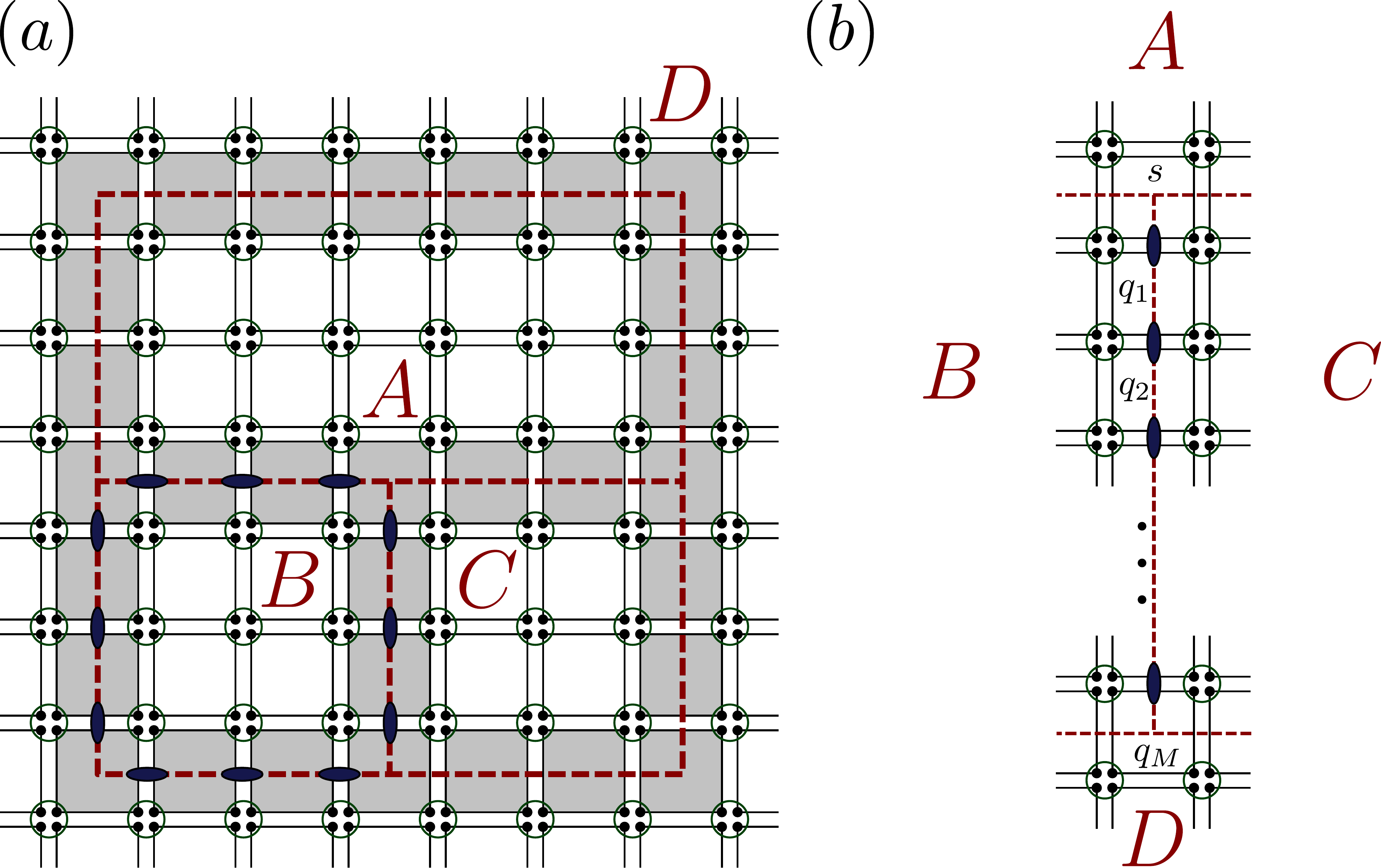}
    \caption{(a) Partitioning of the lattice used in the computation of $\oL_g^r$.
    (b) Labeling of the plaquettes along the boundary between $B$ and $C$. The blue ellipses indicate phases depending on the states of adjacent plaquettes. }
    \label{fig:type-I-rectangular-main}
\end{figure}

We consider a multipartition as in Fig.~\ref{fig:type-I-rectangular-main}(a) to evaluate $\oL_g^r$. 
Note that the entanglement cut passes through plaquettes, i.e., it is defined on the dual lattice.
The key point of the computation is that it is only the GHZ states on these plaquettes that contribute to the multi-entropy. Explicitly, it is useful to write the multi-entropy as,
\begin{align}
	\begin{aligned}
	\oL_{g}^r(\psi) = \Tr\left[ \pi_\alpha^A \pi_\beta^B \pi_\gamma^C  \bigotimes_{b=1}^r (U_g^B)_{1,b} (U_g^B)^\dagger_{2,b}\rho^{\otimes 2r}\right]
    \end{aligned} \, ,
\end{align} 
where,
\begin{align}
	\rho = \bigotimes_p \rho_p, \quad \rho_p = |G|^{-1} \sum_{i_p, i_p'} \ket{i_p i_p i_p i_p} \bra{i_p' i_p' i_p' i_p'},
\end{align}
is the group cohomology ground state, with $p$ running over all plaquettes. Now, we have that,
\begin{align}
	\Tr[\pi_\alpha \rho_p^{\otimes 2r}]  = \Tr[\pi_\beta \rho_p^{\otimes 2r}]  = \Tr[\pi_\gamma \rho_p^{\otimes 2r}] = 1 \, ,
\end{align}
where $\pi_{\alpha,\beta,\gamma}$ act on all four qudits in these expressions.
Let us recall that the partial symmetry action on a region $B$ on the group cohomology ground state only has nontrivial action on the boundary of $B$. It then follows that for a plaquette $p$ not lying along the entanglement cut, $\rho_p$ contributes unity to $\Lambda_g^r$. This allows us to write the order parameter entirely in terms of the GHZ states lying on the entanglement cut:

\begin{align}
    \oL^r_g(\psi) = \sum_{\{\fp_{l,(a,b)},\fp_{l,(a,b)}'\}}  \tilde{\oL}^r_{g}\left(\{\fp_{l,(a,b)}\} \right) S_{g}\left(\{\fp_{l,(a,b)},\fp_{l,(a,b)}'\} \right) \, .
\end{align}
Here, $l$ indexes the plaquettes lying on the entanglement cut---note that this means the only plaquettes which contribute to the order parameter are those lying exactly on the entanglement cut.
Additionally, $\fp_{l,(a,b)}$ and $\fp_{l,(a,b)}'$ index the states appearing in ket and bra of the GHZ state density matrix at plaquette $l$ and with replica index $(a,b)$; that is,
\begin{align}
    \rho_{l,(a,b)} = |G|^{-1} \sum_{\fp_{l,(a,b)}, \fp_{l,(a,b)}'} \ket{\fp_{l,(a,b)}}^{\otimes 4} \bra{\fp_{l,(a,b)}'}^{\otimes 4} \, .
\end{align}
The factor $\tilde{\oL}^r_g$ combines the phase factors arising from acting with the partial symmetry on the boundary of region $B$, while $S_g$ is a product of Kronecker delta functions, coming from the expectation value of the permutation operations and the symmetry action, which relates the different $\fp_{l,(a,b)}$ and $\fp_{l,(a,b)}'$. 

Using these Kronecker delta relations, one finds that the only nontrivial contributions to $\tilde{\oL}^r_g$ come from plaquettes on the boundary between $B$ and $C$; all the other phases cancel out.
One then finds we can write,
\begin{align}
    \begin{split}
    \tilde{\oL}^r_{g} & \left(\{\fp_{l,(a,b)}\} \right)  \\ 
    &= \prod_{b=1}^r \frac{T^g_{q_{1,(1,b)},s_{(1,b)}}}{T^g_{\bg q_{1,(2,b)},\bg s_{(2,b)}}} \left(\prod_{l=1}^{M-1} \frac{T^g_{q_{l+1,(1,b)},q_{l,(1,b)}}}{T^g_{\bg q_{l+1,(2,b)},\bg q_{l,(2,b)}}} \right)
    \end{split}
\end{align}
where we have indexed the plaquettes on the boundary between $B$ and $C$ as in Fig.~\ref{fig:type-I-rectangular-main}(b). After some work using the group cocycle relation of Eq.~\eqref{eq:cocycle}, one finds that the product of phase factors telescopes out to the endpoints of the boundary, yielding,
\begin{align}
    \tilde{\oL}^r_{g}\left(\{\fp_{l,(a,b)}\} \right) &= \prod_{b=1}^r \omega(g,g^b,g)^* = \mathcal{Z}[L(r,1)_g]^* \, .
\end{align}
This is precisely (the complex conjugate of) the TQFT partition function evaluated on the lens space $L(r,1)$ with holonomy $g$ \cite{tantivasadakarn2017} (see Appendix~\ref{app:group-cohomology}). One further verifies that the magnitude of the order parameter does not depend on the choice of symmetry $g$, such that
\begin{align}
    \oL^r_g(\psi) \propto \mathcal{Z}[L(r,1)_g]^*,
\end{align}
where the proportionality constant is an exponentially decaying function of the different lengths of the entanglement cut. This 
confirms that $\oL_g^r$ can be used to probe the Type-I/II invariants; we will comment on the appearance of the complex conjugation below.

\subsection{Type-III Computation}

\begin{figure}
    \centering
    \includegraphics[width=0.9\linewidth]{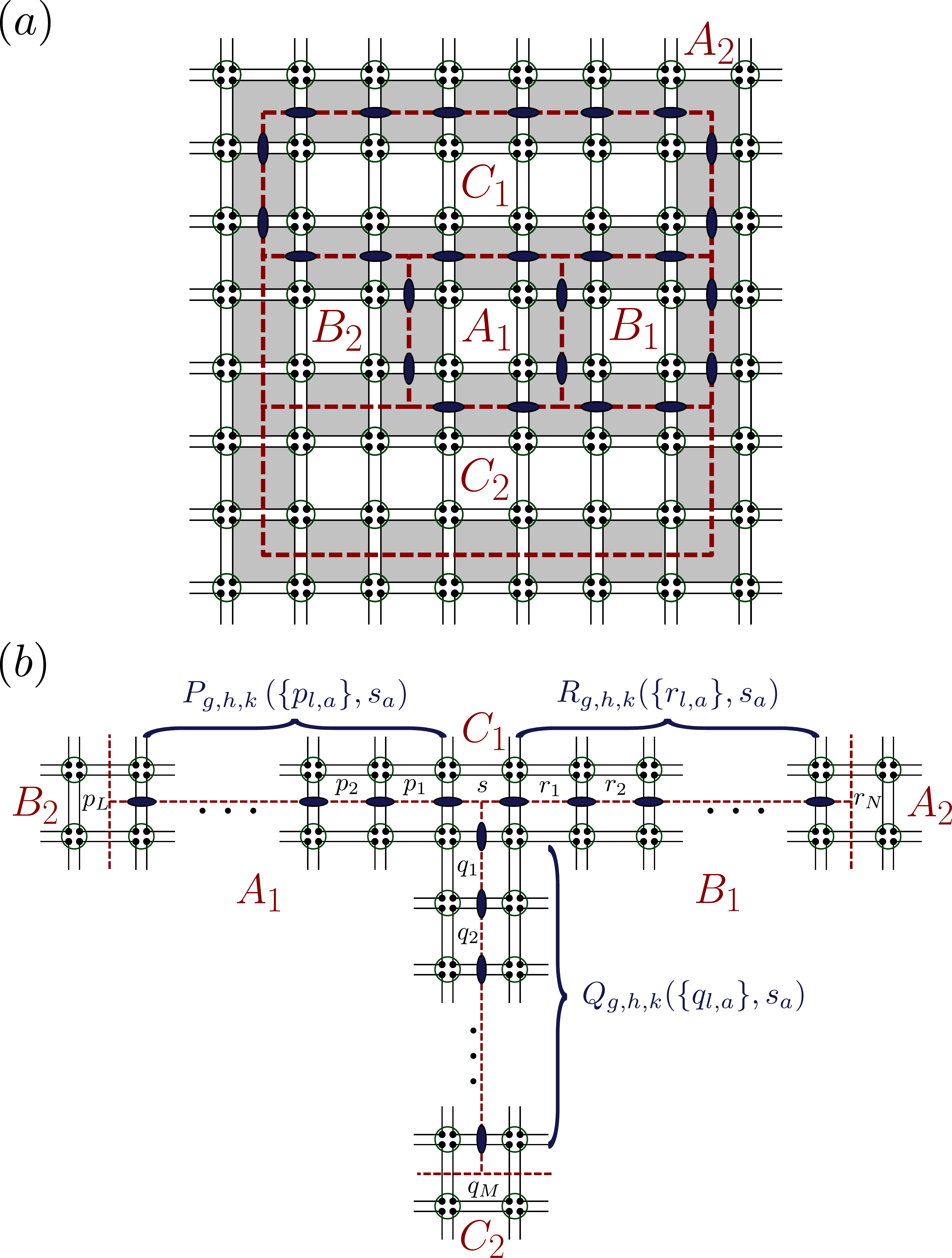}
    \caption{(a) Partitioning of the lattice used in the computation of  $\oT_{g,h,k}$. (b) Labeling of the plaquettes along the boundaries between the $A_1$, $B_1$, and $C_1$ subregions. The blue ellipses indicate phases depending on the states of adjacent plaquettes.}
    \label{fig:type-III-rectangular-main}
\end{figure}

We next turn to the evaluation of $\oT_{g,h,k}$, for which we consider a multipartition as in Fig.~\ref{fig:type-III-rectangular-main}(a). As in the preceding computation, we can express the order parameter evaluated in a lattice group cohomology state as (see Appendix~\ref{app:lattice-details} for more details),
\begin{align}
    \oT_{g,h,k}(\psi) = \sum_{\{\fp_{l,a},\fp_{l,a}'\}}  \tilde{\oT}_{g,h,k}\left(\{\fp_{l,a}\} \right) S_{g,h,k}\left(\{\fp_{l,a},\fp_{l,a}'\} \right) .
\end{align}
Again, $l$ indexes the plaquettes lying on the entanglement cut while $\fp_{l,a}$ and $\fp_{l,a}'$ index the states appearing in ket and bra of the GHZ state density matrix at plaquette $l$ and with replica index $a$. Here, $\tilde{\oT}_{g,h,k}$ groups together the phase factors from the partial symmetry, while $S_{g,h,k}$ groups together the Kronecker deltas arising from the trace. 

Using these Kronecker relations, one may show that the only non-trivial contributions to $\tilde{\oT}_{g,h,k}$ come from plaquettes lying on the boundaries between the three subregions $A_1$, $B_1$, and $C_1$, such that we can write,
\begin{align}
    \notag \tilde{\oT}_{g,h,k}\left(\{\fp_{l,a}\} \right) = &P_{g,h,k}\left(\{p_{l,a}\}, s_a \right) Q_{g,h,k}\left(\{q_{l,a}\}, s_a \right) \\
    & \times R_{g,h,k}\left(\{r_{l,a}\}, s_a \right) .
\end{align}
Here, $P$, $Q$, and $R$ are the phase contributions associated to the $C_1/A_1$, $A_1/B_1$, and $B_1/C_1$ boundaries, respectively. We have indexed the plaquettes lying on these boundaries as in Fig.~\ref{fig:type-III-rectangular-main}(b).

With a bit of work using the cocycle relation, Eq.~\eqref{eq:cocycle}, we find that the phase contributions all along the boundaries reduce to contributions coming from the plaquette at the $A_1 / B_1 / C_1$ trijunction. 
Explicitly, we find that
\begin{align}
    Q_{g,h,k}\left(\{q_{l,a}\}, s_a \right) = \frac{\omega(ak, g, h) \omega(a, h, g) }{\omega(ak, h, g) \omega(a, g, h)}
\end{align}
where $a=\bg\bh\bk \bar s$, with $s$ indexing the state at the plaquette at said trijunction. Additionally,
\begin{align}
P_{g,h,k}\left(\{p_{l,a}\}, s_a \right) &= Q_{k,g,h}\left(\{p_{l,a}\}, s_a \right)  , \\
R_{g,h,k}\left(\{r_{l,a}\}, s_a \right) &= Q_{h,k,g}\left(\{r_{l,a}\}, s_a \right)  .
\end{align}
Again employing the cocycle relation, we may simplify the product of these expressions to find,
\begin{align}
    \begin{split}
    \tilde{\oT}_{g,h,k} & \left(\{p_{l,a}\},\{q_{l,a}\},\{r_{l,a}\}, s_a \right) \\
     = &\frac{\omega(g,h,k)\omega(k,g,h)\omega(h,k,g)}{\omega(g,k,h)\omega(h,g,k)\omega(k,h,g)} = \mathcal{Z}[T^3_{g,h,k}] \, ,
    \end{split}
\end{align}
which is precisely the TQFT partition function on the three-torus with holonomies $g$, $h$, and $k$ along its three non-contractible cycles \cite{tantivasadakarn2017} (see Appendix~\ref{app:group-cohomology}). Moreover, one can confirm that the magnitude of the order parameter does not depend on the choice of $g$, $h$, or $k$ such that,
\begin{align}
    \oT_{g,h,k}(\psi) \propto \mathcal{Z}[T_{g,h,k}^3] 
\end{align}
where, again, the proportionality constant is non-negative and decays exponentially with the lengths of the entanglement cut.
This verifies our claim for the Type-III order parameter. 

Note that $\oL_g^r$ evaluated to the complex conjugate of the expected lens space partition function, whereas $\Omega_{g,h,k}$ evaluated exactly to the three-torus partition function. As described at the end of Section~\ref{sec:tqft}, we believe this stems from a choice of convention in how the orientations of the fluxes in the TQFT are assigned to the partial symmetry operations in the definitions of the order parameters. Though we are unsure how to fix this convention, the above computations are in agreement with the claim that $\oL_g^r$ and $\oT_{g,h,k}$ are able to probe all topological invariants defining a 2D Abelian SPT.

\section{Discussion \label{sec:discussion} }

In this work, we have presented two entanglement order parameters, $\oL_g^r$ and $\oT_{g,h,k}$,
for 2D bosonic SPTs protected by Abelian, on-site symmetries.
Intriguingly, these correspond to symmetry-twisted versions of multipartite entanglement quantities known as multi-entropies.
We demonstrated that they satisfy fundamental consistency checks to serve as order parameters for such SPTs and as probes of symmetry-protected multipartite entanglement. Using TQFT methods, we argued that these order parameters effectively probe the braiding statistics of symmetry defects, which serve as topological invariants for Abelian SPTs.
Finally, we verified the validity of our proposal in a family of fixed-point lattice models realizing all SPTs of interest. 

Beyond being of conceptual interest, our order parameters may have practical value, in that we expect they can be probed in experiment using techniques such as randomized measurements \cite{elben2023}. With an eye towards quantum simulator experiments, it would also be interesting to explore whether the multipartite entanglement of 2D SPTs probed by order parameters has operational utility, analogous to how the entanglement structure of 1D SPTs allows them to serve as resource states for measurement-based quantum computation \cite{else2012}.

By combining partial symmetries and replica permutation operations, one can in principle simulate other spacetime partition functions with non-trivial holonomies and hence construct order parameters for a larger class of SPTs. For instance, Ref.~\cite{tantivasadakarn2017} demonstrated that partition functions extracting topological invariants can be constructed for both bosonic and fermionic SPTs protected by on-site symmetries in 2D, 3D, and 4D; 
our methods can readily be extended to construct order parameters for these classes of SPTs. It would also be interesting to extend our methods to SPTs protected by non-Abelian symmetries, mixed-state SPTs \cite{degroot2022,ma2023,ma2025,manjunath2025,huang2025a,huang2025b}, and symmetry-enriched criticality \cite{scaffidi2017,verresen2021,thorngren2021}.

In particular, 
just as the appropriate spacetime manifolds to consider for Abelian bosonic 1D SPTs and 2D Type-III SPTs are the two-torus and three-torus, respectively, the appropriate manifolds to consider for certain 3D and 4D SPTs 
are the four-torus and five-torus, respectively. Given that the order parameters for the former two are symmetry-twisted versions of bipartite and tripartite R\'enyi multi-entropies, respectively, we expect the latter two should be probed by symmetry-twisted tetrapartite and pentapartite multi-entropies, respectively. This suggests a hierarchy of symmetry-protected multipartite entanglement as a defining feature of SPTs protected by onsite Abelian symmetries. Intriguingly, a different relation between multipartite non-separability and anomalous symmetries (which appear on the boundaries of SPTs) was discussed in Ref. \cite{lessa2025}; it would be interesting to explore if there exists a relation between this and multipartite entanglement of the bulk SPT.
Moreover, while our order parameters demonstrate that SPTs supporting fluxes with a given set of braiding statistics have specific patterns of multipartite entanglement, it would interesting if one could make that connection directly. More precisely, can one show that if a $\mathbb{Z}_N$ flux in a $\mathbb{Z}_N$ symmetric, gapped phase has nontrivial spin, then it must have four-partite entanglement without making use of $\Lambda_g^r$? We leave this to future work.

Now, while we have argued that $\oL_g^r$ and $\oT_{g,h,k}$ probe SPT order, they can also receive contributions from other forms of many-body entanglement, including \emph{chiral} and \emph{topological} order. Indeed, as previously noted, $\oL_{g=e}^r$
was introduced and studied in Refs.~\cite{sheffer2025,sheffer2026} in the context of such states, where it was shown the phase of this order parameter receives universal contributions from the chiral central charge and the topological spins of the anyons supported by the topological order. 
Let us first consider the case where the state in question is in an \emph{invertible phase} (i.e. does not support anyons, but may be chiral). Then, by computing $\oL_g^r$  and normalizing by $\oL_{e}^r$, we can subtract out the contribution from the chiral central charge and extract the Type-I/II invariants. 
In contrast, since $\oT_{e,e,e}$ is manifestly real and non-negative, it cannot depend on the chiral central charge in a universal way---a real number is invariant under time-reversal while the chiral central charge is odd. Thus, the phase of $\oT_{g,h,k}$ itself should still extract the Type-III invariant. Hence, we expect our order parameters to still extract SPT data for invertible phases.

If the state lies in a noninvertible (i.e. topologically ordered) phase, then it is no longer meaningful to define SPT invariants for the state \cite{lu2016,barkeshli2019}. 
Nevertheless, it is natural to expect that $\oL_g^r$ and $\oT_{g,h,k}$ may also serve as useful order parameters for \emph{symmetry-enriched} topological order \cite{barkeshli2019,teo2015} in which, for instance, the anyons carry fractional symmetry numbers. We leave a full investigation of this to future work.

As previously noted, the computation of the order parameters in the lattice group cohomology models effectively reduces to a computation of the anomaly of the effective boundary symmetry. 
With this in mind, it may prove useful to investigate whether our claims in Eq.~\eqref{eq:order-param-claims} could be proven using tensor network methods. Indeed, the corresponding order parameter for 1D SPTs was originally constructed using matrix-product states \cite{haegeman2012}. More precisely, in our case one may hope that one could verify our claims for those states captured by semi-injective PEPS, which have been proposed to describe 2D SPTs \cite{molnar2018}.

More generally, an important question to ask is under what conditions the symmetry-twisted multi-entropies of the form Eq.~\eqref{eq:stme} probe universal data of a gapped phase. More precisely, we have constructed specific examples of symmetry-twisted multi-entropies that probe SPT phases, operating under the philosophy that they should simulate the corresponding path integral on a particular space-time manifold. This led to the choice of a specific set of permutation operators and pattern of partial symmetry operator applications. We then found that, with these choices, these quantities probe symmetry-protected multipartite entanglement in a manner that suggests they should be robust to ``generic" perturbations of the wavefunction and hence probe universal data of the phase of matter. 
Now, it need not be the case that arbitrary choices of permutations and patterns of partial symmetry operator applications will always yield a symmetry-twisted multi-entropy that probes universal data and is robust to perturbations. One might expect that some finite set of constraints imposed on the choice of permutations and partial symmetries should be sufficient to guarantee these features; this question will be answered in part in forthcoming work \cite{gass2026inprep}. 

Looking further afield, the study of symmetry resolved bipartite entanglement has proven fruitful in characterizing a range of equilibrium and non-equilibrium systems \cite{castroalvaredo2024}. As a further generalization to multipartite correlations, we expect symmetry-twisted multi-entropies will likely serve as useful tools in uncovering the universal data hidden in multipartite correlations in more generic many-body systems, beyond the context of topological phases.

\acknowledgments

We thank Vladimir Calvera, Naren Manjunath, Curt von Keyserlingk, Yarden Sheffer, Wilbur Shirley, and Carolyn Zhang for useful discussions. R.S. and R.V. thank KITP for hospitality during the program ``Noise-robust Phases of Quantum Matter", during which part of this work was completed. This research was supported in part by grant NSF PHY-2309135 to the Kavli Institute for Theoretical Physics (KITP).

\bibliography{references}

\newpage 

\appendix

\section{Review of Group Cohomology \label{app:group-cohomology}} 

In this Appendix, we provide a brief review of the salient structures of the cohomology of groups, focusing on $\cH^d[G,U(1)]$ for finite $G$, which provides the classification of bosonic SPTs in $d-1$ spatial dimensions; we refer the reader to Ref.~\cite{chen2013} for a more detailed discussion. First, we define a \emph{$d$-cochain} $\omega(g_1 , g_2, \dots , g_d) \in U(1)$ as a function,
\begin{align}
    \omega : \underbrace{G \times \dots G}_{d\text{ factors}} \to U(1) \, .
\end{align}
The $d$-cochains form a group $\mathcal{C}^d[G,U(1)]$ with group multiplication inherited from $U(1)$, viewed as a group. We next define the \emph{coboundary} operator,
\begin{align}
    d : \mathcal{C}^d[G,U(1)] \to \mathcal{C}^{d+1}[G,U(1)] \, ,
\end{align}
given explicitly by,
\begin{align}
    \begin{split}
    d\omega & (g_1, \dots , g_{d+1}) \\
    & = \omega(  g_2, \dots , g_{d+1}) \omega(g_1, \dots , g_{d})^{(-1)^{d+1}} \\
    &\qquad \times \prod_{i=1}^d \omega(g_1, \dots , g_i g_{i+1} , \dots , g_{d+1} )^{(-1)^i} \, ,
    \end{split} 
\end{align}
which satisfies $d^2 = 1$. A $d$-cochain is a \emph{$d$-cocycle} if it satisfies the \emph{cocycle condition}, $d\omega = 1$. 
For reference, we here write down the cocycle condition for $d=3$, which is relevant to the 2D SPTs of interest:
\begin{align}
	\frac{\omega(g_2, g_3, g_4) \omega(g_1, g_2 g_3, g_4) \omega(g_1, g_2, g_3)}{\omega(g_1 g_2, g_3, g_4)\omega(g_1, g_2, g_3 g_4)} = 1 \, . \label{eq:cocycle}
\end{align}
We frequently employ this relation in our lattice computations.
Conversely, a $d$-cochain is a \emph{$d$-coboundary} if it can be written as $\omega = d\mu$ for some $(d-1)$-cochain $\mu$. The $d^{th}$ cohomology group $\cH^d[G,U(1)]$ is then defined as the quotient of the group of $d$-cocycles by the group of $d$-coboundaries. The equivalence classes $[\omega] \in \cH^d[G,U(1)]$ label distinct SPTs in $d-1$ spatial dimensions.

\subsection{Partition Functions and Topological Invariants for 2D Abelian SPTs}

The group cohomology models introduced in Ref.~\cite{chen2013} are a family of models described by path integrals on discretized Euclidean spacetimes realizing all bosonic SPT phases within the group cohomology classification and in arbitrary dimension, for a given symmetry group $G$. By coupling these models to a background gauge field for $G$ and integrating out the matter fields, one obtains the effective response theory for a given SPT. We refer the reader to, e.g. Refs.~\cite{wang2015,tantivasadakarn2017} for the details of this procedure and here simply state the result.

Working in $d$ spacetime dimensions, the input for the response function is the symmetry group $G$ and a cocycle $\omega(g_1 , \dots , g_d)$, with $g_i \in G$, where $[\omega]  \in \cH^d[G,U(1)]$ determines to which SPT class the theory belongs. We wish to construct the response theory for the SPT on a manifold $M$. To that end, we consider a triangulation of $M$ into a simplicial complex; for instance, in our primary case of interest, $d=3$, we decompose $M$ into tetrahedra, with orientations assigned to each edge such that they do not form closed loops. To each edge, we assign a group-valued degree of freedom, corresponding to the gauge field. We impose the condition that the gauge field configuration is \emph{flat}; that is, the oriented product of group elements around a $d-1$ simplex is the identity element. The corresponding SPT partition function on this manifold is then given by,
\begin{align}
    \mathcal{Z}[\omega;M] = \prod_{[ij \dots l]} \left[ \omega(g_i, g_j , \dots g_l) \right]^{s_{[ij \dots l]}}.
\end{align}
Here, the product is over all simplices, $[ij \dots l]$, each of which we label with an ordered sequence of edges $i < j < \dots < l$. The orientation of the simplex is given by $s_{[ij \dots l]} = \pm 1$. Hence, each simplex $[ij \dots l]$ contributes the $U(1)$-valued weight $\left[ \omega(g_i, g_j , \dots g_l) \right]^{s_{[ij \dots l]}}$, where $g_i$ is the group element at edge $i$. Using the fact that $\omega$ is a cocycle, one can show that this partition function is independent of the choice of simplicial decomposition of $M$, such that it describes a topological theory, and two cocycles yield the same partition function if and only if they lie within the same cohomology class. Hence, a partition function constructed from a cocycle in a given cohomology class yields the response function for SPT phases in that cohomology class.

If $M$ is a closed manifold, the partition function evaluates to a phase, $\mathcal{Z}[\omega;M] \in U(1)$. If $M$ has trivial genus, this phase is trivial. However, if $M$ has non-trivial genus and one chooses a gauge field configuration with non-trivial holonomies (i.e. symmetry flux) along each cycle, this phase can be made non-trivial. As cocycles within the same cohomology class yield the same partition function, this implies these phases may be taken as \emph{topological invariants} for SPTs.

In the case of 2D bosonic Abelian SPTs, it was shown in Ref.~\cite{tantivasadakarn2017} that an exhaustive set of such topological invariants can be obtained in this manner. 
Indeed, as discussed in the main text, for an Abelian symmetry group $G = \prod_i \mathbb{Z}_{N_i}$, there are three ``types" of SPTs, corresponding to the classes of topological indices in Eq.~\eqref{eq:kunneth}. The Type-I and Type-II indices can be extracted from the lens space partition functions, while the Type-III indices can be extracted from the three-torus partition function. We discuss the corresponding partition functions below.

\begin{figure}
    \centering
    \includegraphics[width=\linewidth]{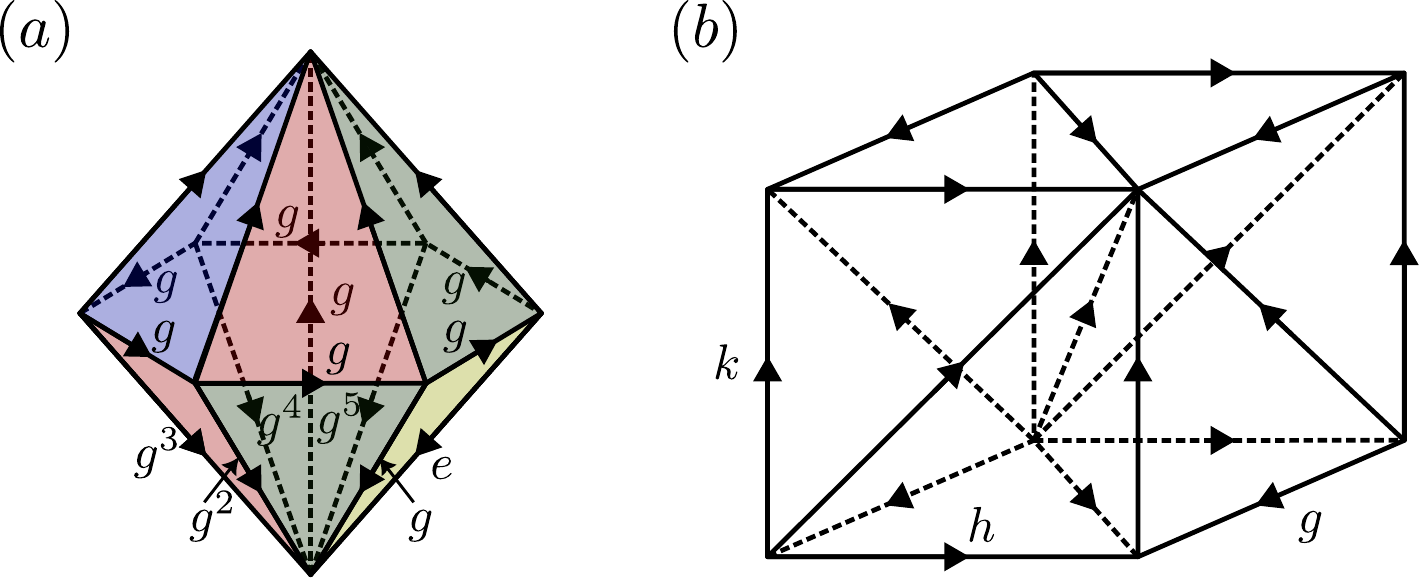}
    \caption{Simplicial decompositions for the (a) Lens space $L(r,1)_g$ with $r=6$ and (b) the three-torus $T^3_{g,h,k}$. Each decomposition involves six tetrahedra. In (a), the colored faces are identified while in (b), opposing faces are identified. The edges are labeled by the value of the gauge field. The values on the unlabeled edges are fixed by the flatness condition and the identification of faces.}
    \label{fig:app-partition-functions}
\end{figure}

\subsubsection{Lens Space}

The lens space $L(r,1)$ may be obtained by taking the three-ball and identifying the northern hemisphere with the southern hemisphere, after a clockwise (when looking down at the north pole) $2\pi / r$ rotation of the former relative to the latter. We illustrate this gluing in Fig.~\ref{fig:app-partition-functions}(a) for the case of $L(r,1)$, where the three-ball is represented as a hexagonal bipyramid. Here, faces of the same color are identified. This space has a single non-contractible cycle corresponding to the great circle connecting the north and south poles, which are identified due to the gluing of the northern and southern hemispheres. Note that a point on the equator with azimuthal angle $\theta$ is identified with the equatorial point at angle $\theta+2\pi/r$. There are thus $r$ line segments on the equator which are identified with one another (corresponding to the six edges of the hexagonal base in Fig.~\ref{fig:app-partition-functions}(a)). Any one of these segments is also a noncontractible cycle, but they are not independent of the great circle connecting the northern and southern poles; indeed, one can check that the latter can be deformed into any one of the former, and thus they are homologically equivalent. Through this equivalence of the great circle to any one of the equatorial segments, we see that the noncontractible cycle has order $r$; that is, traversing the cycle $r$ times in fact yields a contractible path.

For a symmetry group $G$ and $g \in G$ such that $g^r=e$, we denote by $L(r,1)_g$ the lens space with a $g$ holonomy along the great circle. One finds that,
\begin{align}
    \mathcal{Z}[\omega;L(r,1)_g] = \prod_{b=1}^r \omega (g,g^b,g).
\end{align}
This expression may be directly obtained by considering the simplicial decomposition of $L(r,1)$, as shown in Fig.~\ref{fig:app-partition-functions}(a) for $r=6$. Now, let $g_i$ be the generator of the $\mathbb{Z}_{N_i}$ factor of $G$. Then the lens space partition function 
$Z_i \equiv  \mathcal{Z}[\omega;L(N_i,1)_{g_i}]$
extracts an invariant fixing the Type-I index $\mathbb{Z}_{N_i}$ in Eq.~\eqref{eq:kunneth}. 
As shown in Ref.~\cite{tantivasadakarn2017}, this invariant is $N^i$ times the spin of a $\mathbb{Z}_{N_i}$ vortex.
Similarly, for generators $g_i, g_j$ of $\mathbb{Z}_{N_{i,j}}$, respectively, one may compute the lens space partition function 
$Z_{ij} \equiv  \mathcal{Z}[\omega;L(N^{ij},1)_{g_ig_j}]$
with $r=N^{ij}=\mathrm{lcm}(N_i,N_j)$.
This corresponds to $N^{ij}$ times the spin of a composite $\mathbb{Z}_{N_i}$ and $\mathbb{Z}_{N_j}$ vortex.
An invariant fixing the Type-II class $\mathbb{Z}_{N^{ij}}$ in Eq.~\eqref{eq:kunneth} is then given by $Z_{ij} Z_i^{-N^{ij}/N_i} Z_j^{-N^{ij}/N_j}$, which yields the braiding phase of a $\mathbb{Z}_{N_i}$ vortex with a $\mathbb{Z}_{N_j}$ vortex.

\subsubsection{Three-Torus}

Let $T^3_{g,h,k}$ be the three-torus with flux insertions $g$, $h$, and $k$ along the three non-contractible cycles, with the orientation shown in Fig.~\ref{fig:app-partition-functions}(b). One finds,
\begin{align}
    \mathcal{Z}[\omega;T^3_{g,h,k}] = \frac{\omega(g,h,k)\omega(k,g,h)\omega(h,k,g)}{\omega(g,k,h)\omega(h,g,k)\omega(k,h,g)}
\end{align}
Once again, one can read off this expression from the simplicial decomposition given in Fig.~\ref{fig:app-partition-functions}(b). For the generators $g_{i,j,k}$ of $\mathbb{Z}_{N_{i,j,k}}$, respectively, the partition function on $T^3_{g_i,g_j,g_k}$ yields a topological invariant fixing the Type-III index $\mathbb{Z}_{N_{ijk}}$ in Eq.~\eqref{eq:kunneth}. As shown in Ref.~\cite{tantivasadakarn2017}, this invariant is
the braiding phase accrued by braiding a $\mathbb{Z}_{N_i}$ vortex clockwise around a $\mathbb{Z}_{N_j}$ vortex and then a $\mathbb{Z}_{N_k}$ vortex followed by braiding the $\mathbb{Z}_{N_i}$ vortex \emph{counter}-clockwise around the $\mathbb{Z}_{N_j}$ vortex and then the $\mathbb{Z}_{N_k}$ vortex.

\section{Constraining Spurious Contributions \label{app:spurious} }

\begin{figure}
  \centering
    \includegraphics[width=\linewidth]{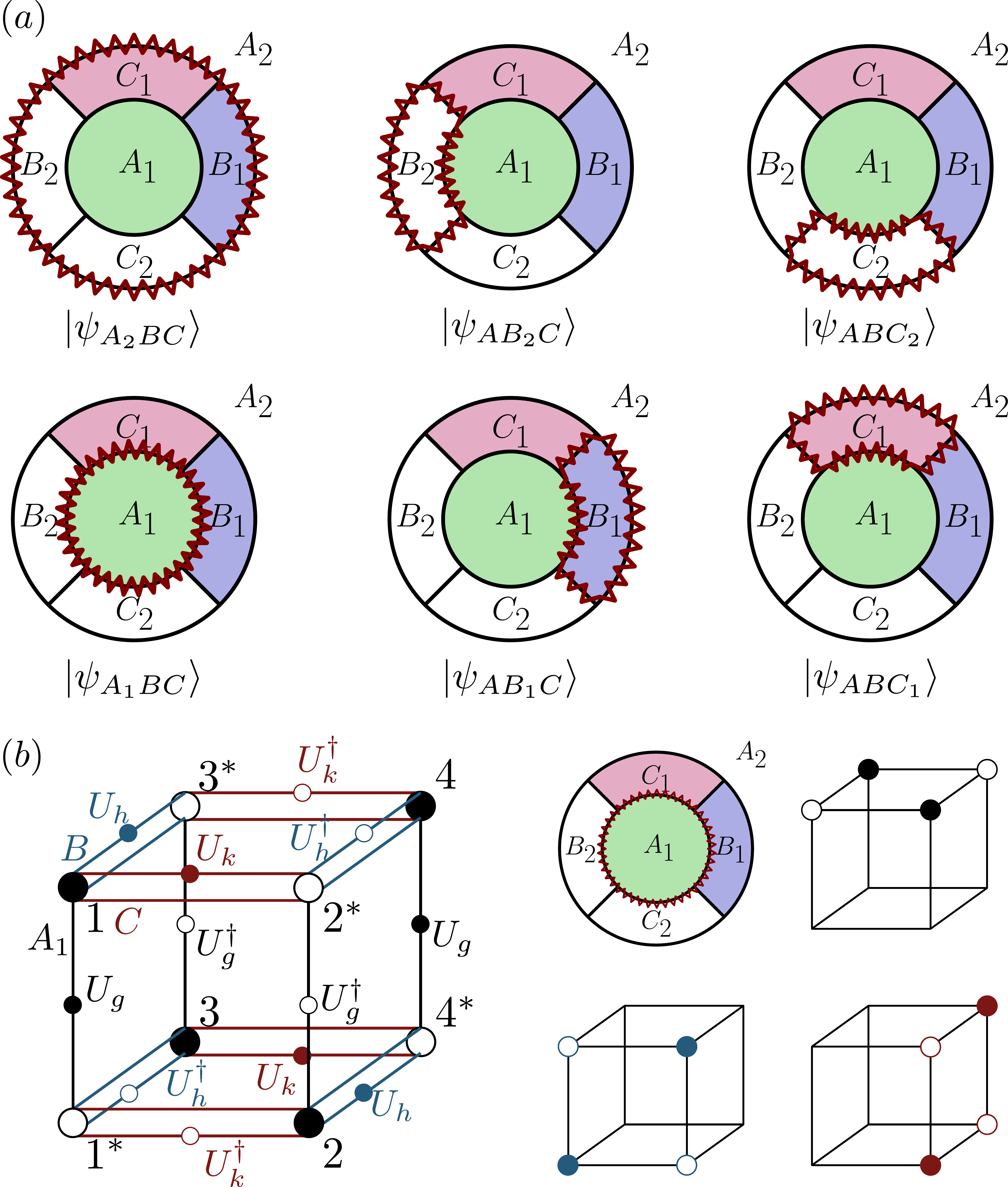}
    \caption{(a) Examples of symmetric 1D states, represented by the jagged red lines dressing the entanglement cuts which \emph{cannot} contribute to the phase of the Type-III order parameter. (b) Tensor network representation for the Type-III order parameter evaluated on a state of the form $\ket{\psi_{A_1 BC}}$. Note that each vertex has two blue and two red legs, corresponding to the $B_{1,2}$ and $C_{1,2}$ subregion indices, respectively. Recall that partial symmetries only act on the $A_1$, $B_1$, and $C_1$ legs. As in Fig.~\ref{fig:type-III-positivity}, the cubes on the right indicate the pattern of global symmetry applications that can be used to eliminate all partial symmetry insertions.}\label{fig:type-III-nonspurious}
\end{figure}

In the main text, we showed that $\oT_{g,h,k}$ 
has a trivial phase on states which entangle at most three adjacent subregions. This allowed us to exclude spurious contributions from 0D entangled states. In this Appendix, we go a step further and show that the Type-III order parameter in fact has a trivial phase when evaluated on any pentapartite entangled state. 
There are six such classes of states, as depicted in Fig.~\ref{fig:type-III-nonspurious}(a), and they include states of the form, for example, $\ket{\psi_{A_1 BC}}$ and $\ket{\psi_{A_2 BC}}$. These describe 1D states that, among others, lie along the boundaries of $A_1$ and $A_2$, respectively, which are precisely those that could yield spurious contributions to, for instance, the topological entanglement entropy \cite{zou2016,williamson2019,kato2020,gass2024}.

To show that these states do not yield spurious contributions to $\oT_{g,h,k}$, we again use the positivity arguments employed in the main text. Let us first focus on a state of the form $\ket{\psi_{A_1 BC}}$; the tensor network representation of the order parameter evaluated on this state is given in Fig.~\ref{fig:type-III-nonspurious}(b). As before, we can arrange the kets and bras to lie on the vertices of a cube. The $A_1$ legs form the vertical edges of the cube. Each vertex now, however, has two blue legs and two red legs, corresponding to the $B_{1,2}$ and $C_{1,2}$ subregions; the edges of the top and bottom faces of the cubes thus consist of pairs of legs. Note that only the $A_1$, $B_1$, and $C_1$ legs have partial symmetries acting on them. In spite of this additional structure, we can again use the gauge symmetry of the tensor network described in Section~\ref{sec:multipartite} of the main text to eliminate all partial symmetry insertions on the legs. The pattern of global symmetries that need to be applied is shown in Fig.~\ref{fig:type-III-nonspurious}(b) and, in fact, corresponds to the same set of gauge transformations used in Fig.~\ref{fig:type-III-positivity}(b). Once the partial symmetries are eliminated, the tensor network has three manifest reflection symmetries, corresponding to the reflection symmetries of the cube. 
Essentially identical considerations imply that $\oT_{g,h,k}$ evaluated on states of the form $\ket{\psi_{AB_1 C}}$ and $\ket{\psi_{ABC_1}}$ is real and non-negative. Likewise, the tensor network representation of $\Omega_{g,h,k}$ evaluated on states of the form $\ket{\psi_{A_2 BC}}$ (i.e. those appearing in the first row of Fig.~\ref{fig:type-III-nonspurious}(a)) has a similar structure, except now partial symmetries act on only two sets of legs. Hence, one needs only two of the three gauge transformations applied in Fig.~\ref{fig:type-III-nonspurious} to eliminate the partial symmetry insertions, after which the tensor network becomes manifestly reflection symmetric.

Altogether, we find that $\oT_{g,h,k}$ only has a nontrivial phase on states which entangle all six subregions, making manifest the sense in which it probes multipartite entanglement. In particular, this rules out spurious contributions from certain 1D entangled states dressing the entanglement cut. Indeed, as shown in Fig.~\ref{fig:type-III-nonspurious}(a), a 1D state dressing the entanglement cut surrounding any given single subregion only entangles five subregions and hence cannot give a nontrivial phase to the Type-III order parameter. Note that spurious contributions to entanglement-based order parameters like the TEE precisely come from such 1D states. 
This does not rule out all spurious 1D entangled states; for instance, a 1D state encircling $A_1$ and $B_1$ can entangle all six regions and hence, in principle, can contribute to the order parameter.
Nevertheless, the fact that we can rule out certain classes of spurious entanglement  gives us additional confidence in the stability of the phase of $\oT_{g,h,k}$ to ``generic" local perturbations.

\section{Details of the Lattice Computations} \label{app:lattice-details}

In this appendix we provide the full details of the computations of the order parameters in the lattice group cohomology models. We recall that the ground states, $\ket{\phi}$ of these models were defined in Eq.~\eqref{eq:grp-coh-state}. Central to the following computations is the action of the partial symmetry $U_g^A$ on some region $A$ acting on $\ket{\phi}$, given by Eq.~\eqref{eq:MPU}. For convenience, we include here the action of the Hermitian conjugate, $(U_g^A)^\dagger$,
\begin{align}
    (U_g^A)^\dagger \ket{\phi} = \left(\sum_{\{p_l \}} \left(T_{\{ \bg p_l\}}^g\right)^* L_{\{ p_l\}}^{\bg} \right) \ket{\phi} \, ,
\end{align}
where $\bg$ indicates the group inverse of $g$.

\begin{widetext}

\subsection{Type-I/II Order Parameter Computation}

\begin{figure}
    \centering
    \includegraphics[width=0.7\linewidth]{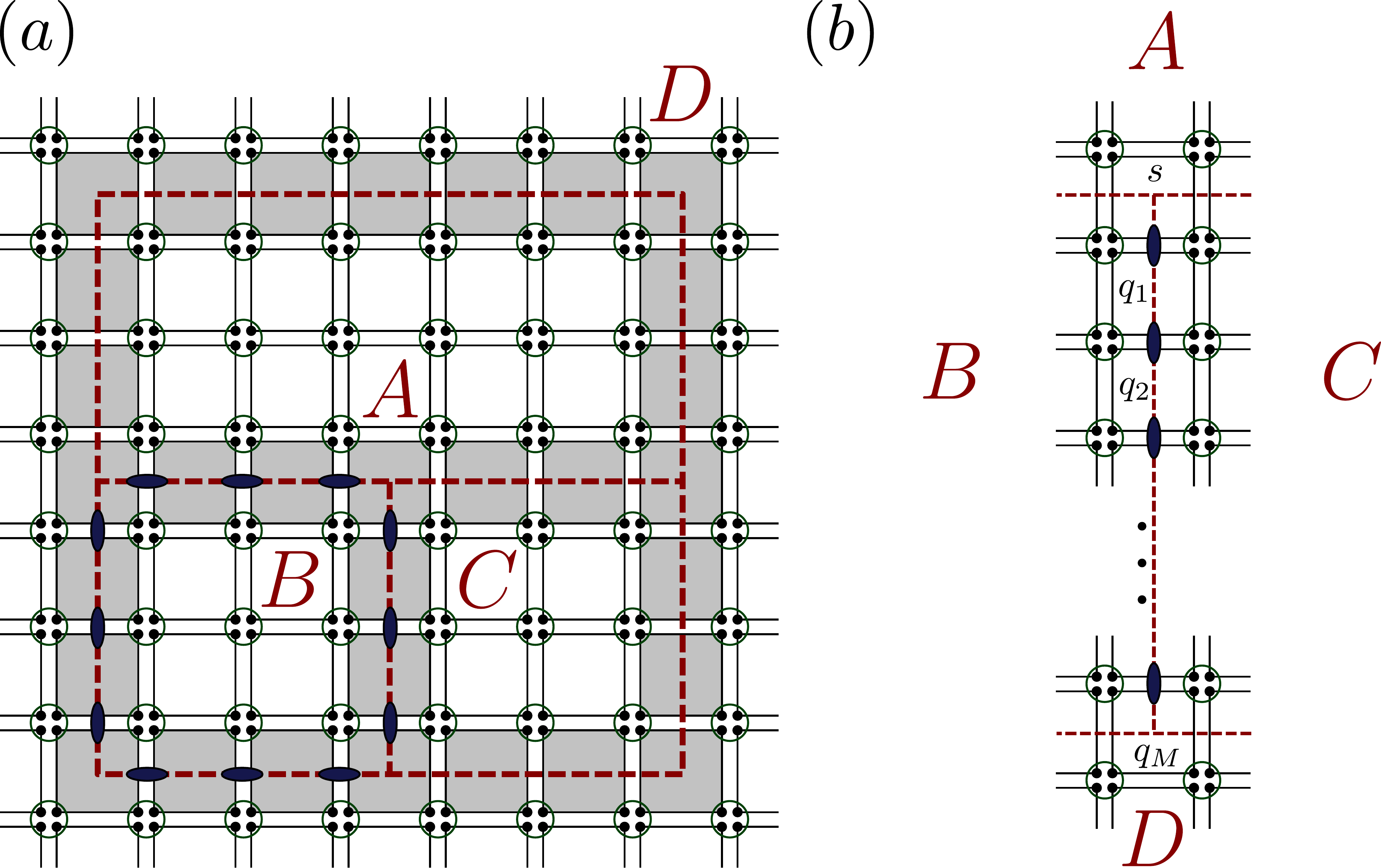}
    \caption{(a) Partitioning of the lattice used in the computation of $\oL_g^r$. The shaded plaquettes are those lying along the entanglement cut and which contribute to the magnitude of the order parameter. The dark blue ellipses indicate phase factors, which depend on the states of two adjacent plaquettes, appearing in Eq.~\eqref{eq:type-I-total-phase}.
    (b) Labeling of the plaquettes along the boundary between $B$ and $C$. These phases, on the boundary between $B$ and $C$ and appearing in Eq.~\eqref{eq:type-I-phase}, are the only ones which do not trivially vanish.}
    \label{fig:type-I-rectangular}
\end{figure}

\subsubsection{Setup}

Let us consider a group cohomology model with symmetry group $G$ on the square lattice. The topology of the lattice---whether it be a disk, torus, etc.---is irrelevant. 
We consider a multipartition like that in Fig.~\ref{fig:type-I-rectangular}.
Let $g \in G$ be a generator of a $\mathbb{Z}_r$ factor of $G$.
We claim that,
\begin{align}
	\oL^r_g(\psi) = e^{i\vartheta_g} \oL^r_e(\psi),
\end{align}
where $\vartheta_g$ is a universal phase corresponding to the (complex conjugate of the) SPT partition function on the Lens space $L(r,1)_g$, which can be used to extract the Type-I and Type-II topological invariants, as described in the main text.
Our goal is to show this and to extract this universal phase.

In order to organize our computation, it will be most convenient to use the representation of the order parameter given in the form,
\begin{align}
	\begin{aligned}
	\oL_{g}^r(\psi) = \Tr\left[ \pi_\alpha^A \pi_\beta^B \pi_\gamma^C  \bigotimes_{b=1}^r (U_g^B)_{1,b} (U_g^B)^\dagger_{2,b}\rho^{\otimes 2r}\right]
    \end{aligned} \, .
\end{align} 
Our computation proceeds in several steps. First, we establish that the only contributions to both the magnitude and phase of the order parameter come from GHZ states crossing the entanglement cuts. Second, we show that the non-trivial contributions to the phase come from GHZ states that cross the boundary between $B$ and $C$. Finally, in the most lengthy step, we show that this phase factor is precisely the desired topological invariant.

To begin, let us recall that the group cohomology density matrix has the form
\begin{align}
	\rho = \bigotimes_p \rho_p, \qquad \rho_p = |G|^{-1} \sum_{i_p, i_p'} \ket{i_p i_p i_p i_p} \bra{i_p' i_p' i_p' i_p'}
\end{align}
where $p$ runs over all plaquettes and $\rho_p$ is a state supported on the four qudits on the corners of the plaquette. One readily verifies that,
\begin{align}
	\Tr[\rho_p^{\otimes 2r}] = \Tr[\pi_\alpha \rho_p^{\otimes 2r}]  = \Tr[\pi_\beta \rho_p^{\otimes 2r}]  = \Tr[\pi_\gamma \rho_p^{\otimes 2r}] = 1 \, ,
\end{align}
where $\pi_{\alpha,\beta,\gamma}$ act on all four qudits in these expressions. Thus, if $p$ is contained entirely within one of the subregions $A$, $B$, $C$, or $D$ (i.e. all four qudits on the corners of $p$ are contained in one of these subregions), then $\rho_p$ does not contribute to the order parameter. Here we also made use of the fact noted in the main text that a partial symmetry operation acting on a group cohomology wavefunction only has nontrivial action on the boundary of its support. 

It then follows that the nontrivial contributions to the phase and magnitude of the order parameter only come from plaquettes crossing the entanglement cuts [see Fig.~\ref{fig:type-I-rectangular}(a)]. With this in mind, it will be convenient for the purposes of organizing our calculation to write the order parameter as, 
\begin{align}
	\oL^r_g(\psi) = \sum_{\{\fp_{l,(a,b)},\fp_{l,(a,b)}'\}}  \tilde{\oL}^r_{g}\left(\{\fp_{l,(a,b)}\} \right) S_{g}\left(\{\fp_{l,(a,b)},\fp_{l,(a,b)}'\} \right) \, .
\end{align}
In $\fp_{l,(a,b)}$, $l$ indexes the plaquettes lying along the entanglement cut while $(a,b)$ denotes the replica index. The density matrix of the plaquette at position $l$ is given by $\rho_l = |G|^{-1}\sum_{\fp_l,\fp_l'} \ket{\fp_l \fp_l \fp_l \fp_l} \bra{\fp_l' \fp_l' \fp_l' \fp_l'}$. We have grouped the phase contributions and the trace into the terms $\tilde{\oL}^r_g$ and $S_g$, respectively. Explicitly, 
\begin{align}
	S_g = \Tr\left[\pi_\alpha^{\partial A} \pi_\beta^{\partial B} \pi_\gamma^{\partial C}  \bigotimes_{b=1}^r (V_g^{\partial B})_{1,b} (V_g^{\partial B})^\dagger_{2,b} \bigotimes_{a=1}^2\bigotimes_{b=1}^r \bigotimes_{l} \rho(\fp_{l,(a,b)},\fp_{l,(a,b)}')\right] , \label{eq:type-I-trace}
\end{align}
where,
\begin{align}
	\rho(a,b) \equiv |G|^{-1} \ket{aaaa}\bra{bbbb} \, ,
\end{align}
and we have defined 
\begin{align}
\pi_{\alpha,\beta,\gamma}^{\partial X} = \bigotimes_{i \in \partial X} \pi_{\alpha,\beta,\gamma}^i,
\end{align}
where the product is over all qudits lying in the boundary plaquettes of $X$ as well as,
\begin{align}
V_g^{\partial X} = \bigotimes_{i\in \partial X \cap X } V_g^i,
\end{align}
where the product is over all qudits lying both in the boundary plaquettes of $X$ as well as in $X$ itself.
Here we have defined $V_g^i$ as $V_g^i \ket{h}_i=\ket{gh}_i$.
Evaluating the trace gives a product of Kronecker delta functions relating the different plaquette indices; rather than writing them each out explicitly, we simply note that they enforce the following equalities, depending on which subregions the plaquette corresponding to the index $\mathfrak{p}_l$ intersects:
\begin{align}
	\begin{aligned}
	\fp& \in A \implies 
		\left\{
	\begin{aligned}
		\fp_{(1,b)} &= \fp_{(1,b-1)}' \\
		\fp_{(2,b)} &= \fp_{(2,b+1)}'
	\end{aligned}\right. \, ,
	\quad
	\fp \in B \implies 
	\left\{
	\begin{aligned}
		g   \fp_{(1,b)} &= \fp_{(2,b)}' \\
		\bg \fp_{(2,b)} &= \fp_{(1,b)}'
	\end{aligned}\right.\, ,
	\\
	\fp& \in C \implies
	\left\{
	\begin{aligned}
		\fp_{(1,b)} &= \fp_{(2,b-1)}' \\
		\fp_{(2,b)} &= \fp_{(1,b+1)}'
	\end{aligned}\right.	 \, ,
	 \quad
	\fp \in D \implies 
		\fp_{(a,b)} = \fp_{(a,b)}' \, .
	\end{aligned}
	 \label{eq:kronecker-deltas-I}
\end{align}
Here we have suppressed the subscript $l$. These constraints are most readily deduced by using a diagrammatic representation of Eq.~\eqref{eq:type-I-trace}. In Fig.~\ref{fig:type-I-kronecker}, we provide such a representation for a contribution to Eq.~\eqref{eq:type-I-trace} for $r=3$ coming from the plaquette located at the trijunction between the subregions $A$, $B$, and $C$.

\begin{figure}
    \centering
    \includegraphics[width=0.65\linewidth]{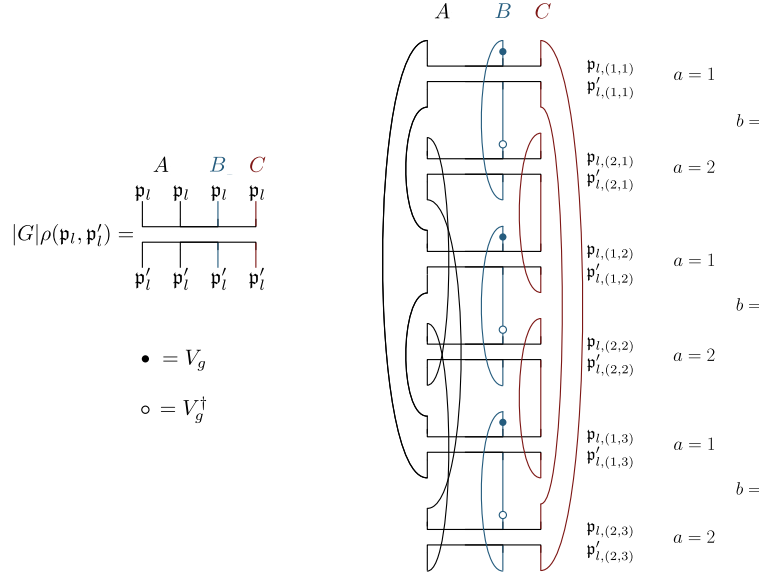}
    \caption{A diagrammatic representation of the contribution to Eq.~\eqref{eq:type-I-trace} for $r=3$ from the plaquette $l$ at the trijunction between the subregions $A$, $B$, and $C$ (see Fig.~\ref{fig:type-I-rectangular}); this plaquette has two qudits in the $A$ subregion, and one in each of $B$ and $C$. (Left) A tensor network depiction of $\rho(\fp_l,\fp_l')$, with the physical indices colored according to the subregion they belong to. We represent the action of $V_g$ and $V_g^\dagger$ with filled-in and empty circles, respectively. (Right) The contribution to Eq.~\eqref{eq:type-I-trace}. Evaluating the contractions yields Kronecker deltas enforcing the $A$, $B$, and $C$ constraints of Eq.~\eqref{eq:kronecker-deltas-I} on $\fp_l$ and $\fp_l'$. To reduce clutter, we have grouped the two physical indices corresponding to the two $A$ qudits into a single index.}
    \label{fig:type-I-kronecker}
\end{figure}

The total phase, $\tilde{\oL}^r_g$, is a product of the $T_{ij}^g$ coming from the partial symmetry action along the boundary of $B$:
\begin{align}
	\tilde{\oL}^r_{g}\left(\{\fp_{l,(a,b)}\}\right) =  \prod_{l \in \partial B}\prod_{b=1}^r \left[ T_{\fp_{l,(1,b)},\fp_{l+1,(1,b)}}^g (T_{\bg \fp_{l,(2,b)},\bg \fp_{l+1,(2,b)}}^g)^* \right]  \label{eq:type-I-total-phase}
\end{align}
The product is over the plaquettes along the boundary of $B$, with the phases coming from the partial symmetry action along said boundary on each replica. 
The index $l$ labels the plaquettes counter-clockwise around the boundary. 

\subsubsection{Computing the Phase}

Now, we show that the phase contributions from plaquettes along the boundary between $B$ and $D$ are trivial. Indeed, consider two neighboring plaquettes, $\fp_l$ and $\fp_{l+1}$ on a portion of the entanglement cut separating $B$ and $D$. For concreteness, let us assume that the bond points from $l$ to $l+1$. Then these will contribute a factor of,
\begin{align}
	 \prod_{b=1}^r  \theta_{\fp_{l,(1,b)},\fp_{l+1,(1,b)}}^g (\theta_{\bg \fp_{l,(2,b)},\bg \fp_{l+1,(2,b)}}^g)^*  \in \tilde{\oL}^r_{g}\left(\{\fp_{l,(a,b)}\}\right) \, ,
\end{align}
to the total phase factor. Now, since these two plaquettes intersect both $B$ and $D$, we have from Eq.~\eqref{eq:kronecker-deltas-I} that,
\begin{align}
	\left\{
	\begin{aligned}
		g   \fp_{(1,b)} &= \fp_{(2,b)}' \\
		\bg \fp_{(2,b)} &= \fp_{(1,b)}'
	\end{aligned}\right. \quad \text{and} \quad\quad
\fp_{(a,b)} = \fp_{(a,b)}' \implies
	\fp_{(2,b)} = g \fp_{(1,b)}
\end{align}
for both $\fp_l$ and $\fp_{l+1}$, where we have suppressed the $l$ subscript in the above. Using these equalities, we have that
\begin{align}
	\theta_{\fp_{l,(1,b)},\fp_{l+1,(1,b)}}^g (\theta_{\bg \fp_{l,(2,b)},\bg \fp_{l+1,(2,b)}}^g)^* = \theta_{\fp_{l,(1,b)},\fp_{l+1,(1,b)}}^g (\theta_{ \fp_{l,(1,b)}, \fp_{l+1,(1,b)}}^g)^* = 1 \, ,
\end{align}
and hence the phase contribution vanishes, as claimed. It is straightforward to see the same is true if the bond points instead from $l+1$ to $l$.

Next, let us turn to the contributions from the boundary between $B$ and $A$. Again, let us consider two neighboring plaquettes,  $\fp_l$ and $\fp_{l+1}$ on a portion of the entanglement cut separating $B$ and $A$. Again, let us first assume the bond points from $l$ to $l+1$. Then this pair of plaquettes will contribute a factor of,
\begin{align}
	 \prod_{b=1}^r  \theta_{\fp_{l,(1,b)},\fp_{l+1,(1,b)}}^g (\theta_{\bg \fp_{l,(2,b)},\bg \fp_{l+1,(2,b)}}^g)^* \in \tilde{\oL}^r_{g}\left(\{\fp_{l,(a,b)}\}\right) \, ,
\end{align}
to the total phase factor. Now, since these two plaquettes intersect both $B$ and $A$, we have from Eq.~\eqref{eq:kronecker-deltas-I} that,
\begin{align}
	\left\{
	\begin{aligned}
		g   \fp_{(1,b)} &= \fp_{(2,b)}' \\
		\bg \fp_{(2,b)} &= \fp_{(1,b)}'
	\end{aligned}\right. \quad \text{and} \quad\quad
		\left\{
	\begin{aligned}
		\fp_{(1,b)} &= \fp_{(1,b-1)}' \\
		\fp_{(2,b)} &= \fp_{(2,b+1)}'
	\end{aligned}\right. \implies
	\fp_{(2,b)} = g \fp_{(1,b+1)} \, .
\end{align}
Hence,
\begin{align}
	 \prod_{b=1}^r  \theta_{\fp_{l,(1,b)},\fp_{l+1,(1,b)}}^g (\theta_{\bg \fp_{l,(2,b)},\bg \fp_{l+1,(2,b)}}^g)^*  = \prod_{b=1}^r  \theta_{\fp_{l,(1,b)},\fp_{l+1,(1,b)}}^g (\theta_{\fp_{l,(1,b+1)}, \fp_{l+1,(1,b+1)}}^g)^* = 1 \, ,
\end{align}
where, in the last equality, we used the fact that $b+r$ is identified with $b$. Again, one readily verifies the same holds if the bond instead points from $l+1$ to $l$. The phase contribution from this portion of the entanglement cut thus is also trivial.

Finally, we turn to the boundary between $B$ and $C$. 
We label the plaquettes on this boundary as in Fig.~\ref{fig:type-I-rectangular}: $q_l$ label the plaquettes on the $B/C$ boundary with $q_M$ being the plaquette at the $BCD$ trijunction, while $s$ labels the plaquette at the $ABC$ trijunction. We then have for the total phase factor, 
\begin{align}
	\tilde{\oL}^r_{g}\left(\{\fp_{l,(a,b)}\} \right) = \prod_{b=1}^r \frac{T^g_{q_{1,(1,b)},s_{(1,b)}}}{T^g_{\bg q_{1,(2,b)},\bg s_{(2,b)}}} \left(\prod_{l=1}^{M-1} \frac{T^g_{q_{l+1,(1,b)},q_{l,(1,b)}}}{T^g_{\bg q_{l+1,(2,b)},\bg q_{l,(2,b)}}} \right) \, . \label{eq:type-I-phase}
\end{align}
Note that the counter-clockwise orientation around the boundary of $B$ corresponds to the direction of \emph{decreasing} $l$ in the plaquette indices $q_l$. 
In order to simplify this expression, we first note that since the plaquette $s$ intersects $A$, $B$, and $C$, the constraints of Eq.~\eqref{eq:kronecker-deltas-I} imply,
\begin{align}
	\left\{
	\begin{aligned}
		s_{(1,b)} = g s_{(1,b-1)} \\
		s_{(2,b)} = gs_{(1,b+1)} 
	\end{aligned}\right.
	 \implies
	\left\{
	\begin{aligned}
	  s_{(1,b)} &= g^{b-1} s_{(1,1)} \equiv g^{b-1} s  \\
	  s_{(2,b)} &= g^{b+1} s
	\end{aligned}\right.
\end{align}
Conversely, the plaquette $q_M$ intersects $B$, $C$, and $D$, and so Eq.~\eqref{eq:kronecker-deltas-I} implies,
\begin{align}
	\left\{
	\begin{aligned}
	q_{M,(1,b)} &= g q_{M,(1,b-1)}  \\
	q_{M,(2,b)} &= gq_{M,(1,b)}
	\end{aligned}\right.
	\implies 
	\left\{
	\begin{aligned}
	q_{M,(1,b)} &= g^{b-1} q_{M,(1,1)} \equiv g^{b-1} q_M  \\
	q_{M,(2,b)} &= g^{b} q_M 
	\end{aligned}
	\right.
\end{align}
Finally, the plaquettes $q_{1 \leq l < M}$ only intersect $B$ and $C$. Hence, the constraints of Eq.~\eqref{eq:kronecker-deltas-I} imply,
\begin{align}
	\left\{
	\begin{aligned}
	q_{l,(1,b)} &= g q_{l,(1,b-1)}  \\
	q_{l,(2,b)} &= gq_{l,(2,b-1)}
	\end{aligned}\right.	
	\implies 
	\left\{
	\begin{aligned}
	q_{l,(1,b)} &= g^{b-1} q_{l,(1,1)} \equiv g^{b-1} q_{l,1} \\
	q_{l,(2,b)} &= g^{b-1} q_{l,(2,1)} \equiv g^{b-1} q_{l,2} 
	\end{aligned}
	\right.
\end{align}
Imposing these constraints, we find for the total phase, Eq.~\eqref{eq:type-I-phase},
\begin{align}
	\begin{split}
	\tilde{\oL}^r_{g}\left(\{\fp_{l,(a,b)}\} \right) 
	&= \prod_{b=1}^r \frac{T^g_{g^{b-1}q_{1,1},g^{b-1}s}}{T^g_{g^{b-1} \bg q_{1,2}, g^{b-1} g s}} \left(\prod_{l=1}^{M-2} \frac{T^g_{g^{b-1} q_{l+1,1}, g^{b-1} q_{l,1}}}{ T^g_{ g^{b-1} \bg q_{l+1,2}, g^{b-1} \bg q_{l,2}}} \right)  \frac{T^g_{g^{b-1} q_M, g^{b-1} q_{M-1,1}}}{T^g_{g^{b-1} q_M, g^{b-1} \bg q_{M-1,2}}} \, . \end{split} \label{eq:type-I-phase-explicit}
\end{align}
Note that each factor in this expression is of the form $T^g_{g^{b-1} h, g^{b-1}k}$, for different choices of $h$ and $k$.

In order to simplify this expression, we make use of the relation,
\begin{align}
	\prod_{b=1}^r \omega(\bar h k, \bar k \bg^b,g) = \prod_{b=1}^r \frac{\omega(\bar h,\bg^b,g)}{\omega(\bar k,\bg^b,g)}, \label{eq:cocycle-identity}
\end{align}
which can be derived using the cocycle relation Eq.~\eqref{eq:cocycle}.
Explicitly, applying this cocycle relation to $(g_1,g_2,g_3,g_4) = (\bh k,  \bk , \bg^b, g)$, one finds
\begin{align}
	\prod_{b=1}^r \omega(\bar h k, \bar k \bg^b, g) = \prod_{b=1}^r \frac{\omega(\bar h,\bg^b,g)}{\omega(\bar k,\bg^b,g)} \frac{\omega(\bar h k, \bar k, \bg^{b-1})}{\omega(\bar h k, \bar k , \bg^b)}.
\end{align}
Using the product over $b$ and the fact that $g^r = e$, we find we can cancel terms to recover Eq.~\eqref{eq:cocycle-identity}.

Now, consider a factor $T^g_{g^{b-1} h , g^{b-1} k}$ appearing in Eq.~\eqref{eq:type-I-phase-explicit} and suppose first that the bond points from $h$ to $k$ (here we abuse notation and use the same index to refer to the state of a qudit on the plaquette and the plaquette itself). Then,
\begin{align}
	\prod_{b=1}^r T^g_{g^{b-1} h , g^{b-1} k} = \prod_{b=1}^r \theta^g_{g^{b-1} h , g^{b-1} k} = \prod_{b=1}^r \omega(\bh k, \bk \bg^b , g) = \prod_{b=1}^r \frac{\omega(\bar h,\bg^b,g)}{\omega(\bar k,\bg^b,g)} \, ,
\end{align}
where we used Eq.~\eqref{eq:cocycle-identity} in the final equality. Now, suppose instead that the bond points from $k$ to $h$. Then,
\begin{align}
	\prod_{b=1}^r T_{g^{b-1} h , g^{b-1} k}^g = \prod_{b=1}^r (\theta^g_{g^{b-1} k , g^{b-1} h})^* = \prod_{b=1}^r \omega(\bk h, \bh \bg^b , g)^*  = \prod_{b=1}^r \left(\frac{\omega(\bar k,\bg^b,g)}{\omega(\bar h,\bg^b,g)}\right)^* = \prod_{b=1}^r \frac{\omega(\bar h,\bg^b,g)}{\omega(\bar k,\bg^b,g)} \, ,
\end{align}
where we used Eq.~\eqref{eq:cocycle-identity} in the penultimate equality. Hence, we see that,
\begin{align}
	\prod_{b=1}^r T^g_{g^{b-1} h , g^{b-1} k} = \prod_{b=1}^r \frac{\omega(\bar h,\bg^b,g)}{\omega(\bar k,\bg^b,g)} \, , \label{eq:T-relation}
\end{align}
\emph{irrespective} of whether the bond points from $h$ to $k$ or \textit{vice versa}.

Now, focusing on the terms appearing in the numerators of Eq.~\eqref{eq:type-I-phase-explicit}, we find, using Eq.~\eqref{eq:T-relation},
\begin{align}
	\notag \prod_{b=1}^r &T^g_{g^{b-1}q_{1,1},g^{b-1}s} \left(\prod_{l=1}^{M-2} T^g_{g^{b-1} q_{l+1,1},g^{b-1} q_{l,1}}\right) T^g_{g^{b-1} q_M, g^{b-1} q_{M-1,1}} \\
	\notag &= \prod_{b=1}^r \frac{\omega(\bq_{1,1}, \bg^b, g)}{\omega(\bar s , \bg^b , g)}  \prod_{l=1}^{M-2} \frac{\omega(\bq_{l+1,1},\bg^b,g)}{\omega(\bq_{l,1}, \bg^b,g)} \frac{\omega(\bq_M,\bg^b,g)}{\omega(\bq_{M-1,1},\bg^b,g)} \\
	&= \prod_{b=1}^r \frac{\omega(\bq_M,\bg^b,g)}{\omega(\bar s , \bg^b , g)} \, .
\end{align}
Note that all the contributions from the plaquettes between the two endpoints of the boundary cancel out. Likewise, we find for the terms in the denominators of Eq.~\eqref{eq:type-I-phase-explicit},
\begin{align}
	\notag \prod_{b=1}^r & T^g_{g^{b-1} \bg q_{1,2}, g^{b-1} g s} \left( \prod_{l=1}^{M-2} T^g_{ g^{b-1} \bg q_{l+1,2}, g^{b-1} \bg q_{l,2}} \right) T^g_{g^{b-1} q_M, g^{b-1} \bg q_{M-1,2}} \\
	\notag &= \prod_{b=1}^r \frac{\omega(g \bq_{1,2},\bg^b,g)}{\omega(\bg \bar s, \bg^b, g)} \prod_{l=1}^{M-2} \frac{\omega(g \bq_{l+1,2}, \bg^b, g)}{\omega(g \bq_{l,2} , \bg^b, g)} \frac{\omega(\bq_M, \bg^b, g)}{\omega(g \bq_{M-1,2}, \bg^b, g)} \\
	&= \prod_{b=1}^r \frac{\omega(\bq_M , \bg^b, g)}{\omega(\bg \bar s, \bg^b, g)} \, .
\end{align}
Once again, we see that the product telescopes out.

Altogether, we find,
\begin{align}
	\tilde{\oL}^r_{g}\left(\{\fp_{l,(a,b)}\} \right) &= \prod_{b=1}^r \frac{\omega(\bg \bar s , \bg^b , g)}{\omega(\bar s, \bg^b, g)} \, . 
\end{align}
Now, we also have that,
\begin{align}
	\prod_{b=1}^r \omega(\bg \bar s, \bg^b, g)  = \prod_{b=1}^r \omega(\bg \bar s, \bg^{b-1}, g) = \prod_{b=1}^r \frac{\omega(\bs , \bg^b , g)}{\omega(g , \bg^b , g)} \, .
\end{align}
In the first equality, we again made use of the fact that $g^r=e$ to shift around the terms in the product. In the second equality, we employed Eq.~\eqref{eq:cocycle-identity} with $h = s$ and $k = \bg$. Hence,
\begin{align}
	\tilde{\oL}^r_{g}\left(\{\fp_{l,(a,b)}\} \right) &= \prod_{b=1}^r \omega(g,\bg^b,g)^* =\prod_{b=1}^r \omega(g,g^b,g)^* = \mathcal{Z}[L(r,1)_g]^*,
\end{align}
which is precisely the (complex conjugate of the) desired lens space partition function, as indicated in the final equality \cite{tantivasadakarn2017}. In the second equality, we simply rearranged terms in the product. We therefore have,
\begin{align}
	\oL^r_g(\psi) = \mathcal{Z}[L(r,1)_g]^* \sum_{\{\fp_{l,(a,b)},\fp_{l,(a,b)}'\}} S_{g}\left(\{\fp_{l,(a,b)},\fp_{l,(a,b)}'\} \right) \, .
\end{align}
The magnitude of the order parameter is simply given by the remaining sum. Evaluating this amounts to counting the number of independent sums over the plaquettes on the entanglement cut, $\{\fp_{l,(a,b)},\fp_{l,(a,b)}'\}$, subject to the constraints imposed by Eq.~\eqref{eq:kronecker-deltas-I}. One finds, 
\begin{align}
	\oL^r_g(\psi) = \mathcal{Z}[L(r,1)_g]^* |G|^{-4(2r-1) -r (|\partial AB| + |\partial AC| + |\partial BD| + |\partial CD|) - 2(r-1) ( |\partial AD| + |\partial BC|) }  \, .
\end{align}
Here, the first term in the exponent comes from the four trijunctions. The remaining terms come from the boundaries between pairs of subregions; for instance, $|\partial AB|$ is the number of plaquettes on the boundary between $A$ and $B$, \emph{not counting} the two plaquettes that also lie on trijunctions. One readily verifies that the magnitude is independent of the choice of the symmetry $g$, as claimed.

\subsection{Type-III Order Parameter Computation}

\begin{figure}
    \centering
    \includegraphics[width=\linewidth]{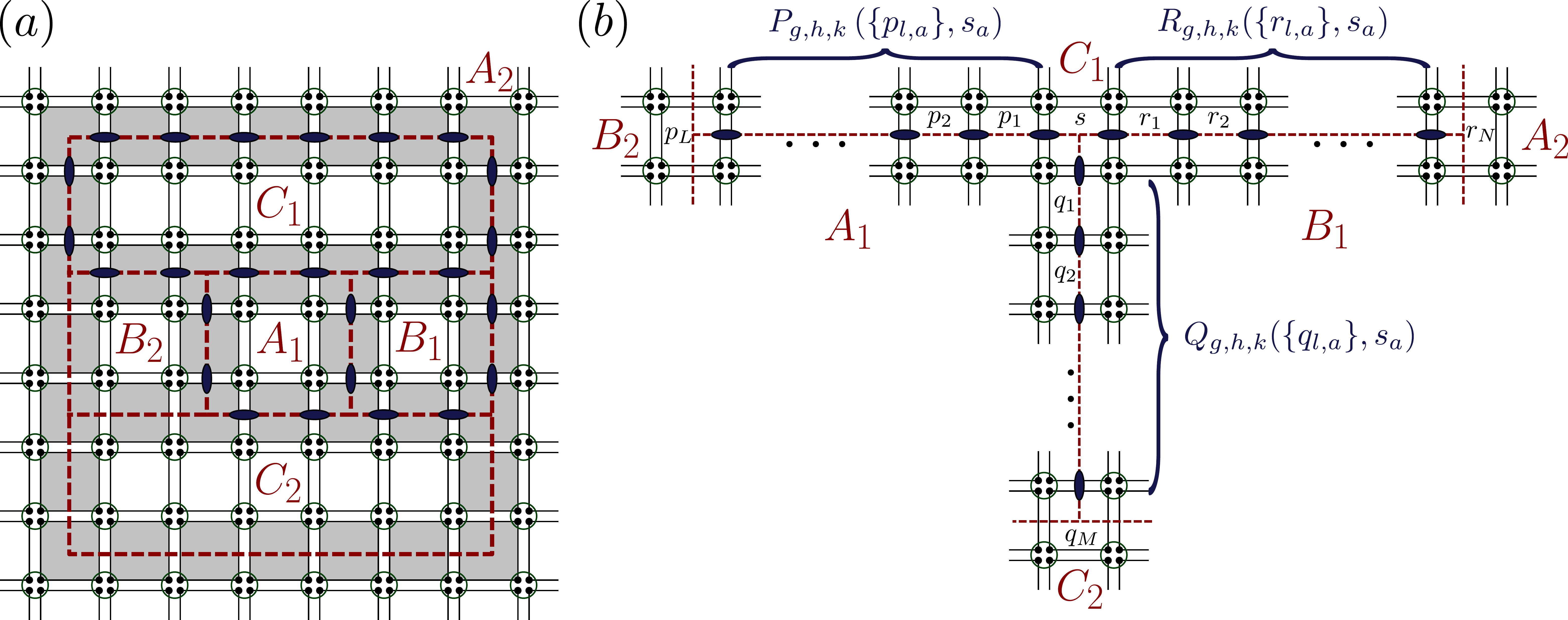}
    \caption{(a) Partitioning of the lattice used in the computation of $\oT_{g,h,k}$. The shaded plaquettes are those lying along the entanglement cut and which contribute to the magnitude of the order parameter. The dark blue ellipses indicate phase factors, which depend on the states of two adjacent plaquettes, that contribute to Eq.~\eqref{eq:type-III-total-phase}. (b) Labeling of the plaquettes along the boundaries between the $A_1$, $B_1$, and $C_1$ subregions and the only phases which do not trivially vanish in Eq.~\eqref{eq:type-III-total-phase}. The phases of each leg of the entanglement cut are grouped together into the three factors of Eq.~\eqref{eq:total-phase-pqr}.}
    \label{fig:type-III-rectangular}
\end{figure}

\subsubsection{Setup}

Let us again consider a square lattice group cohomology model with symmetry group $G$. Let $g,h,k \in G$ be generators of three separate discrete factors $\mathbb{Z}_{N_{1,2,3}}$, respectively, of $G$.
We claim that,
\begin{align}
	\oT_{g,h,k}(\psi) = e^{i\vartheta_{g,h,k}} \oT_{e,e,e}(\psi),
\end{align}
where $\vartheta_{g,h,k}$ is a universal phase corresponding to the SPT partition function on the three-torus with flux insertions, $T^3_{g,h,k}$, which corresponds to the Type-III invariant  under the symmetry $\mathbb{Z}_{N_1} \times \mathbb{Z}_{N_2} \times \mathbb{Z}_{N_3}$.
Our goal is to show this and to extract this universal phase.

In order to organize our computation, it will be most convenient to use the representation of the order parameter given in the form,
\begin{align}
	\begin{aligned}
	\oT_{g,h,k}(\psi)
    &= \Tr\left[ \tau_\beta^B\tau^C_\gamma ( U_g^{A_1} U_h^{B_1} U_k^{C_1}) \otimes ( (U_g^{A_1})^\dagger U_h^{B_1} (U_k^{C_1})^\dagger)
    \otimes ( (U_g^{A_1})^\dagger (U_h^{B_1})^\dagger U_k^{C_1}) 
    \otimes ( U_g^{A_1} (U_h^{B_1})^\dagger (U_k^{C_1})^\dagger) \rho^{\otimes 4}\right] \, .
    \end{aligned}
\end{align} 
As before, our computation proceeds in several steps. First, we establish that the only contributions to both the magnitude and phase of the order parameter come from GHZ states crossing the entanglement cuts. Second, we show that the non-trivial contributions to the phase come from GHZ states that cross the boundaries between $A_1$ and $B_1$, $A_1$ and $C_1$, or $B_1$ and $C_1$. Finally, we show that this phase factor is precisely the Type-III topological invariant.

To begin, we repeat here for convenience the density matrix for a lattice group cohomology model:
\begin{align}
	\rho = \bigotimes_p \rho_p, \qquad \rho_p = |G|^{-1} \sum_{i_p, i_p'} \ket{i_p i_p i_p i_p} \bra{i_p' i_p' i_p' i_p'}
\end{align}
where $p$ runs over all plaquettes and $\rho_p$ is a state supported on the four qudits on the corners of the plaquette. As in the preceding subsection, one readily verifies that,
\begin{align}
	\Tr[\rho_p^{\otimes 4}] = \Tr[\tau_\beta \rho_p^{\otimes 4}]  = \Tr[\tau_\gamma \rho_p^{\otimes 4}]  = 1 \, ,
\end{align}
where $\tau_{\beta,\gamma}$ act on all four qudits. Thus, if $p$ is contained entirely within one of the subregions $A$, $B$, or $C$ (i.e. all four qudits on the corners of $p$ are contained in one of these subregions), then $\rho_p$ does not contribute to the order parameter. Here we also made use of the fact noted above that a partial symmetry operation acting on a group cohomology wavefunction only has nontrivial action on the boundary of its support.

It then follows that the nontrivial contributions to the phase and magnitude of the order parameter only come from plaquettes crossing the entanglement cuts. With this in mind, it will be convenient for the purposes of organizing our calculation to write the order parameter as,
\begin{align}
	\oT_{g,h,k}(\psi) = \sum_{\{\fp_{l,a},\fp_{l,a}'\}}  \tilde{\oT}_{g,h,k}\left(\{\fp_{l,a}\} \right) S_{g,h,k}\left(\{\fp_{l,a},\fp_{l,a}'\} \right) \, .
\end{align}
In $\fp_{l,a}$, $l$ indexes the plaquettes lying along the entanglement cut while $a=1,\dots, 4$ denotes the replica index. The density matrix of the plaquette at position $l$ is given by $\rho_l = |G|^{-1}\sum_{\fp_l,\fp_l'} \ket{\fp_l \fp_l \fp_l \fp_l} \bra{\fp_l' \fp_l' \fp_l' \fp_l'}$. We have grouped the phase contributions and the trace into the terms $\tilde{\oT}$ and $S$, respectively. Explicitly,
\begin{align}
    \begin{aligned}
	S_{g,h,k} = \Tr\Bigg[\tau_\beta^{\partial B} \tau_\gamma^{\partial C} & [ V_g^{\partial A_1} V_h^{\partial B_1} V_k^{\partial C_1}] \otimes [ (V_g^{\partial A_1})^\dagger V_h^{\partial B_1} (V_k^{\partial C_1})^\dagger] \\
    &
    \otimes [ (V_g^{\partial A_1})^\dagger (V_h^{\partial B_1})^\dagger V_k^{\partial C_1}] 
    \otimes [ V_g^{\partial A_1} (V_h^{\partial B_1})^\dagger (V_k^{\partial C_1})^\dagger] \bigotimes_{a=1}^4 \bigotimes_{l} \rho(\fp_{l,a},\fp_{l,a}')\Bigg] \, ,
    \end{aligned} \label{eq:type-III-trace}
\end{align}
where, as before,
\begin{align}
	\rho(a,b) \equiv |G|^{-1} \ket{aaaa}\bra{bbbb} \, .
\end{align}
Evaluating the trace gives a product of Kronecker delta functions relating the different plaquette indices; rather than writing them each out explicitly, we simply note that they enforce the following equalities, depending on which subregions the plaquette corresponding to the index $\mathfrak{p}$ intersects:
\begin{align}
	\begin{split}
	&\fp \in A_1 \implies 
		\left\{
	\begin{aligned}
		g\fp_1 &= \fp_1' \\
		\bg \fp_2 &= \fp_2'  \\
		\bg\fp_3 &= \fp_3'  \\
		g \fp_4 &= \fp_4' 		
	\end{aligned}
	\right. \, , \quad
		\fp \in B_1 \implies 
		\left\{
	\begin{aligned}
		h\fp_1 = \fp_3' \\
		h\fp_2 = \fp_4'  \\
		\bh \fp_3=\fp_1'  \\
		\bh \fp_4 = \fp_2' 		
	\end{aligned}
	\right. \, , \quad
		\fp \in C_1 \implies 
		\left\{
	\begin{aligned}
		k\fp_1 = \fp_2' \\
		\bk \fp_2 = \fp_1'  \\
		k \fp_3 = \fp_4'  \\
		\bk\fp_4 = \fp_3' 		
	\end{aligned}
	\right. \, , \\ 
	&\fp \in A_2 \implies 
		\left\{
	\begin{aligned}
		\fp_1 &= \fp_1' \\
		\fp_2 &= \fp_2'  \\
		\fp_3 &= \fp_3'  \\
		\fp_4 &= \fp_4' 		
	\end{aligned}
	\right. \, , \quad\,\,\,
		\fp \in B_2 \implies 
		\left\{
	\begin{aligned}
		\fp_1 = \fp_3' \\
		\fp_2 = \fp_4'  \\
		\fp_3=\fp_1'  \\
		\fp_4 = \fp_2' 		
	\end{aligned}
	\right. \, , \quad\,\,\,\,
		\fp \in C_2 \implies 
		\left\{
	\begin{aligned}
		\fp_1 = \fp_2' \\
		\fp_2 = \fp_1'  \\
		\fp_3 = \fp_4'  \\
		\fp_4 = \fp_3' 		
	\end{aligned}
	\right.	
	\end{split}
	 \label{eq:kronecker-deltas-III}
\end{align}
Here we have suppressed the subscript $l$. These constraints are most readily deduced by using a diagrammatic representation of Eq.~\eqref{eq:type-III-trace}. In Fig.~\ref{fig:type-III-kronecker}, we provide such a representation for a contribution to Eq.~\eqref{eq:type-III-trace} coming from the plaquette located at the trijunction between the subregions $A_1$, $B_1$, and $C_1$.

\begin{figure}
    \centering
    \includegraphics[width=0.55\linewidth]{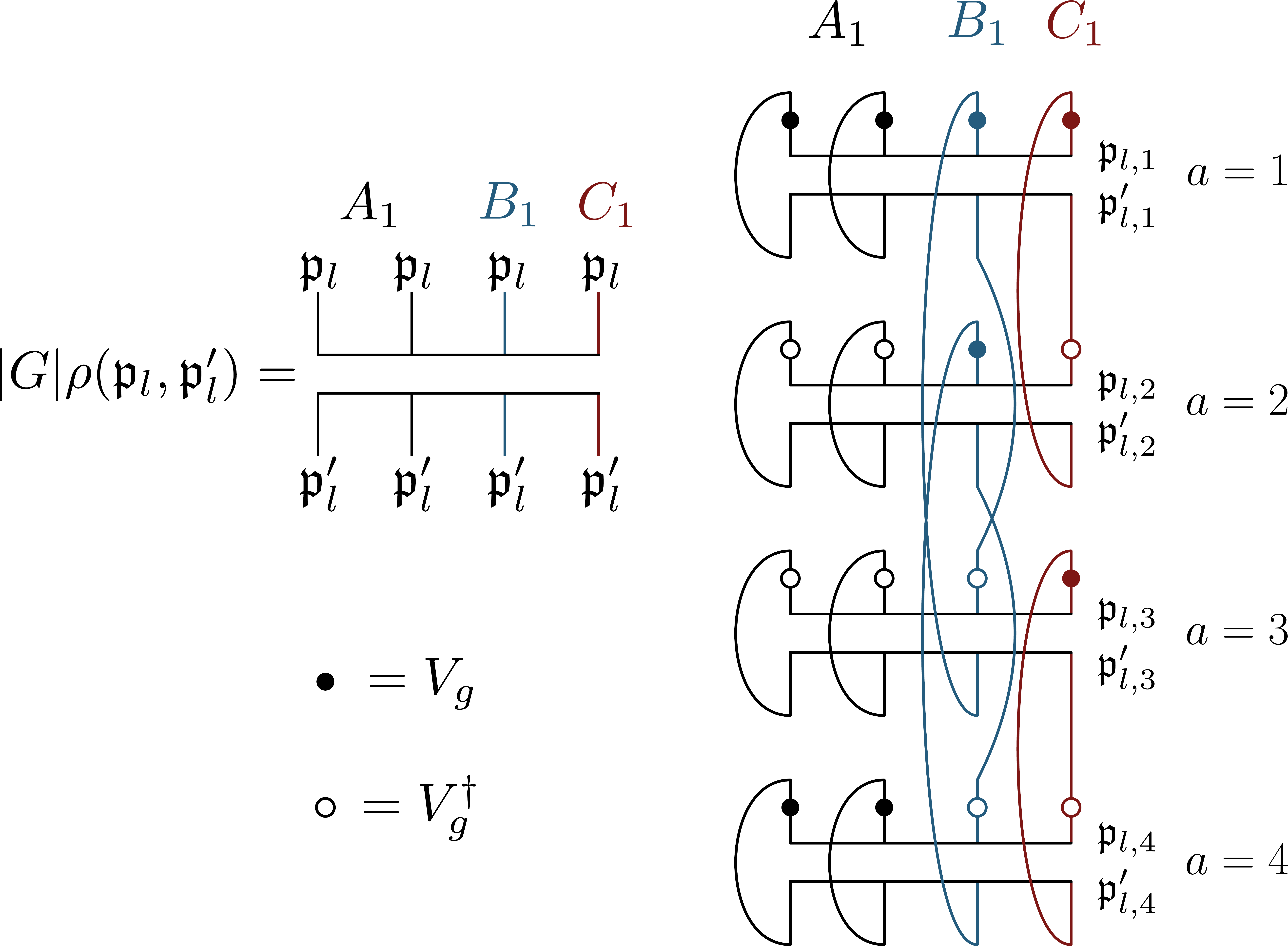}
    \caption{A diagrammatic representation of the contribution to Eq.~\eqref{eq:type-III-trace} from the plaquette $l$ at the trijunction between the subregions $A_1$, $B_1$, and $C_1$ (see Fig.~\ref{fig:type-III-rectangular}); this plaquette has two qudits in the $A_1$ subregion, and one in each of $B_1$ and $C_1$. (Left) A tensor network depiction of $\rho(\fp_l,\fp_l')$, with the physical indices colored according to the subregion they belong to. We represent the action of $V_g$ and $V_g^\dagger$ with filled-in and empty circles, respectively. (Right) The contribution to Eq.~\eqref{eq:type-III-trace}. Evaluating the contractions yields Kronecker deltas enforcing the first line of constraints of Eq.~\eqref{eq:kronecker-deltas-III} on $\fp_l$ and $\fp_l'$.}
    \label{fig:type-III-kronecker}
\end{figure}

The total phase, $\tilde{\oT}_{g,h,k}$, is a product of the $T^g_{ij}$ coming from the partial symmetry actions along the entanglement cuts: 
\begin{align}
	\begin{aligned}
	\tilde{\oT}_{g,h,k}\left(\{\fp_{l,a}\} \right) = \prod_{l \in \partial A_1} & \left( \frac{ T_{\fp_{l,1},\fp_{l+1,1}}^g T_{\fp_{l,4},\fp_{l+1,4}}^g}{ T_{\bg \fp_{l,2},\bg \fp_{l+1,2}}^g  T_{\bg \fp_{l,3},\bg \fp_{l+1,3}}^g} \right) \prod_{l \in \partial B_1} \left( \frac{ T_{\fp_{l,1},\fp_{l+1,1}}^h  T_{\fp_{l,2},\fp_{l+1,2}}^h}{ T_{\bh \fp_{l,3},\bh \fp_{l+1,3}}^h  T_{\bh \fp_{l,4},\bh \fp_{l+1,4}}^h}\right)  \\
	&\times \prod_{l \in \partial C_1} \left( \frac{ T_{\fp_{l,1},\fp_{l+1,1}}^k T_{\fp_{l,3},\fp_{l+1,3}}^k}{ T_{\bk \fp_{l,2},\bk \fp_{l+1,2}}^k  T_{\bk \fp_{l,4},\bk \fp_{l+1,4}}^k}  \right) \, .
	\end{aligned} \label{eq:type-III-total-phase}
\end{align}
The three products correspond to the contributions from the partial symmetry actions acting on $A_1$, $B_1$, and $C_1$ respectively. In a given product, the index $l$ labels the plaquettes in a counter-clockwise fashion around the boundary of the corresponding subregion. Note that, on segments of the entanglement cut where two of the subregion boundaries coincide, these two boundaries and hence the corresponding products of phase factors will have \emph{opposite} orientations.
For instance, on the boundary between $A_1$ and $B_1$, if the partial $g$ symmetry operation contributes a phase of $T_{ij}^g$, then the partial $h$ symmetry operation will contribute a factor of $T_{ji}^h$. 
Within each set of parentheses are the four phase factors arising from the action of the corresponding partial symmetry on each of the four replicas.

We now show that the only nontrivial phase factors come from plaquettes on the boundaries between $A_1$ and $B_1$, $A_1$ and $C_1$, or $B_1$ and $C_1$. 
To begin, consider two neighboring plaquettes, $\fp_l$ and $\fp_{l+1}$ on a portion of the entanglement cut separating $A_1$ and $B_2$, supposing first that the bond points from $l$ to $l+1$. These will contribute a factor of,
\begin{align}
	 \theta_{\fp_{l,1},\fp_{l+1,1}}^g (\theta_{\bg \fp_{l,2},\bg \fp_{l+1,2}}^g)^* (\theta_{\bg \fp_{l,3},\bg \fp_{l+1,3}}^g)^*  \theta_{\fp_{l,4},\fp_{l+1,4}}^g \in \tilde{\oT}_{g,h,k}\left(\{\fp_{l,a}\} \right) \, ,
\end{align}
to the total phase factor. Now, since these two plaquettes intersect both $A_1$ and $B_2$, we have from Eq.~\eqref{eq:kronecker-deltas-III} that,
\begin{align}
\begin{cases}
	g   \fp_1 = \fp_1' \\
	\bg \fp_2 = \fp_2' \\
	\bg   \fp_3 = \fp_3' \\
	g \fp_4 = \fp_4'
\end{cases} \quad \text{and} \quad\quad
\begin{cases}
	\fp_1 = \fp_3' \\
	\fp_2 = \fp_4' \\
	\fp_3 = \fp_1' \\
	\fp_4 = \fp_2'
\end{cases} \implies
\begin{cases}
	\fp_2 = g \fp_4 \\
	\fp_3 = g \fp_1
\end{cases}
\end{align}
for both $\fp_l$ and $\fp_{l+1}$, where we have suppressed the $l$ subscript in the above. Using these equalities, we have that
\begin{align}
	\theta_{\fp_{l,1},\fp_{l+1,1}}^g (\theta_{\bg \fp_{l,2},\bg \fp_{l+1,2}}^g)^* (\theta_{\bg \fp_{l,3},\bg \fp_{l+1,3}}^g)^*  \theta_{\fp_{l,4},\fp_{l+1,4}}^g =
	\theta_{\fp_{l,1},\fp_{l+1,1}}^g (\theta_{\fp_{l,4},\fp_{l+1,4}}^g)^* (\theta_{\fp_{l,1},\fp_{l+1,1}}^g)^*  \theta_{\fp_{l,4},\fp_{l+1,4}}^g = 1 \, ,
\end{align}
as claimed. It is readily confirmed the same holds if instead the bond points from $l+1$ to $l$. On imposing the constraints of Eq.~\eqref{eq:kronecker-deltas-III}, one finds by essentially identical computations that the phase factor contributions from two adjacent plaquettes on a boundary between any subregions $\alpha_1$ and $\beta_2$, with $\alpha , \beta \in \{ A, B, C \}$, will always cancel out. 

With this observation, we can write the total phase $\tilde{\oT}_{g,h,k}$ as a product of three factors coming from just three boundaries.
We label the plaquettes on the boundaries between the $A_1$, $B_1$, and $C_1$ as in Fig.~\ref{fig:type-III-rectangular}: $p_l$ label the plaquettes on the $A_1/C_1$ boundary, $q_l$ on the $A_1/B_1$ boundary, and $r_l$ on the $B_1/C_1$ boundary, while $s$ labels the plaquette at the trijunction. Here the $l$ are numbered such that $p_L$, $q_M$, and $r_N$ are the plaquettes at the $A_1B_2C_1$, $A_1B_1C_2$, and $A_2B_1C_1$ trijunctions, respectively. With this notation, we have for the total phase factor,
\begin{align}
	\tilde{\oT}_{g,h,k}\left(\{\fp_{l,a}\} \right) = P_{g,h,k}\left(\{p_{l,a}\}, s_a \right) Q_{g,h,k}\left(\{q_{l,a}\}, s_a \right) R_{g,h,k}\left(\{r_{l,a}\}, s_a \right) , \label{eq:total-phase-pqr}
\end{align}
where $P$, $Q$, and $R$ correspond to the contributions along the boundaries between $A_1$ and $C_1$; $A_1$ and $B_1$; and $B_1$ and $C_1$, respectively.

\subsubsection{Phase Contribution from the $A_1 / B_1$ Boundary}

Let us now turn to the evaluation of the total phase factor $\tilde{\oT}_{g,h,k}$.
We first focus on the boundary between $A_1$ and $B_1$ and the corresponding factor $Q_{g,h,k}$, which receives phase contributions from the $U_g^{A_1}$ and $U_h^{B_1}$ partial symmetry operators. Explicitly, 
\begin{align}
	\begin{split}
	Q_{g,h,k}\left(\{q_{l,a}\}, s_a \right) = & \frac{ T_{q_{1,1},s_1}^g  T_{q_{1,4},s_4}^g}{ T_{\bg q_{1,2},\bg s_2}^g  T_{\bg q_{1,3},\bg s_3}^g} \frac{ T_{s_1,q_{1,1}}^h   T_{s_2,q_{1,2}}^h}{  T_{\bar h s_3,\bar h q_{1,3}}^h  T_{\bar h s_4,\bar h q_{1,4}}^h}  \prod_{l=1}^{M-1} \frac{ T_{q_{l+1,1},q_{l,1}}^g  T_{q_{l+1,4},q_{l,4}}^g}{ T_{\bg q_{l+1,2},\bg q_{l,2}}^g  T_{\bg q_{l+1,3},\bg q_{l,3}}^g} \frac{ T_{q_{l,1},q_{l+1,1}}^h   T_{q_{l,2},q_{l+1,2}}^h }{ T_{\bar h q_{l,3},\bar h q_{l+1,3}}^h  T_{\bar h q_{l,4},\bar h q_{l+1,4}}^h}
	\end{split}
\end{align}
Note, as emphasized above, $A_1$ and $B_1$ boundaries, and hence the string of phase factors arising from the partial $g$ and $h$ symmetries, have opposite orientation to one another on this portion of the entanglement cut.
Though this appears at first glance to be an unwieldy expression, it will drastically simplify. First, noting that the plaquette $s$ intersects $A_1$, $B_1,$ and $C_1$, the plaquette $q_M$ intersects $A_1$, $B_1$, and $C_2$, while the plaquettes $q_{1 \leq l < M}$ only intersect $A_1$ and $B_1$, one finds that the constraints of Eq.~\eqref{eq:kronecker-deltas-III} imply,   
\begin{align}
\begin{cases}
	s_1 \equiv s \\
	s_2 = gk s \\
	s_3 = gh s \\
	s_4 = hk s
\end{cases} \qquad
\begin{cases}
	q_{M,1} \equiv q_M \\
	q_{M,2} = g q_M \\
	q_{M,3} = gh q_M \\
	q_{M,4} = h q_M
\end{cases} \qquad
\begin{cases}
	q_{l,3} = g h q_{l,1} \equiv g h q_l \\
	q_{l,4} = \bg h q_{l,2} \equiv \bg h q_l' \, ,
\end{cases} \label{eq:Q-constraints}
\end{align}
where we have defined $s \equiv s_1$, $q_{M,1} \equiv q_M$, $q_l \equiv q_{l,1}$, and $q_l' \equiv q_{l,2}$.
Plugging these relations into the expression for $Q$, we obtain, 
\begin{align}
	\begin{split}
	Q_{g,h,k}\left(\{q_{l,a}\}, s_a \right) = & \frac{ T_{q_{1},s}^g  T_{\bg h q_{1}',hks}^g}{ T_{\bg q_{1}',ks}^g  T_{h q_{1},h s}^g} \frac{ T_{s,q_{1}}^h   T_{gks,q_{1}'}^h}{ T_{gs, g q_{1}}^h  T_{k s,\bg q_{1}'}^h}  
	\prod_{l=1}^{M-2} \frac{ T_{q_{l+1},q_{l}}^g  T_{\bg h q_{l+1}',\bg h q_{l}'}^g}{ T_{\bg q_{l+1}',\bg q_{l}'}^g  T_{h q_{l+1},h q_{l}}^g} \frac{ T_{q_{l},q_{l+1}}^h   T_{q_{l}',q_{l+1}'}^h }{  T_{g q_{l},g q_{l+1}}^h  T_{ \bg q_{l}',\bg q_{l+1}'}^h} \\
	&\times \frac{ T_{q_{M},q_{M-1}}^g  T_{h q_{M},\bg h q_{M-1}'}^g}{ T_{q_{M},\bg q_{M-1}'}^g  T_{h q_{M},h q_{M-1}}^g}  \frac{ T_{q_{M-1},q_{M}}^h  T_{q_{M-1}',g q_{M}}^h}{ T_{g q_{M-1}, g q_{M}}^h  T_{\bg q_{M-1}', q_{M}}^h} \, .
	\end{split}
\end{align}
It is convenient to massage this expression into the form, 
\begin{align}
	\begin{split}
	Q_{g,h,k}\left(\{q_{l,a}\}, s_a \right) =& \frac{ T_{s,q_{1}}^h  T_{q_{1},s}^g}{T_{h q_{1},hs}^g  T_{g s, g q_{1}}^h} \frac{ T_{\bg h q_{1}',hks}^g  T_{gks,q_{1}'}^h}{ T_{k s,\bg q_{1}'}^h   T_{\bg q_{1}',ks}^g} 
	\prod_{l=1}^{M-2} \frac{ T_{q_{l},q_{l+1}}^h T_{q_{l+1},q_{l}}^g }{T_{h q_{l+1},h q_{l}}^g   T_{g q_{l}, g q_{l+1}}^h}
	 \frac{  T_{\bg h q_{l+1}',\bg h q_{l}'}^g T_{q_{l}',q_{l+1}'}^h}{ T_{\bg q_{l}',\bg q_{l+1}'}^h T_{\bg q_{l+1}',\bg q_{l}'}^g } \\
	 & \qquad \qquad \qquad \qquad \qquad \times \frac{ T_{q_{M-1},q_{M}}^h T_{q_{M},q_{M-1}}^g  }{ T_{h q_{M},h q_{M-1}}^g  T_{g q_{M-1}, g q_{M}}^h} \frac{  T_{h q_{M},\bg h q_{M-1}'}^g  T_{q_{M-1}',g q_{M}}^h}{ T_{\bg q_{M-1}', q_{M}}^h T_{q_{M},\bg q_{M-1}'}^g } \, .
	 \end{split}
\end{align}
Defining, 
\begin{align}
\boxed{
F_{h,g}(a,b) = \frac{ T_{a,b}^h T_{b,a}^g }{ T_{h b,h a}^g  T_{g a, g b}^h} } \, ,
\end{align}
we may write the above more compactly as,
\begin{align}
	Q_{g,h,k}\left(\{q_{l,a}\}, s_a \right) = \frac{F_{h,g}(s,q_1)}{F_{h,g}(ks, \bg q_1')} \prod_{l=1}^{M-2} \frac{F_{h,g}(q_l,q_{l+1})}{F_{h,g}(\bg q_l' , \bg q_{l+1}')} \frac{F_{h,g}(q_{M-1},q_M)}{F_{h,g}(\bg q_{M-1}',q_M)} \, .
\end{align}

In order to make progress, it will prove useful to study the properties of $F_{h,g}(a,b)$. First, let us consider the case where the bond points from $a$ to $b$, such that
\begin{align}
	F_{h,g}(a,b) = \frac{ \theta_{a,b}^h  \theta_{h a, h b}^g}{ \theta_{a, b}^g  \theta_{g a, g b}^h}\, , \quad \text{(bond points from $a$ to $b$)}.
\end{align}
In order to simplify this expression, it is first convenient to make note of the relation,
\begin{align}
	\theta_{a,b}^h \theta_{h a, h b}^g = \omega(\bar a b, \bar b\bh, h) \omega(\bar a b, \bg \bh \bar b, g) = \frac{\omega(\bg \bh \bar a, g, h) \omega(\bar a b, \bg \bh \bar b, gh)}{\omega(\bg \bh \bar b, g, h)} \, ,
\end{align}
which follows from applying the cocycle relation Eq.~\eqref{eq:cocycle},
to $(g_1,g_2,g_3,g_4) = (\bar a b, \bg \bh \bar b, g, h)$. 
Likewise, by exchanging $g$ with $h$, we obtain
\begin{align}
	\theta_{a,b}^g \theta_{g a, g b}^h = \frac{\omega(\bg \bh \bar a, h, g) \omega(\bar a b, \bg \bh \bar b, gh)}{\omega(\bg \bh \bar b, h, g)} \, ,
\end{align}
and thus
\begin{align}
	F_{h,g}(a,b) = \frac{\omega(\bg \bh \bar a, g, h)\omega(\bg \bh \bar b, h, g)}{\omega(\bg \bh \bar b, g, h)\omega(\bg \bh \bar a, h, g)} \, , \quad \text{(bond points from $a$ to $b$)} \, . 
\end{align}
Now suppose the bond points from $b$ to $a$, such that,
\begin{align}
	F_{h,g}(a,b) = \left(\frac{ \theta_{b,a}^h  \theta_{h b, h a}^g}{ \theta_{b, a}^g  \theta_{g b, g a}^h} \right)^* \, , \quad \text{(bond points from $b$ to $a$)} .
\end{align}
It follows immediately from the above that
\begin{align}
	F_{h,g}(a,b) = \left( \frac{\omega(\bg \bh \bar b, g, h)\omega(\bg \bh \bar a, h, g)}{\omega(\bg \bh \bar a, g, h)\omega(\bg \bh \bar b, h, g)} \right)^* \, ,\quad \text{(bond points from $b$ to $a$)} ,
\end{align}
which is in fact the \emph{same} expression as for when the bond points from $a$ to $b$. We thus conclude,
\begin{align}
	\boxed{
	F_{h,g}(a,b) = \frac{\omega(\bg \bh \bar a, g, h)\omega(\bg \bh \bar b, h, g)}{\omega(\bg \bh \bar b, g, h)\omega(\bg \bh \bar a, h, g)}} \, , \label{eq:F-relation}
\end{align}
irrespective of the orientation of the bond connecting the plaquettes $a$ and $b$. 

Having made these observations about $F_{h,g}$, we may return to the explicit evaluation of $Q_{g,h,k}$. Let us first focus on the phases in the product over the intermediate plaquettes $l$. Using Eq.~\eqref{eq:F-relation}, we find that the product over $l$ therefore telescopes out,
\begin{align}
	\prod_{l=1}^{M-2} F_{h,g}(q_l,q_{l+1}) = \prod_{l=1}^{M-2} \frac{\omega(\bg \bh \bar q_l, g, h) \omega(\bg \bh \bar q_{l+1}, h, g)}{\omega(\bg \bh \bar q_{l+1}, g, h) \omega(\bg \bh \bar q_l, h, g)} = \frac{\omega(\bg \bh \bar q_1, g, h) \omega(\bg \bh \bar q_{M-1}, h, g)}{\omega(\bg \bh \bar q_{M-1}, g, h) \omega(\bg \bh \bar q_1, h, g)} \, ,
\end{align}
Similarly,
\begin{align}
	\prod_{l=1}^{M-2} F_{h,g}(\bg q_l' , \bg q_{l+1}')^{-1}= 
	\prod_{l=1}^{M-2} \left( \frac{\omega(\bh \bar q_l', g, h) \omega(\bh \bar q_{l+1}', h, g)}{\omega(\bh \bar q_{l+1}', g, h) \omega(\bh \bar q_l', h, g)} \right)^{-1} = \frac{\omega(\bh \bar q_{M-1}', g, h) \omega(\bh \bar q_1', h, g)}{\omega(\bh \bar q_1', g, h) \omega(\bh \bar q_{M-1}', h, g)} \, .
\end{align}
Note that all the phase factors from the intermediate plaquettes cancel out, leaving only the contributions from the endpoints of the boundary between $A_1$ and $B_1$. 

Moving on to the contributions from the endpoints, we again use Eq.~\eqref{eq:F-relation} to find,
\begin{align}
	\frac{F_{h,g}(q_{M-1},q_M)}{F_{h,g}(\bg q_{M-1}',q_M)} = 
	\frac{\omega(\bg \bh \bar q_{M-1}, g, h) }{\omega(\bg \bh \bar q_{M-1}, h, g)}
	\frac{\omega(\bh \bar q_{M-1}', h, g)}{\omega(\bh \bar q_{M-1}', g, h) }
\end{align}
Combining these results so far, we find,
\begin{align}
	Q_{g,h,k}\left(\{q_{l,a}\}, s_a \right) = \frac{F_{h,g}(s,q_1)}{F_{h,g}(ks, \bg q_1')}  \frac{\omega(\bg \bh \bar q_1, g, h) \omega(\bh \bar q_1', h, g)}{\omega(\bg \bh \bar q_1, h, g) \omega(\bh \bar q_1', g, h)} \, .
\end{align}
Again using Eq.~\eqref{eq:F-relation}, we have that
\begin{align}
	\frac{F_{h,g}(s,q_1)}{F_{h,g}(ks, \bg q_1')} = \frac{\omega(\bg \bh \bar s, g, h) \omega(\bg \bh \bar q_{1}, h, g)}{\omega(\bg \bh \bar q_{1}, g, h) \omega(\bg \bh \bar s, h, g)} \frac{\omega(\bh \bar q_{1}', g, h) \omega(\bg\bh \bk \bar s, h, g)}{\omega(\bg \bh \bk \bar s, g, h) \omega(\bh \bar q_{1}', h, g)} \, .
\end{align}
Altogether,
\begin{align}
	\begin{split}
	Q_{g,h,k}\left(\{q_{l,a}\}, s_a \right) 
	&= \frac{\omega(\bg \bh \bar s, g, h) \omega(\bg\bh \bk \bar s, h, g) }{\omega(\bg \bh \bar s, h, g) \omega(\bg \bh \bk \bar s, g, h)} \, .
	\end{split}
\end{align}
We see that every phase factor contribution from the boundary, except that at the trijunction between $A_1$, $B_1$, and $C_1$, has canceled out. For later convenience, we define $a = \bg \bh \bk \bar s$, so that
\begin{align}
	\boxed{
	Q_{g,h,k}\left(\{q_{l,a}\}, s_a \right) = \frac{\omega(ak, g, h) \omega(a, h, g) }{\omega(ak, h, g) \omega(a, g, h)} } \, .
\end{align}
In particular, $Q\left(\{q_{l,a}\}, s_a \right)$ in fact only depends on $s$.

\subsubsection{Phase Contribution from the $A_1 / C_1$ Boundary}

The computations of the phase factor contributions from the remaining two boundaries proceed similarly. Let us now turn to the boundary between $A_1$ and $C_1$. We have,
\begin{align}
	P_{g,h,k}\left(\{p_{l,a}\}, s_a \right) = & \frac{T_{s_1,p_{1,1}}^g T_{s_4,p_{1,4}}^g}{T_{\bg s_2, \bg p_{1,2}}^g T_{\bg s_3,\bg p_{1,3}}^g} \frac{T_{p_{1,1},s_1}^k T_{p_{1,3},s_3}^k}{T_{\bk p_{1,2},\bk s_2}^k T_{\bk p_{1,4},\bk s_4}^k} \prod_{l=1}^{L-1} \frac{T_{p_{l,1},p_{l+1,1}}^g T_{p_{l,4},p_{l+1,4}}^g}{T_{\bg p_{l,2},\bg p_{l+1,2}}^g T_{\bg p_{l,3},\bg p_{l+1,3}}^g} \frac{T_{p_{l+1,1},p_{l,1}}^k T_{p_{l+1,3},p_{l,3}}^k}{T_{\bk p_{l+1,2},\bk p_{l,2}}^k T_{\bk p_{l+1,4},\bk p_{l,4}}^k}
\end{align}
Once again, note that the orientations of the phase factors arising from the partial $g$ and $k$ symmetries are \emph{opposite} to one another.
As before, the plaquette $s$ intersects $A_1$, $B_1,$ and $C_1$, whereas the plaquette $p_L$ intersects $A_1$, $B_2$, and $C_1$, while the plaquettes $p_{1 \leq l < L}$ only intersect $A_1$ and $C_1$; hence, the constraints of Eq.~\eqref{eq:kronecker-deltas-III} imply,
\begin{align}
\begin{cases}
	s_1 \equiv s \\
	s_2 = gk s \\
	s_3 = gh s \\
	s_4 = hk s
\end{cases} \qquad
\begin{cases}
	p_{L,1} \equiv p_L \\
	p_{L,2} = gk p_L \\
	p_{L,3} = g p_L \\
	p_{L,4} = k p_L
\end{cases} \qquad
\begin{cases}
	p_{l,2} = g k p_{l,1} \equiv gk p_l \\
	p_{l,3} = g \bk p_{l,4} \equiv g \bk p_l' \, .
\end{cases} \label{eq:P-constraints}
\end{align}
This yields,
\begin{align}
    \begin{split}
	P_{g,h,k}\left(\{p_{l,a}\}, s_a \right) =& \frac{T_{s,p_{1}}^g T_{hk s,p_{1}'}^g}{T_{k s, k p_{1}}^g T_{hs ,\bk p_{1}'}^g} \frac{T_{p_{1},s}^k T_{g\bk p_{1}',ghs}^k}{T_{g p_{1},gs}^k T_{\bk p_{1}',hs}^k} \prod_{l=1}^{L-2} \frac{T_{p_{l},p_{l+1}}^g T_{p_{l}',p_{l+1}'}^g}{T_{k p_{l}, k p_{l+1}}^g T_{\bk p_{l}',\bk p_{l+1}'}^g} \frac{T_{p_{l+1},p_{l}}^k T_{g\bk p_{l+1}',g\bk p_{l}'}^k}{T_{g p_{l+1},g p_{l}}^k T_{\bk p_{l+1}',\bk p_{l}'}^k}   \\
	&\qquad\times \frac{T_{p_{L-1},p_{L}}^g T_{p_{L-1}',kp_{L}}^g}{T_{k p_{L-1}, k p_{L}}^g T_{\bk p_{L-1}',p_{L}}^g} \frac{T_{p_{L},p_{L-1}}^k T_{g p_{L},g\bk p_{L-1}'}^k}{T_{g p_{L},g p_{L-1}}^k T_{p_{L},\bk p_{L-1}'}^k} \, .
    \end{split}
\end{align}
As before, it is convenient to rearrange the terms in this expression as,
\begin{align}
	\notag P_{g,h,k}\left(\{p_{l,a}\}, s_a \right) = &\frac{T_{s,p_1}^g T_{p_1,s}^k}{T_{gp_1,gs}^k T_{ks,kp_1}^g} \frac{T_{g\bk p_1',ghs}^k T_{hks,p_1'}^g}{T_{hs,\bk p_1'}^g T_{\bk p_1',hs}^k} 
	\prod_{l=1}^{L-2} \frac{T_{p_l,p_{l+1}}^g T_{p_{l+1},p_l}^k}{T_{gp_{l+1},gp_l}^k T_{kp_l,kp_{l+1}}^g} \frac{T_{g\bk p_{l+1}',g\bk p_l'}^k T_{p_l',p_{l+1}'}^g }{T_{\bk p_l', \bk p_{l+1}'}^g T_{\bk p_{l+1}',\bk p_l'}^k} \\
	\notag &\qquad\times	\frac{T_{p_{L-1},p_{L}}^g T_{p_{L},p_{L-1}}^k}{T_{g p_{L},g p_{L-1}}^k T_{k p_{L-1}, k p_{L}}^g} \frac{T_{g p_{L},g\bk p_{L-1}'}^k T_{p_{L-1}',kp_{L}}^g}{T_{\bk p_{L-1}',p_{L}}^g T_{p_{L},\bk p_{L-1}'}^k} \\
	=& \frac{F_{g,k}(s,p_1)}{F_{g,k}(hs, \bk p_1')} \prod_{l=1}^{L-2} \frac{F_{g,k}(p_l,p_{l+1})}{F_{g,k}(\bk p_l' , \bk p_{l+1}')} \frac{F_{g,k}(p_{L-1},p_L)}{F_{g,k}(\bk p_{L-1}',p_L)} \, .
\end{align}
On inspection of this expression, we see that,
\begin{align}
	P_{g,h,k}\left(\{p_{l,a}\}, s_a \right) = Q_{k,g,h}\left(\{p_{l,a}\}, s_a \right) \, ,
\end{align}
which is to say the expression for $P$ is identical to that for $Q$, under the formal substitutions, $(g,h,k) \to (k,g,h)$, $q_l \to p_l$, $q_l' \to p_l'$, and $M \to L$. We may then immediately conclude,
\begin{align}
	\boxed{
	P_{g,h,k}\left(\{p_{l,a}\}, s_a \right) = \frac{\omega(ah,k,g)}{\omega(ah,g,k)}\frac{\omega(a,g,k)}{\omega(a,k,g)} } \, ,
\end{align}
where we recall we previously set $a = \bg \bh \bk \bar s$.

\subsubsection{Phase Contribution from the $B_1 / C_1$ Boundary}

Finally, the treatment of the boundary between $B_1$ and $C_1$ largely follows the same manipulations, aside from some initial massaging of the phase factors. Explicitly, 
\begin{align}
	\begin{split}
	R_{g,h,k}\left(\{r_{l,a}\}, s_a \right) = \frac{T_{r_{1,1},s_1}^h T_{r_{1,2}, s_2}^h}{T_{\bh r_{1,3}, \bh s_3}^h T_{\bh r_{1,4}, \bh s_4}^h} \frac{T_{s_1,r_{1,1}}^k T_{s_3,r_{1,3}}^k}{T_{\bk s_2,\bk r_{1,2}}^k T_{\bk s_4,\bk r_{1,4}}^k} \prod_{l=1}^{N-1} \frac{ T_{r_{l+1,1},r_{l,1}}^h  T_{r_{l+1,2},r_{l,2}}^h}{T_{\bh r_{l+1,3},\bh r_{l,3}}^h T_{\bh r_{l+1,4},\bh r_{l,4}}^h}  \frac{T_{r_{l,1},r_{l+1,1}}^k T_{r_{l,3},r_{l+1,3}}^k}{T_{\bk r_{l,2},\bk r_{l+1,2}}^k T_{\bk r_{l,4},\bk r_{l+1,4}}^k}   \, .
	\end{split}
\end{align}
As before the plaquette $s$ intersects $A_1$, $B_1,$ and $C_1$, whereas the plaquette $r_N$ intersects $A_2$, $B_1$, and $C_1$, while the plaquettes $r_{1 \leq l < N}$ only intersect $B_1$ and $C_1$; hence, the constraints of Eq.~\eqref{eq:kronecker-deltas-III} enforce,
\begin{align}
\begin{cases}
	s_1 \equiv s \\
	s_2 = gk s \\
	s_3 = gh s \\
	s_4 = hk s
\end{cases} \qquad
\begin{cases}
	r_{N,1} \equiv r_N \\
	r_{N,2} = k r_N \\
	r_{N,3} = h r_N \\
	r_{N,4} = hk r_N
\end{cases} \qquad
\begin{cases}
	r_{l,4} = h k r_{l,1} \equiv hk r_l \\
	r_{l,2} = \bh k r_{l,3} \equiv \bh k r_l' \, .
\end{cases} \label{eq:R-constraints}
\end{align}
This yields,
\begin{align}
	R_{g,h,k}\left(\{r_{l,a}\}, s_a \right) = & \frac{T_{r_{1},s}^h  T_{\bh k r_{1}', gk s}^h}{T_{\bh r_{1}', g s}^h T_{k r_{1}, k s}^h}   \frac{T_{s,r_{1}}^k T_{ghs,r_{1}'}^k}{T_{gs,\bh r_{1}'}^k T_{hs,h r_{1}}^k}  \prod_{l=1}^{N-2} \frac{ T_{r_{l+1},r_{l}}^h  T_{\bh k r_{l+1}', \bh k r_{l}'}^h }{ T_{\bh r_{l+1}',\bh r_{l}'}^h T_{k r_{l+1},k r_{l}}^h }   \frac{T_{r_{l},r_{l+1}}^k T_{r_{l}',r_{l+1}'}^k}{T_{\bh r_{l}',\bh r_{l+1}'}^k T_{h r_{l},h r_{l+1}}^k}  \\
	\notag &\qquad \times \frac{T_{r_{N},r_{N-1}}^h  T_{k r_{N}, \bh k r_{N-1}'}^h}{  T_{r_{N},\bh r_{N-1}'}^h T_{k r_{N},k r_{N-1}}^h}\frac{T_{r_{N-1},r_{N}}^k T_{r_{N-1}',h r_{N}}^k}{T_{\bh r_{N-1}',r_{N}}^k T_{h r_{N-1},h r_{N}}^k} \, .
\end{align}
As before, we rearrange terms to write,
\begin{align}
	R_{g,h,k}\left(\{r_{l,a}\}, s_a \right) = &  \frac{T_{r_{1},s}^h T_{s,r_{1}}^k}{T_{h s,h r_{1}}^k T_{k r_{1}, k s}^h}  \frac{T_{ghs,r_{1}'}^k T_{\bh k r_{1}', gk s}^h}{T_{\bh r_{1}', g s}^h T_{gs,\bh r_{1}'}^k}
	 \prod_{l=1}^{N-2} \frac{T_{r_{l+1},r_{l}}^h T_{r_{l},r_{l+1}}^k}{T_{h r_{l},h r_{l+1}}^k T_{k r_{l+1},k r_{l}}^h}
	  \frac{T_{r_{l}',r_{l+1}'}^k T_{\bh k r_{l+1}', \bh k r_{l}'}^h}{T_{\bh r_{l+1}',\bh r_{l}'}^h T_{\bh r_{l}',\bh r_{l+1}'}^k} \\
	\notag &\qquad \times
	   \frac{T_{r_{N},r_{N-1}}^h T_{r_{N-1},r_{N}}^k}{T_{h r_{N-1},h r_{N}}^k T_{k r_{N},k r_{N-1}}^h} 
	   \frac{T_{r_{N-1}',h r_{N}}^k T_{k r_{N}, \bh k r_{N-1}'}^h}{T_{r_{N},\bh r_{N-1}'}^h T_{\bh r_{N-1}',r_{N}}^k } \\
	  \notag =& \frac{F_{h,k}(r_1,s)}{F_{h,k}(\bh r_1',gs)} \prod_{l=1}^{N-2} \frac{F_{h,k}(r_{l+1},r_l)}{F_{h,k}(\bh r_{l+1}',\bh r_l' )} \frac{F_{h,k}(r_N,r_{N-1})}{F_{h,k}(r_N, \bh r_{N-1}')}
\end{align}
This is not quite in the form of $Q$---note that the ordering of the arguments of the $F_{h,k}$ is reversed relative to those of $Q$. However, we see from Eq.~\eqref{eq:F-relation} that we have, 
\begin{align}
    F_{h,k}(a,b) = F_{k,h}(b,a) \, .
\end{align}
Applying this relation, we can rewrite $R$ as,
\begin{align}
	\begin{split}
	R_{g,h,k}\left(\{r_{l,a}\}, s_a \right) =& \frac{F_{k,h}(s,r_1)}{F_{k,h}(gs, \bh r_1')} \prod_{l=1}^{N-2} \frac{F_{k,h}(r_l,r_{l+1})}{F_{k,h}(\bh r_l', \bh r_{l+1}')} \frac{F_{k,h}(r_{N-1},r_N)}{F_{k,h}(\bh r_{N-1}',r_N)} \, .
	\end{split}
\end{align}
From this expression, we observe that,
\begin{align}
	R_{g,h,k}\left(\{r_{l,a}\}, s_a \right) = Q_{h,k,g}\left(\{r_{l,a}\}, s_a \right) 
\end{align}
and so we immediately deduce,
\begin{align}
	\boxed{
	R_{g,h,k}\left(\{r_{l,a}\}, s_a \right) = \frac{\omega(ag,h,k)}{\omega(ag,k,h)}\frac{\omega(a,k,h)}{\omega(a,h,k)} } \, , 
\end{align}
where we recall $a = \bg \bh \bk \bar s$.

\subsubsection{Combining all Phase Contributions}

We have simplified the phase $\tilde{\oT}_{g,h,k}$ to the point that it has a dependence on just the plaquette of qudits at the $A_1,B_1,C_1$ trijunction, labelled by $s$. We can massage this expression to remove this final dependence on this plaquette. We have that,
\begin{align}
	\begin{split}
	\tilde{\oT}_{g,h,k}\left(\{p_{l,a}\},\{q_{l,a}\},\{r_{l,a}\}, s_a \right) &= \frac{\omega(ah,k,g)}{\omega(ah,g,k)}\frac{\omega(a,g,k)}{\omega(a,k,g)}
	\frac{\omega(ak, g, h) \omega(a, h, g) }{\omega(ak, h, g) \omega(a, g, h)}
	 \frac{\omega(ag,h,k)}{\omega(ag,k,h)}\frac{\omega(a,k,h)}{\omega(a,h,k)} \\
	 &= \left( \frac{\omega(a,k,h)}{\omega(a,k,g)}\frac{\omega(ak,g,h)}{\omega(ak,h,g)} \right) \left( \frac{\omega(a,h,g)}{\omega(a,h,k)}  \frac{\omega(ah,k,g)}{\omega(ah,g,k)} \right) \left( \frac{\omega(a,g,k)}{\omega(a,g,h)}\frac{\omega(ag,h,k)}{\omega(ag,k,h)} \right)
	 \end{split}
\end{align}
where, in the second line, we have grouped together terms in a suggestive manner. Indeed, applying the cocycle relation, Eq.~\eqref{eq:cocycle} on $(a,k,g,h)$, we find
\begin{align}
	\frac{\omega(ak,g,h)}{\omega(a,k,g)} = \omega(k,g,h)\frac{\omega(a,kg,h)}{\omega(a,k,gh)} \, .
\end{align}
Similarly, the cocycle condition on $(a,k,h,g)$ yields,
\begin{align}
	\frac{\omega(ak,h,g)}{\omega(a,k,h)} = \omega(k,h,g)\frac{\omega(a,kh,g)}{\omega(a,k,hg)},
\end{align}
from which we find,
\begin{align}
	\frac{\omega(a,k,h)}{\omega(a,k,g)}\frac{\omega(ak,g,h)}{\omega(ak,h,g)} = \frac{\omega(k,g,h)}{\omega(k,h,g)} \frac{\omega(a,kg,h)}{\omega(a,kh,g)} \, .
\end{align}
By cyclically permuting $g$, $h$, and $k$ in the above expression, we also find,
\begin{align}
	\frac{\omega(a,h,g)}{\omega(a,h,k)}\frac{\omega(ah,k,g)}{\omega(ah,g,k)} = \frac{\omega(h,k,g)}{\omega(h,g,k)} \frac{\omega(a,hk,g)}{\omega(a,hg,k)} \quad \text{and} \quad
	\frac{\omega(a,g,k)}{\omega(a,g,h)}\frac{\omega(ag,h,k)}{\omega(ag,k,h)} = \frac{\omega(g,h,k)}{\omega(g,k,h)} \frac{\omega(a,gh,k)}{\omega(a,gk,h)} \, .
\end{align}
Substituting these expressions into $\tilde{\oT}$, we find,
\begin{align}
	\tilde{\oT}_{g,h,k}\left(\{p_{l,a}\},\{q_{l,a}\},\{r_{l,a}\}, s_a \right) = \frac{\omega(g,h,k)\omega(k,g,h)\omega(h,k,g)}{\omega(g,k,h)\omega(h,g,k)\omega(k,h,g)} = \mathcal{Z}[T^3_{g,h,k}] \, ,
\end{align}
which is precisely the value of the TQFT partition function on the three-torus with symmetry flux insertions $g$, $h$, and $k$ through the three cycles \cite{tantivasadakarn2017}. Hence,
\begin{align}
	\oT_{g,h,k}(\psi) = \mathcal{Z}[T^3_{g,h,k}] \sum_{\{\fp_{l,a},\fp_{l,a}'\}}  S_{g,h,k}\left(\{\fp_{l,a},\fp_{l,a}'\} \right)  \, .
\end{align}
The magnitude of the order parameter is given by the sum over the plaquette states, $\fp_{l,a}$, $\fp_{l,a}'$, subject to the constraints of Eqs.~\eqref{eq:Q-constraints}, \eqref{eq:P-constraints}, and \eqref{eq:R-constraints} imposed by $S_{g,h,k}$. This amounts to counting the number of independent sums over $\fp_{l,a}$, $\fp_{l,a}'$ subject to these constraints. One readily finds,
\begin{align}
	\boxed{
    \oT_{g,h,k}(\psi) = \mathcal{Z}[T^3_{g,h,k}] |G|^{-|\partial A| -|\partial B| -|\partial C|} } \, ,
\end{align}
where $|\partial A|$ counts the number of plaquettes lying on the boundary of $A$, and likewise for $\partial B$ and $\partial C$. Note that plaquettes are double-counted here, so that a plaquette on the boundary between, say, $A_1$ and $C_2$ contributes $|G|^{-2}$ and a plaquette at a trijunction contributes $|G|^{-3}$. It is readily seen that the magnitude is the same for any choice of $g$, $h$, and $k$. 

\end{widetext}

\end{document}